\documentclass[twocolumn,a4paper,aps,prb,superscriptaddress,nofootinbib,longbibliography]{revtex4-2}
\usepackage{amsmath,amssymb,amsfonts}
\usepackage{braket}
\usepackage[pdftex]{graphicx,color}
\usepackage{epstopdf}
\graphicspath{{./figures/}}
\usepackage[latin1]{inputenc}
\usepackage{times}
\usepackage{mathrsfs}
\usepackage{float}
\usepackage[normalem]{ulem}
\usepackage{empheq}
\usepackage[sort&compress]{natbib}
\usepackage[colorlinks,bookmarks=true,citecolor=blue,linkcolor=blue,urlcolor=blue,breaklinks=true]{hyperref}
\usepackage{simplewick}
\usepackage[latin1]{inputenc}
\usepackage[vcentermath]{youngtab}
\usepackage{theorem}

\newcommand{\be}{\mathrm{e}}

\begin{document}
\title{
Lieb-Schultz-Mattis  constraints for the insulating phases of  the  one-dimensional SU($N$) Kondo lattice model
}
\author{Philippe Lecheminant}
\affiliation{Laboratoire de Physique Th\'eorique et
 Mod\'elisation, CNRS, CY Cergy Paris Universit\'e, Cergy-Pontoise, France}
\author{Keisuke Totsuka}
\affiliation{Center for Gravitational Physics and Quantum Information, 
Yukawa Institute for Theoretical Physics, Kyoto University, Kyoto 606-8502, Japan}
\date{\today}
\begin{abstract}
The nature of the insulating phases of the SU($N$)-generalization of the one-dimensional Kondo lattice model is investigated by means of non-perturbative approaches.  By extending the Lieb-Schultz-Mattis (LSM) argument to multi-component fermion systems with translation and global SU($N$) symmetries, we derive two indices which depend on the filling and the ``SU($N$)-spin'' (representation) of the local moments. These indices strongly constrain possible insulating phases; for instance, when the local moments transform in the $N$-dimensional (defining) representation of SU($N$), 
a featureless Kondo insulator is possible only at filling $f= 1-1/N$.  
To obtain further insight into the insulating phases suggested by the LSM argument, 
we derive low-energy effective theories by adding an antiferromagnetic Heisenberg exchange interaction among the local moments 
[the SU($N$) Kondo-Heisenberg model].  
A conjectured global phase diagram of the SU($N$) Kondo lattice model as a function of the filling and the Kondo coupling is then 
obtained by a combination of different analytical approaches.  
\end{abstract}
\maketitle
\section{Introduction}
\label{sec:intro}

Heavy-fermion materials have attracted much interest over the years as an example of strongly correlated systems which harbor novel phases of matter and quantum phase transitions \cite{Coleman-book-15,Coleman-Julich-lecture-15}.  
In these systems, the interplay between localized magnetic moments from immobile $d$ or $f$-electrons and itinerant conduction electrons is usually described by the Kondo-lattice model (KLM) \cite{Doniach-77,Hewson-book-93}.  In such a model, a lattice of localized magnetic moments  interacts with tight-binding electrons through an antiferromagnetic exchange interaction, the Kondo coupling $J_{\text{K}}$.  The KLM represents the minimal model to investigate the competition between magnetic ordering due to the Ruderman-Kittel-Kasuya-Yosida (RKKY) interaction and the Kondo screening \cite{Doniach-77,Hewson-book-93,Coleman-book-15}.  
At small $J_{\text{K}}$, the long-range RKKY interaction among the localized spins, mediated by conduction electrons, is expected to produce an antiferromagnetic ordered state, whereas for large $J_{\text{K}}$ a paramagnetic insulating phase emerges. The latter is the Kondo insulating phase which is best visualized as a collection of spin-singlet states made between the localized spins and conduction electrons. 

More exotic Kondo insulators have been discussed recently through the interplay between correlation and spin-orbit coupling with the stabilization of a topological Kondo insulating phase \cite{Dzero-S-G-C-10,Dzero-X-G-C-review-16}.  
The strong hybridization between spin-orbit coupled localized $f$-electrons and itinerant $d$-orbital electrons leads to the formation of topological insulating phase with protected metallic Dirac surface states. In this respect, the Kondo insulator SmB$_6$ compound has been predicted to host a three-dimensional topological insulator state with metallic protected Dirac surface states at the X points on the (001) surface (see, e.g., Refs. \onlinecite{Dzero-X-G-C-review-16,Li-S-K-A-20} for a review).  
Other possible candidates of topological Kondo insulators are YbB$_{12}$ and FeSi compounds \cite{Hagiwara-et-al-16,breindel-et-al-23}.
 Another mecanism to stabilize new exotic Kondo insulating phases is to study the generalization of the KLM where the lattice localized spins is replaced by a two-dimensional ${\mathbb Z}_2$ quantum spin liquid such as the Kitaev model on the honeycomb lattice \cite{Kitaev-06} or its variants \cite{Yao-Z-K-2009,Yao-L-2011}. The resulting Kondo-Kitaev models describe various novel quantum phases of matter as topological superconductivity, odd-frequency pair-density wave, and a Kondo phase with order fractionalization \cite{Choi-K-R-K-18,Seifert-M-V-18,Farias-C-M-P-20, Carvalho-T-F-M-21,Tsvelik-C-22,Coleman-P-T-22}.

Here, we will  explore another route to stabilize unconventional Kondo insulating phases by enlarging the SU(2) spin symmetry of the Kondo interaction to SU($N$). This SU($N$) generalization of the KLM  has been originally introduced in the early eighties mainly as a mathematical convenience by furnishing a small parameter $1/N$ which facilitates a controlled large-$N$ expansion about the limit $N \rightarrow \infty$ \cite{Read-N-83,Coleman-83,Read-N-D-84,Auerbach-L-86}. There are now strong physical motivations to study the SU($N$)-symmetric KLM. First of all, ultracold atomic gases 
of alkaline-earth and ytterbium fermions make it possible to simulate SU($N$) Kondo physics in a very controlled fashion \cite{Gorshkov-et-al-10}. 
These atoms have a long-lived singlet ground state $g$ ($^{1} S_0$) and a metastable triplet excited state $e$ ($^{3} P_0$) 
in which the electronic state is decoupled almost perfectly from the nuclear one thereby leading to nuclear-spin-independent atomic collisions. 
This leads then to the experimental realization of fermions with an SU($N$) symmetry 
where $N \leq 2I+1$ ($I$ being the nuclear spin) (see, e.g., Refs. \onlinecite{Cazalilla-R-14} and \onlinecite{Capponi-L-T-16} for reviews).  
Several proposals to realize the SU($N$) KLM exploit a state-dependent optical lattice to selectively localize the $e$ atoms 
whereas the $g$ atoms remain mobile 
thereby playing the role of the conduction fermions \cite{Gorshkov-et-al-10,Foss-Feig-H-R-10,Nakagawa-K-15,Isaev-R-15}.
Some experimental investigations have been made to explore this heavy-fermion physics with two-orbital alkaline-earth fermions \cite{Riegger-D-H-F-B-F-18,Ono-A-H-S-T-21}.   

A second more recent motivation to study SU($N$) heavy fermions problems is the work of Song and Bernevig \cite{Song-B-22}  which describes the physics of twisted bilayer graphene as a topological heavy fermion problem with the hybridization of flat band f-electrons with a topological band of conduction electrons.  This leads to the prediction that the magic-angle twisted bilayer graphene could be described as a SU(8) or SU(4)-symmetric KLM depending on the energy scale \cite{Chou-D-23,Lau-C-23,Hu-B-T-23,Hu-et-al-23,Hu-B-T-23,Zhou-S-23}.

In this paper, we consider the SU($N$) generalization of the SU(2) KLM  in the simplest situation, namely in one dimension 
to determine its insulating phases when $N >2$. The lattice Hamiltonian of the SU($N$)-KLM is defined as follows:
\begin{equation}
\begin{split}
& \mathcal{H}_{\text{KLM}} = \mathcal{H}_{\text{hop}} + \mathcal{H}_{\text{K}}  \; ,  \\ 
& \mathcal{H}_{\text{hop}} :=   - t \sum_{i} \sum_{\alpha=1}^{N} 
 \left( c_{\alpha,\,i}^\dag c_{\alpha,\,i+1}  + \text{H.c.}\right)   \\
&\mathcal{H}_{\text{K}} := J_{\text{K}}  \sum_{i}\left( 
\sum_{A=1}^{N^{2}-1}\hat{s}_{i}^{A} S_{i}^{A}
\right),   \\
\end{split}
\label{eqn:SUN-KLM}
\end{equation} 
where the model consists of two parts which describe respectively the hopping ($\mathcal{H}_{\text{hop}}$) 
of the $N$-component lattice fermion $c_{\alpha,\,i}$ ($\alpha = 1, \ldots, N$) and the Kondo interaction ($\mathcal{H}_{\text{K}}$) 
between the electronic spin density 
\begin{equation}
\hat{s}_{i}^{A}= \sum_{\alpha,\beta=1}^{N} c^{\dagger}_{\alpha,\,i} T^{A}_{\alpha\beta} c_{\beta,\,i} 
\label{eqn:SUN-fermion-spin}
\end{equation}
[with $T^A$ being the SU($N$) generators in the defining representation that are normalized as: $\text{Tr} (T^A T^B) = \delta^{AB}/2$] 
and the localized SU($N$) spin moments $S_{i}^{A}$.    
Although we can in principle think of any irreducible representations [i.e., SU($N$) ``spins''] for $S_{i}^{A}$, 
we mainly consider for simplicity the $N$-dimensional defining representation $\mathbf{N}$ [i.e., $S=1/2$ in SU(2); a physical explanation of 
the $\mathbf{N}$ representation is given in Appendix~\ref{sec:SUN-defining-rep}].    
Here we do not specify the origin of $\alpha\, (=1,\ldots, N)$ that labels different species of fermions; it may come from $N=2I+1$ 
different nuclear-spin states when $c_{\alpha}$ describes fermions of alkaline-earth-like atoms \cite{Cazalilla-H-U-09,Gorshkov-et-al-10}, 
or from spin-orbit-coupled $J$-multiplets ($N=2J+1$) in heavy-fermion systems \cite{Coqblin-S-69,Coleman-book-15}.   

\begin{figure}[htb]
\begin{center}
\includegraphics[width=\columnwidth,clip]{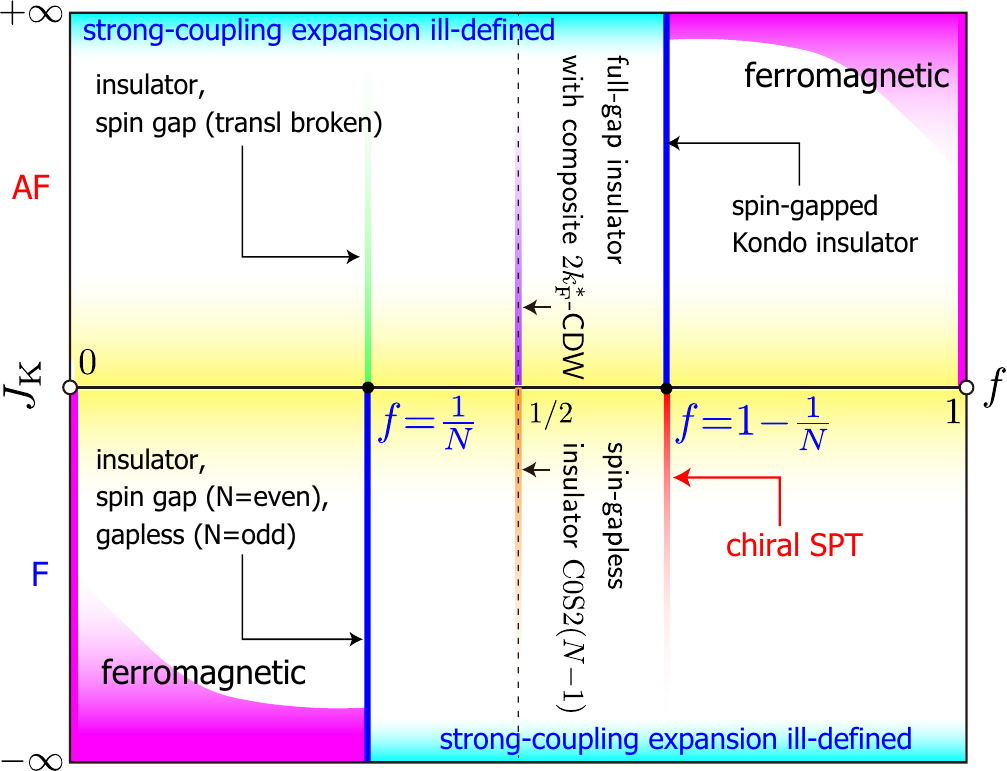}
\end{center}
\caption{ 
A conjectured phase diagram of SU($N$) Kondo lattice model \eqref{eqn:SUN-KLM} in 1D derived from different analytical approaches.  
Translation-invariant insulators are possible only at commensurate fillings $f=m/N$ ($m=1,\ldots,N-1$), whose properties strongly 
depend on $f$ and (the parity of) $N$.  The ground state at $f=1-1/N$ is a featureless full-gap Kondo insulator when $J_{\text{K}}>0$, 
while an inversion-breaking (chiral) SPT phase appears for $J_{\text{K}} <0$.   
The insulating phase at $f=1/N$ ($J_{\text{K}}<0$) exhibits very different behaviors depending on the parity of $N$ which can be 
captured by an effective non-linear sigma model on the flag manifold. The insulator on the $J_{\text{K}}>0$ side has a spin gap 
and breaks translation regardless of the parity of $N$.  
The ferromagnetic phases (highlighted in magenta) extend to the lower-density (when $J_{\text{K}}<0$) 
or higher-density ($J_{\text{K}}>0$) side of these insulating phases.   
The descriptions for $f=1/2$ ($N$-even) are valid for $N \geq 6$. 
\label{fig:SUN-KLM-phase-diag}
}
\end{figure}

The antiferromagnetic ($J_{\text{H}} > 0$) Heisenberg exchange interaction among the localized spins 
could also be added to define the SU($N$)-Kondo-Heisenberg model (KHM):
\begin{equation}
\begin{split}
& \mathcal{H}_{\text{KHM}} = \mathcal{H}_{\text{hop}} + \mathcal{H}_{\text{K}} + \mathcal{H}_{\text{H}}\\
& =   - t \sum_{i} \sum_{\alpha=1}^{N} 
   \left( c_{\alpha,\,i}^\dag c_{\alpha,\,i+1}  + \text{H.c.}\right)   
+ J_{\text{K}}  \sum_{i}\left( 
\sum_{A=1}^{N^{2}-1}\hat{s}_{i}^{A} S_{i}^{A}
\right)  \\
& \phantom{=} 
+ J_{\text{H}}  \sum_{i}\left( 
\sum_{A=1}^{N^{2}-1} S_{i}^{A} S_{i+1}^{A}
\right) 
   \; .
\end{split}  
\label{eqn:SUN-KHM}
\end{equation} 
These two models conserve $(N-1)$ quantities associated to the global SU($N$)-symmetry [generalization of the total $S^{z}$ in SU(2)] as well as 
the total electron number 
\[
\sum_{i=1}^{L}\sum_{\alpha=1}^{N} \hat{n}_{\alpha,i} \quad (  \hat{n}_{\alpha,i} := c_{\alpha,\,i}^\dag c_{\alpha,\,i} ) \; .
\]
From this, we define {\em filling} $f$ ($0 \leq f \leq 1$) as: 
\begin{equation}
f := \frac{1}{NL} \sum_{i=1}^{L}\sum_{\alpha=1}^{N} \hat{n}_{\alpha,i} \; .
\end{equation}   
One of the important differences from the usual $N=2$ case is that the model \eqref{eqn:SUN-KLM} or \eqref{eqn:SUN-KHM} does {\em not} possess 
the particle-hole symmetry\footnote{%
The particle-hole transformation takes the model \eqref{eqn:SUN-KLM} at the filling $f$ to {\em another} model with the local moments in the conjugate representation at $1-f$.} 
except when the local moments $S_{i}^{A}$ are in the self-conjugate representations \cite{Totsuka-23}.   
This is clearly seen in the asymmetry of Fig.~\ref{fig:SUN-KLM-phase-diag} with respect to the half-filling $f=1/2$.

The zero-temperature phase diagram of the one-dimensional (1D) KLM or KHM is well understood in the $N=2$ case 
(see, e.g., Refs.~\onlinecite{Tsunetsugu-S-U-97,Shibata-U-99,Gulacsi-review-04} for reviews of the 1D KLM). 
On top of dominant ferromagnetic or paramagnetic metallic phases, there are several insulating phases depending on the filling and the sign of $J_{\text{K}}$.  
At half-filling $f=1/2$, an insulating Kondo-singlet phase is stabilized for an antiferromagnetic Kondo coupling ($J_{\text{K}} > 0$) 
in both models  where each localized spins binds with a conduction electron 
into a spin singlet \cite{Tsunetsugu-S-U-97,Zachar-K-E-96,Sikkema-A-W-97,Fujimoto-K-97,Shibata-U-99,Zachar-T-01,Chen-S-K-C-24}.  
For a ferromagnetic Kondo interaction ($J_{\text{K}} < 0$), the insulating ground state is replaced by 
a symmetry-protected topological (SPT) phase which is equivalent to the Haldane phase of the spin-1 antiferromagnetic Heisenberg 
spin chain \cite{Tsunetsugu-H-U-S-92,Fujimoto-K-97}.
At quarter-filling $f=1/4$, the insulating phase is dimerized exhibiting the coexistence of local-spin dimerization 
and Peierls-like ordering for the conduction electrons \cite{Xavier-P-M-A-03,Xavier-M-08,Huang2020},  
whereas, at filling $f=3/8$, a charge-density wave (CDW) is stabilized \cite{Huang-S-T-19}. 

In stark contrast to $N=2$, very little is known for the global phase diagram of the KLM (\ref{eqn:SUN-KLM}) or KHM  (\ref{eqn:SUN-KHM}) with $N>2$ when the localized spin operators $S_{i}^{A}$ on the $i$-th site transforms in the ${\bf N}$ of the SU($N$) group.
A recent strong-coupling analysis of the SU($N$) KLM in Ref.~\cite{Totsuka-23} reveals a rich phase diagram depending on the electronic filling $f$ 
when $|J_{\text{K}}|$ is sufficiently large.
A ferromagnetic metallic phase emerges in the KLM in the low-density (respectively high-density) regime 
when $J_{\text{K}} < 0$  (respectively $J_{\text{K}} > 0$).  Moreover, two insulating phases were identified 
by means of the strong-coupling expansion.  The first one, at filling $f = 1 -1/N$ with sufficiently strong antiferromagnetic $J_{\text{K}}$,  
is a fully gapped SU($N$) Kondo-singlet phase which is a generalization of the similar one found at $f=1/2$ in the ordinary ($N=2$) KLM.  
In this phase, $N-1$ conduction electrons form a {\em site-centered} SU($N$) spin-singlet with the localized magnetic moment on the same site.
The second insulating phase is found at another filling $f=1/N$ and for a sufficiently strong ferromagnetic $J_{\text{K}}$.  
When $N$ is odd, the spin degrees of freedom are gapless whereas they are fully gapped in the even-$N$ case \cite{Totsuka-23}. 
For other commensurate fillings, including the half-filled case ($f=1/2$), no conclusion can be derived from the strong-coupling approach.

In this paper, we determine the nature of the insulating phases of the SU($N$) KLM  (\ref{eqn:SUN-KLM}) and SU($N$) KHM (\ref{eqn:SUN-KHM}) for general commensurate filling $f=m/N$ ($m=1, \ldots, N-1$) by means of non-perturbative approaches. Symmetries together with the filling fraction impose strong non-perturbative constraints on the possible phases realized in a microscopic lattice model  as emphasized 
by the Lieb-Schultz-Mattis (LSM) theorem and its generalization \cite{Lieb-S-M-61,Affleck-L-86,Oshikawa-00,Hastings-04,Watanabe-P-V-Z-15,Ogata-T-19}.  One of the important messages from the LSM theorem is the impossibility in one dimension to get featureless insulating phase for noninteger fillings in quantum systems with translation invariance and global U(1) symmetry, 
enforcing gapless or gapped symmetry-broken ground states as the only possible infrared (IR) behaviors \cite{Yamanaka-O-A-97}.  
We extend the LSM argument to fermionic systems with translation and global SU($N$) symmetries, i.e., the SU($N$) KLM 
and its variant SU($N$) KHM.  
We find that fully gapped translationally-invariant insulators are possible only for a filling  $f= 1 - 1/N$. For other commensurate fillings, a variety of different insulating phases with gapless spin degrees of freedom or multiple ground states with broken translation symmetry is predicted 
from the LSM argument depending on $N$ and the filling $f$ as summarized in Table~\ref{tab:KHM-phases-by-LSM}. 
In the case of the KHM,  a low-energy approach can be derived by exploiting the existence of a spin-exchange $J_{\text{H}} \ne 0$ between the SU($N$) localized spins to derive a continuum description. The interplay between the global internal SU($N$) and lattice translation symmetries 
in this field theory leads to a non-perturbative index that enables us to constrain possible low-energy field theories 
via the 't Hooft anomaly matching condition \cite{Yao-H-O-19}.
To be specific, we identify an index $\mathcal{I}_{1} = f + 1/N$ (mod $1$) which excludes, e.g., featureless spin-gapped insulator 
when $\mathcal{I}_{1}$ is not an integer, in full agreement with the LSM approach.   

Being independent of the details of the models, the LSM argument does not tell much about the actual phase structures and 
the properties of the ground states. To go further, we carry out careful low-energy field-theory analyses guided by the 't Hooft anomaly matching 
condition to identify various insulating phases in the weak-coupling regions of the SU($N$) KHM as shown in Table~\ref{tab:KHM-phases}.  
The principal phases include the followings (see also Fig.~\ref{fig:SUN-KLM-phase-diag}).  

At filling $f=1-1/N$, we find two fully gapped translationally-symmetric insulators, i.e., 
(i) the SU($N$)-singlet Kondo insulator for $J_{\text{K}} >0$ [see Fig.~\ref{fig:SU4-chiral-SPT-vs-Kondo-ins}(a)] and 
(ii) the SPT phase that spontaneously breaks inversion symmetry (dubbed a chiral SPT phase) for $J_{\text{K}} <0$ 
[Fig.~\ref{fig:SU4-chiral-SPT-vs-Kondo-ins}(b)].   
As the Kondo insulator appears quite naturally also in the strong-coupling region, we expect 
that it persists for all $J_{\text{K}} \,(>0)$.  In contrast, as the Kondo coupling alone does not stabilize any particular SU($N$) spin 
at each site when $J_{\text{K}}$ is strongly negative, presumably the chiral SPT phase might exist only at weak couplings.  

At filling $f=1/N$, on the other hand, the ground state depends not only on the sign of $J_{\text{K}}$ but also on the parity of $N$. 
When $N$ is even, the ground state is a full-gap spin-singlet insulator with broken translation symmetry, whose structure differs  
according to the sign of $J_{\text{K}}$ (see Figs.~\ref{fig:SUN-GS-positive-Jk} and \ref {fig:SUN-GS-negative-Jk}).  
When $N$ is odd, the system is insulating regardless of the sign of $J_{\text{K}}$, 
whereas the nature is very different for $J_{\text{K}}>0$ and $J_{\text{K}} <0$; 
when $J_{\text{K}}$ is ferromagnetic, the spin sector remains gapless (the same universality class as the integrable SU($N$) Heisenberg spin chain 
\cite{Sutherland-75}), while we find a fully gapped phase with coexisting valence-bond-solid and charge-density-wave (CDW) 
orders when $J_{\text{K}}>0$.  

For generic commensurate fillings $f = m/N, \; m\ne 1, N-1$, the insulating phases can be spin-gapless (when 
$J_{\text{K}} < 0$; as in the $f = 1/N$ case) or fully gapped ($J_{\text{K}}  > 0$). 
In the latter case, the insulating phase spontaneously breaks the translation symmetry leading to ground-state degeneracy which depends on $N$ 
(see Table \ref{tab:KHM-phases-by-LSM}).  There, a long-range composite-CDW is stabilized 
which is associated to the hybridization between the itinerant electron and a spin-polaron bound state formed by the electrons  
and the localized spin moments.  
The characteristic momentum of the  order parameter takes a renormalized value 
$2k^{*}_{\text{F}} =  \tfrac{2m\pi}{Na_0} +  \tfrac{2\pi}{N a_0} $ since the localized-spin fluctuations (with momentum $\tfrac{2\pi}{N a_0} $) 
now participate in the formation of this composite object.  
In the half-filled ($f=1/2$) case and even $N$ ($N \geq 6$), the ground-state degeneracy of the fully gapped composite CDW phase 
for $J_{\text{K}}  > 0$ depends on the parity of $N/2$. When $N/2$ is odd (respectively even), the ground-state degeneracy is $N/2$ 
(respectively $N$).  For a ferromagnetic Kondo coupling ($J_{\text{K}} < 0$), the insulating phase is spin gapless
with $2(N-1)$ gapless modes.   
For $N=4$, we do not have any decisive conclusion for the ground state so far. 

For the other rational fillings $f=p/q$ [$p$ and $q\,(\neq N)$ are coprime integers], the system is either a metal or 
an insulator with $q/\text{gcd}(N,q)$ [with $\text{gcd}(N,q)$ denoting the greatest common divisor between $N$ and $q$] 
degenerate ground states associated with spontaneously broken translation symmetry.  
Note that when this happens, opening of a charge gap and breaking of translation symmetry must occur {\em simultaneously}.  

Combining the results obtained in this paper and those from the strong-coupling analyses \cite{Totsuka-23}, 
we conjecture the global phase diagram of the SU($N$) KLM \eqref{eqn:SUN-KLM} and KHM \eqref{eqn:SUN-KHM} 
shown in Fig.~\ref{fig:SUN-KLM-phase-diag}.  
Of course, the detailed structure of the phase diagram (precise locations of the boundaries, etc.) will be different for the models \eqref{eqn:SUN-KLM} 
and \eqref{eqn:SUN-KHM}.  However, since most of our arguments based on non-perturbative indices rely only on the kinematical information 
(e.g., the type of SU($N$) moments, fermion filling, etc.) and are independent of the details of the Hamiltonian, we believe that 
the proposed phase diagram (Fig.~\ref{fig:SUN-KLM-phase-diag}) correctly captures the structure common to the two models.

 The rest of the paper is structured as follows. In Sec.~\ref{sec:LSM}, we present our LSM argument on the lattice which gives the constraint 
 to get a translational-invariant featureless insulating phase for the  SU($N$) KLM and SU($N$) KHM. Its field-theory interpretation 
 as an anomaly matching mecanism is investigated in Sec.~\ref{sec:Anomaly} for the SU($N$) KHM.  
 In Sec.~\ref{sec:Weakcoupling}, we analyse the weak-coupling approach to the insulating phases of the latter model 
 to find explicit realization of the possible phases predicted by the LSM theorem. Finally, a summary of the main results 
 is given in Sec.~\ref{sec:Conclusion} together with several technical Appendixes.

\section{Lieb-Schultz-Mattis argument}
\label{sec:LSM}
In this section, the LSM approach is applied to the SU($N$) KLM \eqref{eqn:SUN-KLM} and KHM
\eqref{eqn:SUN-KHM} to derive non-perturbative constraints on the properties of the insulating phases these models host 
by exploiting the SU($N$) and translational symmetries.  
See Ref.~\cite{Hazra-C-21} for a similar approach (i.e., flux insertion) to the Fermi-volume problem of SU($N$) fermion 
systems. 
\subsection{Constructing twist operators}
\label{sec:twist-op}
\subsubsection{Fermion Twist}
\label{sec:fermion-twist}

As twist operations acting on the itinerant fermions ($c^{\dagger}_{\alpha,j}$) and the local moments commute with each other, 
we construct them separately, and then glue them in such a way that they in total create finite-energy excitations. 
Let us begin with the LSM twist $\widehat{\mathcal{U}}_{\alpha}^{(\text{f})}$ for the itinerant fermions. 
For the fermion part, we require that the transformed fermions preserve periodic boundary condition: 
$\widehat{\mathcal{U}}_{\alpha}^{(\text{f})}{}^{\dagger} c^{\dagger}_{\beta,j+L} \widehat{\mathcal{U}}_{\alpha}^{(\text{f})} 
= \widehat{\mathcal{U}}_{\alpha}^{(\text{f})}{}^{\dagger} c^{\dagger}_{\beta,j} \widehat{\mathcal{U}}_{\alpha}^{(\text{f})}$.  
Below, we take  
\begin{equation}
\widehat{\mathcal{U}}_{\alpha}^{(\text{f})} := 
\exp\left\{ i \frac{2\pi}{L} \sum_{j=1}^{L} j \, \hat{n}_{\alpha,j} \right\} \quad (\alpha=1,\ldots, N) 
\label{eqn:U-fermion-twist}
\end{equation}
which transform the fermions as:
\begin{equation}
\widehat{\mathcal{U}}_{\alpha}^{(\text{f})}{}^{\dagger}
c_{\beta,j}^{\dagger} \, 
\widehat{\mathcal{U}}_{\alpha}^{(\text{f})} 
= \be^{-i \frac{2\pi}{L} \delta_{\alpha \beta} j }  c_{\beta,j}^{\dagger}  
\label{eqn:c-on-U-fermion-twist}
\end{equation}
as the set of ``elementary'' twists and consider generic twists of the form 
$(\widehat{\mathcal{U}}_{1}^{(\text{f})})^{m_{1}} \cdots (\widehat{\mathcal{U}}_{N}^{(\text{f})} )^{m_{N}}$ 
specified by the set of integers $(m_{1}, \ldots, m_{N})$ with negative $m_{\alpha} \, (<0)$ being understood 
as $(\widehat{\mathcal{U}}_{\alpha}^{(\text{f})})^{m_{\alpha}} 
:= (\widehat{\mathcal{U}}_{\alpha}^{(\text{f})\dagger})^{|m_{\alpha}|}$.  

Using \eqref{eqn:c-on-U-fermion-twist}, we see that the (non-hermitian) SU($N$) ``spin'' of the itinerant fermions  
$\widehat{\mathcal{S}}^{\mu\nu}_{j} := c_{\mu,j}^{\dagger} c_{\nu,j}$ ($\mu,\nu=1,\ldots, N$), 
that are related to the usual hermitian generators $\widehat{S}^{A}$ as:
$\widehat{\mathcal{S}}^{\mu\nu}_{j} = [S^{A}]_{\nu\mu} \widehat{S}_{j}^{A} + \frac{1}{N} \hat{n}_{j}$,  
transform like:
\begin{equation}
\widehat{\mathcal{U}}_{\alpha}^{(\text{f})}{}^{\dagger} \widehat{\mathcal{S}}^{\mu\nu}_{j} \widehat{\mathcal{U}}_{\alpha}^{(\text{f})}  
= \exp\left\{ -i \frac{2\pi}{L} j (\delta_{\alpha\mu} - \delta_{\alpha\nu} ) \right\} \widehat{\mathcal{S}}^{\mu\nu}_{j}   \; .
\label{eqn:twist-fermion-spin}
\end{equation}
Note that the diagonal generators $\widehat{\mathcal{S}}^{\mu\mu}_{j} = \hat{n}_{\mu,j}$ are invariant under the twist 
as expected.  
By construction, the (second-quantized) SU($N$) spin $\widehat{\mathcal{S}}^{\mu\nu}_{j}$ automatically satisfies 
the periodic boundary condition even after the twist: 
$\widehat{\mathcal{U}}_{\alpha}^{(\text{f})}{}^{\dagger} 
\widehat{\mathcal{S}}^{\mu\nu}_{j+L} \widehat{\mathcal{U}}_{\alpha}^{(\text{f})}  
= \widehat{\mathcal{U}}_{\alpha}^{(\text{f})}{}^{\dagger} 
\widehat{\mathcal{S}}^{\mu\nu}_{j} \widehat{\mathcal{U}}_{\alpha}^{(\text{f})}$.  
\subsubsection{Spin Twist}
\label{sec:spin-twist}
Now let us consider twist operations for the localized spins.  
As has been discussed in the context of SU($N$) spin chains in Ref.~\cite{Affleck-L-86}, 
the $\be^{2\pi i Q_{\alpha,j}}$ generator must commute with all the SU($N$) generators, i.e.,  
$\be^{-2\pi i Q_{\alpha,j}} \mathcal{S}_{j}^{\mu\nu} \be^{2\pi i Q_{\alpha,j}} = \mathcal{S}_{j}^{\mu\nu}$ 
in order for the periodic boundary condition $\mathcal{S}_{j+L}^{\mu\nu}=\mathcal{S}_{j}^{\mu\nu}$ to be preserved by the twist.  
Then, the Schur's lemma dictates that $\be^{2\pi i Q_{\alpha,j}}$ must be a scalar matrix for a fixed 
local SU($N$) moment.   The simplest choice that gives the tightest constraint is  
\begin{equation}
\mathcal{U}_{\alpha}^{(\text{s})}\! (\theta^{(\text{s})}_{\alpha}) := 
\exp\left\{ i \frac{\theta^{(\text{s})}_{\alpha}}{L} \sum_{j=1}^{L} j \, Q_{\alpha,j} \right\} \quad (\alpha=1,\ldots, N)   \; ,
\label{eqn:SUN-LSM-spin-twist-tentative}
\end{equation}
where $Q_{\alpha,j}$ is defined, in the $N$-dimensional defining representation $\mathbf{N}$, by:
\begin{equation}
\begin{split}
& Q_{\alpha,j} := (1/N)\mathbf{1} - \mathbf{e}_{\alpha} \quad \left( [\mathbf{e}_{\alpha}]_{mn} = \delta_{m\alpha} \delta_{n\alpha}   
\right)   \\
&  \text{Tr}(Q_{\alpha,j} ) = 0  \; .
\end{split}
\label{eqn:def-Qalpha}
\end{equation}
The angle $\theta^{(\text{s})}_{\alpha}$ is an integer multiple of $2\pi$ and is to be fixed later.  

Since 
\begin{equation}
[ Q_{\alpha,j} , \mathcal{S}^{\mu\nu}_{j} ] = - (\delta_{\alpha\mu} - \delta_{\alpha\nu} )  \mathcal{S}^{\mu\nu}_{j}  \; ,
\end{equation}
the local spin operators transform like:
\begin{equation}
\mathcal{U}_{\alpha}^{(\text{s})} \! (\theta^{(\text{s})}_{\alpha})^{\dagger}
\mathcal{S}^{\mu\nu}_{j}\, 
\mathcal{U}_{\alpha}^{(\text{s})} \! (\theta^{(\text{s})}_{\alpha}) 
=  \exp\left\{ i \frac{\theta^{(\text{S})}_{\alpha} }{L} j (\delta_{\alpha\mu} - \delta_{\alpha\nu} ) \right\}  \mathcal{S}^{\mu\nu}_{j}  
\label{eqn:twist-local-spin}
\; ,
\end{equation}
which coincides with the twisted fermion spin \eqref{eqn:twist-fermion-spin} aside from the minus sign in the exponent.   
From Eq.~\eqref{eqn:twist-local-spin}, it is obvious that the transformed spins 
$\mathcal{U}_{\alpha}^{(\text{S})} \! (\theta^{(\text{S})}_{\alpha})^{\dagger}
\mathcal{S}^{\mu\nu}_{j}\, 
\mathcal{U}_{\alpha}^{(\text{S})} \! (\theta^{(\text{S})}_{\alpha})$ satisfy the periodic boundary condition 
as they should be. 
\subsection{Twist on Hamiltonian}
\label{sec:twisted-Hamiltonian}
To estimate the excitation energies, we first apply the elementary twist \eqref{eqn:U-fermion-twist} to the hopping term:%
\footnote{%
In the summation of the last line, the term $j=L$ is missing since $\be^{-2\pi i \hat{n}_{\alpha,L}} c^{\dagger}_{\alpha,L}
\be^{2\pi i \hat{n}_{\alpha,L}} = c^{\dagger}_{\alpha,L}$, etc. 
}
\begin{equation}
\begin{split}
& \widehat{\mathcal{U}}_{\alpha}^{(\text{f})}{}^{\dagger} \mathcal{H}_{\text{hop}} \, \widehat{\mathcal{U}}_{\alpha}^{(\text{f})} 
- \mathcal{H}_{\text{hop}}   \\
&=  i \frac{2\pi}{L} \sum_{j =1}^{L} \left[ h^{(\text{hop})}_{j,j+1} \, , \, \frac{1}{2} (\hat{n}_{\alpha,j+1} - \hat{n}_{\alpha,j}) \right]  
+ \text{O}(L^{-1}) \; .
\end{split}
\label{eqn:SUN-hopping-term-variation}
\end{equation}
The precise form of the $\text{O}(L^{-1})$-terms is given by: 
\begin{equation}
\frac{1}{2} t \left( \frac{2\pi}{L} \right)^{2} 
\sum_{i=1}^{L} \sum_{\alpha=1}^{N}  \left(c_{\alpha,\,i}^\dag c_{\alpha,\,i+1}  + \text{H.c.}\right)  \; ,
\end{equation}
which, by translational symmetry, is expected to be of the order $\text{O}(L^{-1})$.  
In deriving \eqref{eqn:SUN-hopping-term-variation}, we have used an identity
\[
\begin{split}
& j \hat{n}_{\alpha,j} + (j+1) \hat{n}_{\alpha,j+1} \\
& = \left( j+\frac{1}{2} \right)(\hat{n}_{\alpha,j+1} + \hat{n}_{\alpha,j}) - \frac{1}{2}(\hat{n}_{\alpha,j} - \hat{n}_{\alpha,j+1})
\end{split}
\]
and $\left[ h_{\text{hop}}^{j,j+1} , (\hat{n}_{\alpha,j+1} + \hat{n}_{\alpha,j}) \right]=0$.   
Note that, due to translation invariance, the leading term in Eq.~\eqref{eqn:SUN-hopping-term-variation} 
is $\text{O}(L^{0})$, in general.  

The variation of the Heisenberg part $\mathcal{H}_{\text{H}}$ 
due to the spin twist $\widehat{\mathcal{U}}_{\alpha}^{(\text{s})}\! (\theta^{(\text{s})}_{\alpha}=2\pi m^{\text{(s)}})$ 
can be calculated similarly: 
\begin{equation}
\begin{split}
& \widehat{\mathcal{U}}_{\alpha}^{(\text{s})}\! (2\pi m^{\text{(s)}})^{\dagger} \, 
 \mathcal{H}_{\text{H}} \, \widehat{\mathcal{U}}_{\alpha}^{(\text{s})} \! (2\pi m^{\text{(S)}})
- \mathcal{H}_{\text{H}}  \\
& = 
 i \frac{2\pi}{L} m^{\text{(s)}} \sum_{k=1}^{L} \left[ h^{(\text{H})}_{k,k+1} \, , \, \frac{1}{2} (Q_{k+1}-Q_{k}) \right]  
 + \text{O}(L^{-1}) \;  ,
\end{split}
\label{eqn:SUN-Heisenberg-term-variation}
\end{equation}
where the integer $m^{\text{(s)}}$ has been introduced to take into account the relative phase between the fermion 
and spin twists.   At this point, the integer $m^{\text{(s)}}$ seems arbitrary. 
However, as is shown in Appendix~\ref{sec:LSM-for-Kondo}, 
it is fixed to $m^{\text{(s)}} = -1$ by requiring that the LSM twist should create $\text{O}(L^{-1})$ excitations.  
Therefore, we are lead to considering the following combination as the elementary twist:
\begin{equation}
\begin{split}
\widehat{\mathcal{U}}_{\alpha} & := 
\widehat{\mathcal{U}}_{\alpha}^{(\text{f})}  \widehat{\mathcal{U}}_{\alpha}^{(\text{s})} (- 2 \pi )  
= \exp\left\{  i \frac{2\pi}{L} \sum_{j=1}^{L} j (\hat{n}_{\alpha,j} - Q_{\alpha,j} ) \right\}  \\
& \quad (\alpha=1,\ldots, N) \; ,
\end{split}
\label{eqn:elementary-twist-SUN-KHM}
\end{equation}
which leaves the Kondo coupling invariant:
\begin{equation}
\widehat{\mathcal{U}}_{\alpha}{}^{\dagger} \mathcal{H}_{\text{K}} \, \widehat{\mathcal{U}}_{\alpha}
- \mathcal{H}_{\text{K}}   =  0 \; .
\label{eqn:SUN-Kondo-term-variation}
\end{equation}
Generic twists are given by the combinations of the form:    
\begin{equation}
\widehat{\mathcal{U}}_{\{ m_{i} \}} := (\widehat{\mathcal{U}}_{1})^{m_{1}} \cdots (\widehat{\mathcal{U}}_{N} )^{m_{N}}
\label{eqn:LSM-twist-generic}
\end{equation} 
that are specified by the set of integers $(m_{1}, \ldots, m_{N})$ with a negative $m_{\alpha} \, (<0)$ being understood 
as $(\widehat{\mathcal{U}}_{\alpha})^{m_{\alpha}} 
:= (\widehat{\mathcal{U}}_{\alpha}^{\dagger})^{|m_{\alpha}|}$.  

Combining all the above results \eqref{eqn:SUN-hopping-term-variation}, \eqref{eqn:SUN-Heisenberg-term-variation}, 
and \eqref{eqn:SUN-Kondo-term-variation}, we obtain the variation of the Kondo-Heisenberg Hamiltonian \eqref{eqn:SUN-KHM} 
due to the elementary twist $\widehat{\mathcal{U}}_{\alpha}$:
\begin{equation}
\begin{split}
& \delta_{\alpha} \mathcal{H}_{\text{KHM}} 
:= \widehat{\mathcal{U}}_{\alpha}^{\dagger} \mathcal{H}_{\text{KHM}} \, \widehat{\mathcal{U}}_{\alpha} 
- \mathcal{H}_{\text{KHM}} \\
& = i \frac{2\pi}{L} \sum_{j=1}^{L}  \left[ h_{j,j+1}^{(\text{KHM})}  , 
\frac{1}{2}\left\{ (\hat{n}_{\alpha,j+1} - Q_{\alpha,j+1} ) - ( (j+1) \to j ) \right\} \right]  \\
& \phantom{=} + \text{O}(L^{-1})  \; .
\end{split}
\label{eqn:variation-SUN-KHM}
\end{equation}
The variation due to generic twists is given similarly.  

When the ground state $| \text{g.s.} \rangle$ (with the energy $E_{\text{g.s.}} $) 
of $\mathcal{H}_{\text{KHM}}$ and the twisted state 
$\widehat{\mathcal{U}}_{\alpha}| \text{g.s.} \rangle$ are orthogonal 
$\langle  \text{g.s.} |\widehat{\mathcal{U}}_{\alpha}| \text{g.s.} \rangle =0$, we expect that 
$\widehat{\mathcal{U}}_{\alpha}| \text{g.s.} \rangle$ is made of excited states.   Let 
$E_{0}^{\text{(exc)}}$ be the energy of the lowest state in the sector to which 
$\widehat{\mathcal{U}}_{\alpha}| \text{g.s.} \rangle$ belongs.  
Then, by the variational principle, $\langle  \text{g.s.} | \delta_{\alpha} \mathcal{H}_{\text{KHM}} | \text{g.s.} \rangle$ gives 
an upper bound on the excitation gap $\Delta_{\alpha} E$:
\begin{equation}
\begin{split}
\Delta_{\alpha} E &= E_{0}^{\text{(exc)}} - E_{\text{g.s.}}   \\
& \leq \langle  \text{g.s.} | 
\widehat{\mathcal{U}}_{\alpha}^{\dagger} \mathcal{H}_{\text{KHM}} \, \widehat{\mathcal{U}}_{\alpha} 
  | \text{g.s.} \rangle -  E_{\text{g.s.}} 
= \langle  \text{g.s.} | \delta_{\alpha} \mathcal{H}_{\text{KHM}} | \text{g.s.} \rangle  \; .
\end{split}
\end{equation}
If $| \text{g.s.} \rangle$ is reflection symmetric, we can expect that the ground-state expectation value 
of the leading $\text{O}(L^{0})$ term in \eqref{eqn:variation-SUN-KHM} vanishes and 
$\langle \text{g.s.} | \delta_{\alpha} \mathcal{H}_{\text{KHM}} | \text{g.s.}\rangle= \text{O($L^{-1}$)}$, which means that 
the gap to the lowest excited state is bounded by $1/L$.   
If we assume low-energy Luttinger-liquid description, we can explicitly give the expression of the energy increase $\Delta_{\alpha} E$ 
which is proportional to $1/L$ (see Appendix~\ref{sec:continuum-LSM}).  
One way to tell if $| \text{g.s.} \rangle$ and $\widehat{\mathcal{U}}_{\alpha}| \text{g.s.} \rangle$ are orthogonal 
to each other or not is to calculate the crystal momentum carried by the twisted state $\widehat{\mathcal{U}}_{\alpha}| \text{g.s.} \rangle$.  
It is straightforward to generalize the above to more general twists $\widehat{\mathcal{U}}_{\{ m_{i} \}}$.  
\subsection{Momentum counting}
\label{sec:LSM-momenum-counting}
The crystal momentum $k_{\alpha}$ of the twisted state $\widehat{\mathcal{U}}_{\{ m_{\alpha} \}} |\text{g.s.} \rangle$ 
can be found by calculating the eigenvalue of the one-site translation $\mathrm{T}_{a_{0}}$ 
($\mathrm{T}_{a_{0}} S_{k}^{A} \mathrm{T}_{a_{0}}^{\dagger} = S_{k+1}^{A}$):
\begin{equation}
\begin{split}
& \be^{ i k_{\alpha} } \widehat{\mathcal{U}}_{\alpha} | \text{g.s.} \rangle 
= \mathrm{T}_{a_{0}}^{\dagger} \, \widehat{\mathcal{U}}_{\alpha} | \text{g.s.} \rangle   \\
& = \mathrm{T}_{a_{0}}^{\dagger} \, \widehat{\mathcal{U}}_{\alpha} \mathrm{T}_{a_{0}} \mathrm{T}_{a_{0}}^{\dagger}  | \text{g.s.} \rangle
= \be^{ i k_{0}} \mathrm{T}_{a_{0}}^{\dagger} \, \widehat{\mathcal{U}}_{\alpha} \mathrm{T}_{a_{0}} | \text{g.s.} \rangle  \; ,
\end{split}
\end{equation}
where $k_{0}$ is the ground-state momentum: $\mathrm{T}_{a_{0}}^{\dagger} | \text{g.s.}\rangle = \be^{ i k_{0}}  | \text{g.s.}\rangle$.  
Using the method used in, e.g., Ref.~\cite{Affleck-L-86}, we obtain the following result:
\begin{equation}
\be^{i (k_{\alpha}  -k_{0}) } =
\be^{- 2\pi i Q_{\alpha,1} } \exp \left\{  
- i \frac{2\pi}{L} \sum_{j=1}^{L} \left( \hat{n}_{\alpha,j}  - Q_{\alpha,j} \right) \right\}     
\; . 
\label{eqn:momentum-shift-Ualpha}
\end{equation}
As is shown in Ref.~\cite{Affleck-L-86}, the operator $\be^{- 2\pi i Q_{\alpha,1} }$ commuting with all the SU($N$) generators 
is a phase determined solely by the ``spin'' of the local moments, i.e., 
$\be^{- 2\pi i Q_{\alpha,1} } = \be^{- i\frac{2\pi}{N} n_{\text{yng}}}$ with $n_{\text{yng}}$ being the number of boxes 
in the Young diagram specifying the local moments ($n_{\text{yng}}=1$ here). 
For further calculations, it is convenient to represents the generators $Q_{\alpha,j}$ of the local moment 
in terms of fixed number ($n_{\text{yng}}$) of fermions $d^{(\text{s})\dagger}_{\alpha,j}$: 
$Q_{\alpha,j} = n_{\text{yng}} /N - \hat{n}^{(\text{s})}_{\alpha,j}$ ($\hat{n}^{(\text{s})}_{\alpha,j} := d^{(\text{s})\dagger}_{\alpha,j}d^{(\text{s})}_{\alpha,j}$) 
\footnote{%
Physically, this seems quite reasonable. The $d^{(\text{s})}_{\alpha}$ fermion may correspond to the $f$-electron (in the heavy-fermion setting) 
or an almost immobile alkaline-earth-like fermion in the metastable excited state when the models \eqref{eqn:SUN-KLM} 
and \eqref{eqn:SUN-KHM} are realized with ultracold gases.}.  
Then, the momentum shift in Eq.~\eqref{eqn:momentum-shift-Ualpha} reads:
\begin{equation}
\delta k_{\alpha} := k_{\alpha}  -k_{0} = - \frac{2\pi}{L} \sum_{j=1}^{L} \left( \hat{n}_{\alpha,j}  + \hat{n}^{(\text{s})}_{\alpha,j} \right)
\end{equation}
where the first and second terms act on the conduction electrons and local moments, respectively.  
However, due to the Kondo coupling, $\sum_{j}\hat{n}_{\alpha,j}$ and $\sum_{j}Q_{\alpha,j}$ are not conserved {\em separately}, 
and it is convenient to move to a basis in which the charge and SU($N$) parts are separated.  

The SU($N$) symmetry of the Hamiltonian guarantees that the $N$ color-resolved total fermion numbers $\mathcal{N}_{\alpha}$  
are all conserved (note that each local SU($N$) moment can be regarded as made of a single localized fermion  
$1= \sum_{\alpha=1}^{N} n_{\alpha,j}^{(\text{s})}$; see Appendix~\ref{sec:Young-diag} for how an SU($N$) local moment is 
constructed from fermions):
\begin{subequations}
\begin{equation}
\begin{split}
& \mathcal{N}_{\alpha} := \sum_{j=1}^{L} \left( \hat{n}_{\alpha,j} + n_{\alpha,j}^{(\text{s})} \right) 
=: n_{\alpha} L  \; (\alpha=1,\ldots, N) \; .  
\label{eqn:SUN-KHM-conserved-1}
\end{split}
\end{equation}
Instead, we may use the total fermion number $\mathcal{N}$ and the total SU($N$) weight $\vec{\Lambda}_{\text{tot}}$:
\begin{align}
& \mathcal{N} := \sum_{\alpha=1}^{N} \mathcal{N}_{\alpha}  
= \sum_{j=1}^{L} \left\{ \sum_{\alpha=1}^{N} \hat{n}_{\alpha,j} + 1 \right\} =: n L 
 \label{eqn:SUN-KHM-conserved-2}    \\
& \vec{\Lambda}_{\text{tot}} := \sum_{\alpha=1}^{N} \mathcal{N}_{\alpha} \vec{\mu}_{\alpha} 
= \sum_{j=1}^{L} \sum_{\alpha=1}^{N} \left(\hat{n}_{\alpha,j} + n_{\alpha,j}^{(\text{s})} \right) \vec{\mu}_{\alpha} 
=: \vec{\lambda}_{\text{tot}} L   \; ,
\label{eqn:SUN-KHM-conserved-3}
\end{align}
\end{subequations}
where $\vec{\mu}_{\alpha}$ are the $\alpha$-th weights in the $N$-dimensional defining representation (${\tiny \yng(1)}$) and 
satisfy $\vec{\mu}_{\alpha}{\cdot} \vec{\mu}_{\beta} = (\delta_{\alpha\beta}-1/N)/2$;    
the conservation of $\mathcal{N}$ \eqref{eqn:SUN-KHM-conserved-2} and $\vec{\Lambda}_{\text{tot}}$ 
 \eqref{eqn:SUN-KHM-conserved-3} imply Eq.~\eqref{eqn:SUN-KHM-conserved-1}, and {\em vice versa}.    
We solve \eqref{eqn:SUN-KHM-conserved-2} and \eqref{eqn:SUN-KHM-conserved-3} for $n_{\alpha}$ to obtain:
 \begin{equation}
n_{\alpha} = n/N + 2 \vec{\mu}_{\alpha} {\cdot} \vec{\lambda}_{\text{tot}} 
=  (f + 1 /N) + 2 \vec{\mu}_{\alpha} {\cdot} \vec{\lambda}_{\text{tot}}   \; .
\label{eqn:fermion-density-color-resolved}
\end{equation} 
In SU(2), Eq.~\eqref{eqn:fermion-density-color-resolved} expresses 
$n_{\uparrow}$ and $n_{\downarrow}$ in terms of the fermion filling $f$ and magnetization $m=\lambda_{\text{tot}}/\sqrt{2}$.  

Now Eq.~\eqref{eqn:fermion-density-color-resolved} enables us to rewrite the momentum shift \eqref{eqn:momentum-shift-Ualpha} 
in terms of the filling $f$ and the SU($N$) ``magnetization'' $\vec{\lambda}_{\text{tot}}$ as:
\begin{equation}
\delta k_{\alpha} 
= -2 \pi 
\left\{
 (f + 1 /N) + 2 \vec{\mu}_{\alpha} {\cdot} \vec{\lambda}_{\text{tot}}  \right\} \; .
 \end{equation}
 It is straightforward to generalize the above to the case of generic twists $\widehat{\mathcal{U}}_{\{ m_{\alpha} \}}$:
\begin{equation}
\begin{split}
& \delta k_{\{ m_{\alpha} \}}  = \sum_{\alpha=1}^{N} m_{\alpha} \delta k_{\alpha}   \\
 &= -2 \pi 
\left\{
 (f + 1 /N) M
 + 2  \vec{\lambda}_{\text{tot}} {\cdot} \left( \sum_{\alpha=1}^{N} \overline{m}_{\alpha} \vec{\mu}_{\alpha} \right)  \right\}  
 \; ,
 \end{split}
 \label{eqn:momentum-shift-SUN-KHM-generic}
\end{equation}
where we have introduced the average ($M/N$) and the zero-mean ($\overline{m}_{\alpha}$) parts of $m_{\alpha}$:
\begin{equation}
\begin{split}
& m_{\alpha} = M/N + \overline{m}_{\alpha} \quad  \left(\alpha=1,\ldots, N \right)  \\
& M := \sum_{\alpha=1}^{N} m_{\alpha}  \, , \quad   
\sum_{\alpha=1}^{N} \overline{m}_{\alpha} = 0  
\end{split}
\label{eqn:average-and-zero-sum}
\end{equation}
(note $\sum_{\alpha=1}^{N} \vec{\mu}_{\alpha}  = \vec{0}$).   
The equation \eqref{eqn:momentum-shift-SUN-KHM-generic} is the central result of this section.  
If $\delta k_{\{ m_{\alpha} \}} \neq 0$ (mod $2\pi$), we can use the variational argument in Sec.~\ref{sec:twisted-Hamiltonian} to 
show that there are low-lying excitations with energies $\text{O}(L^{-1})$.   

Here we would like to stress that the momentum \eqref{eqn:momentum-shift-SUN-KHM-generic} carried by low-energy excitations is determined by 
the ``effective'' filling $f_{\text{eff}} := f + 1 /N$   
that includes contributions of {\em both} the itinerant fermions ($f$) and the local moments ($1/N$),   
and by the SU($N$) magnetization density $\vec{\lambda}_{\text{tot}}$ [i.e., the set of the $(N-1)$ Cartan eigenvalues per site].  

With $M$ and $\overline{m}_{\alpha}$ defined in \eqref{eqn:average-and-zero-sum}, 
generic twists $\widehat{\mathcal{U}}_{\{ m_{\alpha} \}}$ in \eqref{eqn:LSM-twist-generic} can be written as:
\begin{equation}
\begin{split}
& \widehat{\mathcal{U}}_{\{ m_{\alpha} \}}   \\
& = \exp\left\{  i \frac{2\pi}{L} \sum_{j=1}^{L} j  \left[ 
\frac{M}{N}  \hat{n}_{j} 
- \sum_{\alpha=1}^{N} \overline{m}_{\alpha}  \left( \widehat{Q}_{\alpha,j}  + Q^{(\text{S})}_{\alpha,j} \right)
\right]   \right\}   \\
& = \exp\left\{  i \frac{2\pi}{L} \sum_{j=1}^{L} j  \left[ 
\frac{M}{N}  \hat{n}_{j} 
+ 2 \sum_{\alpha=1}^{N} \left( \overline{m}_{\alpha} \vec{\mu}_{\alpha} \right) \cdot \vec{\lambda}_{j} 
\right]   \right\}  \; ,
\end{split}
\label{eqn:LSM-twist-generic-2}
\end{equation}
from which we see that twists with $\overline{m}_{\alpha}=0$ never change the SU($N$)-spin-dependent part 
(i.e., $\widehat{\mathcal{U}}_{\{ m_{\alpha} \}}^{\dagger} \widehat{\mathcal{S}}_{\mu\nu} \widehat{\mathcal{U}}_{\{ m_{\alpha} \}} 
= \widehat{\mathcal{S}}_{\mu\nu}$, 
 $\widehat{\mathcal{U}}_{\{ m_{\alpha} \}}^{\dagger}\mathcal{S}_{\mu\nu} \widehat{\mathcal{U}}_{\{ m_{\alpha} \}} 
= \mathcal{S}_{\mu\nu}$), while those with at least one of $\overline{m}_{\alpha}$ is non-zero create 
spin-charge entangled excitations, in general.   
The simplest of such twists is $\widehat{\mathcal{U}}_{(1,0,\ldots,0)}$ which will play an important role in the next section.  
However, the spin-charge-entangled appearance of $\widehat{\mathcal{U}}_{(1,0,\ldots,0)}$  
does not necessarily mean that it creates excitations in {\em both} the spin and charge sectors.  
In fact, as is discussed in Appendix~\ref{sec:continuum-LSM} using low-energy description, 
as far as spin-charge separation occurs at low energies, the spin-charge entangled twist $\widehat{\mathcal{U}}_{(1,0,\ldots,0)}$ creates 
only spin excitations in the charge-ordered insulators (e.g., Mott and CDW states), while $\widehat{\mathcal{U}}_{(1,\ldots,1)}$ 
never excites the charge-ordered ground states \footnote{%
This may be most easily seen in an extreme situation in which the local fermion number $ \hat{n}_{j}=\sum_{\alpha} \hat{n}_{\alpha,j}$ is constant 
all over the lattice. In this case, the charge part $\exp\left\{  i \frac{2\pi}{L} \frac{M}{N}  \sum_{j=1}^{L} j   \hat{n}_{j}   \right\} $
of \eqref{eqn:LSM-twist-generic-2} just adds a phase, while the second $Q$-dependent part creates spin excitations.}.  
Therefore, in these cases, $\widehat{\mathcal{U}}_{(1,0,\ldots,0)}$ probes only the spin sector.  

\subsection{Predictions for low-energy physics}
\label{sec:LSM-predictions}
In this subsection, we use the results of the previous subsections to predict the low-energy properties 
of the Kondo-Heisenberg Hamiltonian $\mathcal{H}_{\text{KHM}}$ for various filling fractions $f$.  
Specifically, by searching for the values of $f$ at which gapless ground states are expected, we find 
where we can expect (partially) gapped ground states.  
Also, to compare the results with 
those of field-theory arguments given in the next section, we use a short-hand notation C$m$S$n$, 
that was introduced in the context of fermionic ladder models in Ref.~\onlinecite{Balents-F-96},  which means 
that there are $m$ gapless branches in the charge (``C'') sector and $n$ in the SU($N$) spin (``S'') sector 
(a ground state with finite gaps to all excitations is denoted by C$0$S$0$).  
To be specific, let us focus on the simplest case with the local moments in the $N$-dimensional representation ${\tiny \yng(1)}$  
and assume that the ground state is SU($N$)-singlet, i.e., $\vec{\lambda}_{\text{tot}}= \vec{0}$.  
Then, the momentum shift \eqref{eqn:momentum-shift-SUN-KHM-generic} 
due to $\widehat{\mathcal{U}}_{\{ m_{\alpha} \}}$ depends only on the charge part of the twist 
(the zero-mean part $\{ \overline{m}_{\alpha} \}$ that acts on the spin sector does not appear in the momentum shift 
of spin-singlet ground states):
\begin{equation}
\delta k_{\{ m_{\alpha} \}} 
= -2 \pi  (f + 1/N) M   \; .
\end{equation}  
Note that $\delta k_{\{ m_{\alpha} \}}$ depends on $\{ m_{\alpha} \}$ only through the sum $M=\sum_{\alpha} m_{\alpha}$.  
\subsubsection{Possibility of unique full-gap insulator}
\label{sec:unique-full-gap-ins}
We begin by examining the possibility of a unique (i.e., non-degenerate) gapped translationally-invariant ground states. 
The tightest condition is obtained for, e.g., the choice $(m_{1}, \ldots, m_{N})=(1,0,\ldots, 0)$ ($M=1$) 
that generates a charge-spin entangled twist \footnote{%
When the ground state is ``magnetized'' (i.e., $\vec{\lambda}_{\text{tot}} \neq \vec{0}$), 
$\delta k_{(1,0,\ldots, 0)}$ receives the correction $-2\pi \vec{\mu}_{1} {\cdot} \vec{\lambda}_{\text{tot}}$ 
from the magnetization.}:
\begin{equation}
\delta k_{(1,0,\ldots, 0)} 
= - 2 \pi  (f + 1/N)     \; .
\end{equation} 
From this, we introduce the first index:
\begin{equation}
\mathcal{I}_{1} := f + 1/N \; (\text{mod $1$})  \; . 
\label{eqn:1st-LSM-index}
\end{equation}
When $\mathcal{I}_{1}  \notin \mathbb{Z}$, the LSM argument implies either (i) a gapless ground 
state [as the twist $(1,0,\ldots,0)$ affects both charge and spin, we do not care about which sector is gapless] 
or (ii) mutiple ground states with spontaneously broken translational symmetry appear in the limit $L \to \infty$.   
For filling $f$ satisfying the above condition (i.e., $\mathcal{I}_{1} = f+1/N \notin \mathbb{Z}$), the possibility of 
a non-degenerate gapped translationally-invariant ground state is excluded.  
Therefore, a unique fully-gapped ($\text{C}0\text{S}0$) ground state is allowed only when $f+1/N \in \mathbb{Z}$, i.e., 
at filling 
\begin{equation}
f= 1-1/N  \;  .
\label{eqn:unique-full-gap-ins}
\end{equation}  
In fact, it is known \cite{Totsuka-23} that a uniform spin-gapped Kondo-insulator is formed at $f= 1-1/N$ 
at least when $J_{\text{K}} \gg t, J_{\text{H}}$.   
The above argument states that this is the only featureless spin-gap Kondo insulator in the KLM 
\eqref{eqn:SUN-KLM}  and KHM \eqref{eqn:SUN-KHM}.    
\subsubsection{Other insulating phases}
\label{sec:charge-gap}
To explore the possibility of insulating phases for other fillings, let us consider 
the simplest charge-only twist $\widehat{\mathcal{U}}_{(1,\ldots, 1)}$  
\begin{equation}
 (m_{1} , \ldots, m_{N} ) = (1, \ldots, 1) \; .
\end{equation}
According to the general formula \eqref{eqn:momentum-shift-SUN-KHM-generic}, it induces a momentum shift 
\begin{equation}
\delta k_{(1,\ldots,1)} = - 2 \pi  ( N f + 1) 
\; ,
\label{eqn:momentum-shift-charge-twist}
\end{equation}
which leads us to defining the second index:
\begin{equation}
\mathcal{I}_{2} := Nf    \; (\text{mod $1$})  \; .
\label{eqn:2nd-LSM-index}
\end{equation}
It is important to note that although the twists $(1,\ldots,1)$ and $(N,0,\ldots,0)$ create excitations at the same momentum 
$\delta k_{(1,\ldots,1)}=\delta k_{(N,0,\ldots, 0)}= N \times \delta k_{(1,0,\ldots, 0)}$, the natures of 
the excited states are very different \footnote{%
When SU($N$) magnetization $\vec{\lambda}_{\text{tot}}$ is finite, they create excitations at different momenta 
[see Eq.~\eqref{eqn:momentum-shift-SUN-KHM-generic}]}.  

The index $\mathcal{I}_{2}$ restricts the possibility of $\text{C}0\text{S}n$-type ($n \neq 0$) insulators without translation-symmetry-breaking order 
(e.g., charge-disproportionation).  
When spin and charge are entangled (as in higher-dimensional metals), the twist $\widehat{\mathcal{U}}_{(1 ,\ldots, 1)}$ 
no longer probes only the charge sector selectively.  
Nevertheless, non-zero values of $\mathcal{I}_{1}$ [we exclude $f=(N-1)/N$ considered 
already above] and translation symmetry forbid the gap opening in {\em both} the spin and charge sectors \footnote{%
If at least one of the two is gapped, the combined excitations must be gapped, too, which contradicts with the non-zero index $\mathcal{I}_{1}$.}.  
Therefore, let us assume spin-charge separation to explore the possibility of translation-invariant insulators.  
Then, in order for a finite charge gap, the condition $\mathcal{I}_{2}=0$, i.e., 
\begin{equation}
f = m / N \quad  (m=1,\ldots, N-2)   
\label{eqn:charge-gap-condition}
\end{equation}
must be satisfied [the cases $m=0$ and $N$ correspond to trivial (carrierless and fully-occupied, respectively) insulators] 
\footnote{%
Note that {\em spin-only} twists with $M=\sum_{\alpha}m_{\alpha} =0$ do not lead to any meaningful statements for 
the spin-singlet ground states.}.    
As far as translation symmetry is preserved, the spin sector must be gapless (note that in the presence of spin-charge separation, 
$\mathcal{I}_{1} \neq 0$ implies the existence of gapless spin excitations; see Appendix~\ref{sec:continuum-LSM} for how the LSM twists act 
in spin-charge-separated systems); 
a finite spin gap is necessarily accompanied by some sort of symmetry-breaking order in the spin sector. 
We shall call the special fillings \eqref{eqn:charge-gap-condition} {\em commensurate}.  
For other rational fillings, translation-invariant insulators are forbidden and, when the system becomes insulating, 
both the spin and charge sectors necessarily break translation symmetry.
(An example of this is the spin-charge dimerized insulator with algebraic spin correlation 
found in the SU(2) KLM at $f=1/4$ \cite{Xavier-P-M-A-03,Xavier-M-08,Huang2020})

The filling $f=1/N$ [$m=1$ in \eqref{eqn:charge-gap-condition}] is of particular interest, since a non-trivial insulator whose low-energy spin sector 
is described by the SU($N$) Heisenberg model:
\begin{equation}
\mathcal{H}_{\text{eff}} = 
\left( \frac{t^{2}}{2 |J_{\text{K}}|} + \frac{1}{4}J_{\text{H}} \right) \sum_{i} S^{A}_{i} ({\tiny \yng(2)}) S^{A}_{i+1} ({\tiny \yng(2)})   
\label{eqn:eff-Heisenberg-sym-rank-2}
\end{equation} 
is expected \cite{Totsuka-23} at strong coupling $|J_{\text{K}}| \gg t, J_{\text{H}}$ ($J_{\text{K}} <$0; when $J_{\text{K}}>0$, 
the strong-coupling expansion does not lead to any useful conclusions).  
According to recent   field theoretical arguments 
\cite{Yao-H-O-19,Wamer-L-M-A-20,Wamer-A-PRB20}, 
the ground state of the above model is gapless when $N=\text{odd}$, and gapped with broken translation symmetry 
when $N=\text{even}$ (except for $N=2$).   
Let us consider this situation in the light of the LSM argument.   
As has been discussed above, at $f=1/N$, $\mathcal{I}_{2}=0$ (mod $1$) and a charge gap can open without breaking translation.  
The fate of the spin sector is interesting.    The $(1,0,\ldots,0)$ twist 
tells that the entire system is gapless (when the ground state is unique) or has a (spin) gap over  
mutiply degenerate ground states with broken translation symmetry.  
Suppose that we have an insulating ground state that has no charge modulation, etc.  
Then, the translation-symmetry breaking occurs in the spin sector.  
Again, the $\widehat{\mathcal{U}}_{(1,0,\ldots, 0)}$ twist can tell how many degenerate ground states exist 
in the spin-gapped situation. 
The momentum shift 
\[
\delta k_{(1,0,\ldots, 0)} 
= - 2 \pi  (1/N + 1/N) = - \frac{2 \pi}{\left( \frac{N}{2} \right) }
\]  
suggests a reasonable scenario that there are $N/2$ degenerate ground states ($N$ necessarily is 
even) on which the system can hop from one to another 
by the repeated application of the $(1,0,\ldots,0)$ twist (after $\widehat{\mathcal{U}}_{(1,0,\ldots, 0)}^{N/2}$, 
the system returns to the original ground state).  
This agrees with the prediction $\text{GSD}=N/\text{gcd}(N,2)=N/2$ for the pure spin model.  

Clearly, this simple story breaks down when $N= \text{odd}$ and we expect $N$ degenerate ground states to occur 
when a spin gap is finite.   
In fact, recent analytical and numerical studies \cite{Lecheminant-15,Wamer-L-M-A-20,Nataf-G-M-21} show that 
the ground state of the effective spin model \eqref{eqn:eff-Heisenberg-sym-rank-2} for the uniform insulating 
state (one itinerant fermion at each site) remains gapless when $N=\text{odd}$.   Therefore, the gapless option seems to be chosen 
when $J_{\text{K}}<0$.  

Next, let us examine the possibility of opening the charge gap at half-filling $f=1/2$ (not necessarily commensurate):
\begin{equation}
\delta k_{(1,\ldots,1)}/ (2 \pi )  = -  ( N / 2 + 1)   \; ,
\end{equation}
while keeping translation symmetry.  
If $N=\text{odd}$, the LSM argument tells that there must be gapless excitations 
at $k = -  2\pi ( N / 2 + 1)$ created by the $(1,\ldots,1)$ twist.  When spin-charge separation occurs, 
this immediately implies that the charge sector remains gapless for odd-$N$ 
as the LSM twist $\widehat{\mathcal{U}}_{(1 ,\ldots, 1)}$ acts only on the charge sector.   
When spin and charge are coupled, on the other hand, the absence of the LSM gap implies that none of spin and charge is gapped.   
Therefore, we generically expect [combining the analysis of the twist $\widehat{\mathcal{U}}_{(1 ,0,\ldots, 0)}$] 
that {\em both} spin and charge are gapless unless lattice-translation symmetry is broken.  

Therefore, in order for the charge sector to have a gap (without breaking translation symmetry; we do not care 
about the spin sector), $N / 2  \in \mathbb{Z}$, 
i.e., translation-symmetric insulators are possible only when $N = \text{even}$ (and presumably, spin-charge separation is required).    
Even when this is the case, the conclusion from the twist $\widehat{\mathcal{U}}_{(1,0,\ldots, 0)}$ 
tells us that the {\em entire} system should remain gapless at $f=1/2$ ($N \neq 2$ is assumed) unless the translation is broken. 
Therefore, the symmetric insulating ground states allowed for $f=1/2$, $N=\text{even}$ are of the type C0S$n$ 
($n \neq 0$, i.e., at least one of the $N$ spin channels is gapless) as in the usual SU(2) Hubbard model at half-filling.  

If we allow degenerate ground states due to spontaneous translation-symmetry breaking, 
full-gap insulators ($\text{C}0\text{S}0$) are possible even at $f=1/2$ 
regardless the parity of $N$. The number of the degenerate ground states may be estimated 
by looking at the momentum shift due to a single twist:
\[
\delta k_{(1,0,\ldots,0)} = - 2 \pi (1/2+1/N) = -2\pi \frac{N+2}{2N} \; .
\] 
The simplest scenario would be that everytime when the twist $\widehat{\mathcal{U}}_{(1,0,\ldots, 0)}$ is applied, 
a new degenerate ground state is generated.  Therefore, the smallest period of the sequence 
\begin{equation*}
\begin{split}
& 0 \to -2\pi \frac{N+2}{2N} \to -2\pi \frac{N+2}{2N}\times 2 \to -2\pi \frac{N+2}{2N} \times 3 \\
& \; \to
\cdots \to 0 \quad (\text{mod } 2\pi) 
\end{split}
\end{equation*}
gives the ground-state degeneracy (GSD) in the $\text{C}0\text{S}0$ phase \footnote{%
Precisely speaking, the number of degenerate ground states can be any integer multiple of this period 
when some additional discrete symetries (other than lattice translation) are broken simultaneously.}.   
The answer is: 
\begin{equation}
\text{GSD} = 
\begin{cases}
2N  & \text{when $N=$odd} \\
N  & \text{when $N=4p$ ($p \in \mathbb{Z}$)} \\
N/2  & \text{when $N=4p+2$} \; .
\end{cases}
\label{eqn:LSM-GSD-half-filling}
\end{equation}

Before concluding this section, a few comments are in order. 
First, all the above arguments do not assume a particular form of the Hamiltonian and the results are applicable to 
any one-dimensional lattice Hamiltonian [including the models \eqref{eqn:SUN-KLM} and \eqref{eqn:SUN-KHM}] 
consisting of $N$-component fermions and localized SU($N$) moments in 
the SU($N$) ``spin'' ${\tiny \yng(1)}$ that couple to the fermion part; depending on the detail of the Hamiltonian, 
one of the options is chosen among several possibilities that the LSM argument suggests (see Table \ref{tab:KHM-phases-by-LSM}).  

Also, it is straightforward to generalize the treatment to a general SU($N$) spin specified by a Young diagram with 
$n_{\text{yng}}$ boxes (the treatment here is mostly for $n_{\text{yng}}=1$; see Appendix~\ref{sec:Young-diag} for more details on 
the SU($N$) representations and the Young diagrams).   Repeating the same steps, we obtain the two LSM indices: 
\begin{equation}
\begin{split}
& \mathcal{I}_{1} := f + n_{\text{yng}}/N \; (\text{mod $1$})  \\
& \mathcal{I}_{2} := N f + n_{\text{yng}} \; (\text{mod $1$}) \; .
\label{eqn:LSM-index-gen}
\end{split}
\end{equation}
Now the featureless Kondo insulators are possible only at filling: 
\begin{equation}
f = 1+\lfloor  n_{\text{yng}}/N \rfloor - n_{\text{yng}}/N   \; ,
\label{eqn:filling-Kondo-ins-general}
\end{equation} 
where $\lfloor x \rfloor$ denotes the largest integer that does not exceed $x$.  
For instance, in the case of half-filling ($f=1/2$) and the SU($N$) local moment which transforms in the 
antisymmetric self-conjugate representation $n_{\text{yng}} = N/2$ ($N$ even)
\[   \text{\scriptsize $N/2$} \left\{ 
{\tiny \yng(1,1,1,1)  }
\right.  \; ,
\]
featureless Kondo insulators 
can be stabilized as has been shown in Refs.~\onlinecite{Raczkowski-A-20,Raczkowski-A-24}.

\begin{widetext}
\begin{center}
\begin{table}[htb]
\caption{\label{tab:KHM-phases-by-LSM} Properties of insulating phases of the SU($N$) Kondo lattice model \eqref{eqn:SUN-KLM} 
or the SU($N$) Kondo-Heisenberg Hamiltonian \eqref{eqn:SUN-KHM} predicted by the LSM argument (GSD and gcd respectively stand 
for the ground-state degeneracy and the greatest common divisor).  For filling $f=1-1/N$, 
no useful constraint is obtained from the LSM argument about the nature of the insulators that spontaneously break translation symmetry 
(SSB insulators).}
\begin{ruledtabular}   
\begin{tabular}{lcc}
filling ($f$) &  featureless insulator  & SSB insulator  \\
\hline
generic & forbidden & forbidden  \\
\hline
rational ($f= p/q \neq m/N$) & forbidden & possible (C$0$S$0$/C$0$S$n$) \\
\hline
commensurate $m/N$ ($m \neq 1,N-1$) & spin gapless (C$0$S$n$)  
& $\text{GSD} = N/\text{gcd}(N,m+1)$ (for C$0$S$0$)  \\
\hline
$1-1/N$ & full-gap insulator (C$0$S$0$) possible & ---   \\
\hline
$1/N$ &  spin gapless (C0S$n$) & 
$\text{GSD} = N/\text{gcd}(N,2)$ (for C$0$S$0$)
\\
\hline
$1/2$ (half-filling)
&  
\begin{tabular}{ll}
$N=\text{even}$: &  spin-gapless (C$0$S$n$)  \\
$N=\text{odd}$:  & forbidden  
\end{tabular} 
& 
\begin{tabular}{ll}
full-gap insulator (C$0$S$0$) must break translation: \\
$\text{GSD} = 2N$ ($N=$odd), $N$ ($N=4\mathbb{Z}$), $N/2$ ($N=4\mathbb{Z}+2$) 
\end{tabular} 
\end{tabular}
\end{ruledtabular}
\end{table}
\end{center}
\end{widetext}

\section{Predictions from mixed global anomalies}
\label{sec:Anomaly}
In the previous section, we have used the LSM argument to obtain constraints on the nature of the ground state that depend only on 
a set of kinematical information (e.g., $N$ and filling $f$) and does not depend on the details of the models (such as the strength and 
sign of the Kondo coupling $J_{\text{K}}$).  Of course, the actual ground state depends on the values of, e.g., $J_{\text{K}}/t$ and 
 $J_{\text{H}}/t$, and detailed model-dependent analyses are required to map out the ground-state phases.  
In the following sections, we investigate the ground state of the SU($N$) KHM \eqref{eqn:SUN-KHM} 
directly in the continuum limit.  To this end, we first construct the low-energy effective Hamiltonian for the KHM. 
In contrast to the KLM \eqref{eqn:SUN-KLM} in which there is no direct interaction among the local moments, 
the existence of the Heisenberg exchange interaction $J_{\text{H}}$ provides us with a good starting point for a field theory analysis 
of the model \eqref{eqn:SUN-KHM}.  
Below, we implicitly assume that the Kondo interaction is sufficiently weak so that the system first flows towards 
a conformal field theory (CFT) fixed point which we derive in the next section [see Fig.~\ref{fig:RG-invariant}(a)].  

In the following sections, we also frequently use the fact that certain non-perturbative indices must be preserved 
all along the renormalization-group (RG) flow toward low-energies to restrict possible phases. 
Specifically, we interpret the LSM index $\mathcal{I}_{1}$ in terms of the 't Hooft anomaly of the effective theories and use the anomaly-matching argument. 
The analysis of the  global anomalies of the underlying field theory will then give some non-perturbative constraints on the possible phases, 
which should be compared to the ones obtained in Sec.~\ref{sec:LSM} from the LSM theorem.  

\subsection{Continuum-limit description}
\label{sec:Contlimit}
The starting point is the continuum description of the lattice fermion operator
$c_{\alpha,\,i}$ of the SU($N$) KHM (\ref{eqn:SUN-KHM}) in terms of $N$ left-right moving Dirac fermions \cite{Gogolin-N-T-book,Giamarchi-book-04}:
\begin{equation}
c_{\alpha,\,n} \rightarrow \sqrt{a_0} \left(L_{ \alpha}(x)
\be^{-i k_{\text{F}} x} + R_{\alpha}(x) \be^{i k_{\text{F}} x} \right),
\label{contlimitDirac}
\end{equation}
where $k_{\text{F}} a_{0}= \pi f = \pi m/N$ ($0 \leq m \leq N$) is the Fermi momentum and $x= n a_0$, with $a_0$ being the lattice spacing.     
The Hamiltonian density for the hopping part $\mathcal{H}_{\text{hop}}$ of the lattice Hamiltonian \eqref{eqn:SUN-KHM} is equivalent to that 
of $N$ identical left-right moving Dirac fermions:
\begin{equation}
  \mathcal{H}_{\text{hop}} =-i v_{\text{F}} \left(: \! R_{\alpha} ^\dag \partial_x R_{ \alpha} ^{\phantom \dag} \! : - 
  : \! L_{\alpha}^\dag \partial_x L_{\alpha}^{\phantom \dag} \! : \right) ,
\label{HamcontDirac}  
\end{equation}  
where $v_{\text{F}} =  2t a_0$ is the Fermi velocity, the symbol $: {\cdots} :$ denotes  the normal ordering 
with respect to the Fermi sea, and summation over repeated indices is implied in the following.  
The non-interacting part \eqref{HamcontDirac} enjoys continuous U($N$)$|_\text{L}$ $\otimes$ U($N$)$|_\text{R}$ symmetry 
which results from its invariance under independent unitary transformations on the left and right Dirac fermions. 
It is then very helpful to express the Hamiltonian \eqref{HamcontDirac} directly in terms 
of the currents associated to these continuous symmetries.
To this end, we introduce the U(1)$_{\text{c}}$ charge current  and
the SU($N$)$_{1,\text{f}}$ (the subscript ``f'' means fermion) current which underlie the CFT of massless $N$ 
Dirac fermions \cite{Gogolin-N-T-book}:
\begin{equation}
\begin{split}
& j_{\text{c}, \text{L}} = \,  : \! L_{\alpha}^\dagger L_{\alpha} \! :  \quad\textrm{U(1)$_\text{c}$ charge current} \\
&  J_{\text{f}, \text{L}}^{A} =   L_{\alpha}^\dagger T^{A}_{\alpha\beta} L_{\beta}    \quad
	  \; \text{fermion $ \text{SU($N$)}_{\text{f}}$ currents}, 
	\end{split}
	\label{U(2N)currents}
\end{equation}
with $\alpha,\beta = 1, \ldots, N$, and we have similar definitions for the right currents $j_{\text{c}, \text{R}}$ and $J_{\text{f}, \text{R}}^{A}$.\footnote{%
Note that normal ordering is not necessary for the SU($N$) currents as the divergences from the operator products on the right-hand side cancel each other due to $\text{Tr} (T^{A}) =0$.}
In Eq. (\ref{U(2N)currents}),  $T^A$ are the SU($N$) generators that have appeared in Eq.~\eqref{eqn:SUN-fermion-spin}. 
The non-interacting model (\ref{HamcontDirac}) can then be written in terms of these currents
 (the so-called Sugawara construction of the corresponding CFT) \cite{DiFrancesco-M-S-book,Gogolin-N-T-book,James-K-L-R-T-18}:
\begin{equation}
\begin{split}
\mathcal{H}_\text{hop} =& \frac{\pi v^{\text{(f)}}_\text{c}}{N}
 \left[ : j^2_{\text{c}, \text{R}} : + : j^2_{\text{c}, \text{L}} :
\right]   \\
&+ \frac{2\pi v^{\text{(f)}}_\text{s}}{N + 1} \left[ : J^A_{\text{f}, \text{R}} J^A_{\text{f}, \text{R}}: + : J^A_{\text{f}, \text{L}} J^A_{\text{f}, \text{L}}: 
\right] ,
\end{split}
\label{contfreehambis}
\end{equation}
with $v^{\text{(f)}}_\text{c}$ and $v^{\text{(f)}}_\text{s}$ denoting the characteristic velocities for the charge and spin sectors, respectively 
[the superscript ``(f)'' implies the itinerant fermions and the subscripts ``c/s'' are used to denote the charge/spin sectors].    
The second term is the Hamiltonian of the level-$1$ SU($N$) Wess-Zumino-Novikov-Witten (WZNW) CFT. 
In what follows, we frequently use the notation $\text{SU($N$)}_{k}$ to denote the level-$k$ SU($N$) WZNW CFT.  
To distinguish between the $\text{SU($N$)}_{1}$ CFT resulting from the itinerant fermions and that from the local moments, 
we also use the notations $\text{SU($N$)}_{1,\text{f}}$ and $\text{SU($N$)}_{1,\text{s}}$, respectively.

The continuum description of the fermionic SU($N$) spin operator \eqref{eqn:SUN-fermion-spin} at site $n$ can be derived 
using Eq.~\eqref{contlimitDirac}:
\begin{equation}
\hat{s}_{n}^{A}/a_0 \simeq J^{A}_{\text{f}, \text{L}}  +  J^{A}_{\text{f}, \text{R}}  + \be^{i 2 k_\text{F} x} L^{\dagger}_{\alpha} T^A_{\alpha\beta}  R_{\beta} 
+ \text{H.c.} 
\label{spinelecop}
\end{equation}
 It is then useful to introduce a bosonic charge field $\Phi_{\text{c}}$ and an SU($N$)$_{1,\text{f}}$ WZNW
 $g_\text{f}$ field with the scaling dimension $(N-1)/N$  
 to get a non-Abelian bosonized description of the $2k_\text{F}$ part of (\ref{spinelecop}) \cite{Affleck-NP86,James-K-L-R-T-18}:
\begin{equation}
\hat{s}_{n}^{A}/a_0  \simeq J^{A}_{\text{f}, \text{L}}  +  J^{A}_{\text{f}, \text{R}}  
+ i C \be^{i 2 k_{\text{F}} x}  \be^{i \sqrt{ 4 \pi/N} \Phi_{\text{c}}}   \; {\rm Tr} ( g_{\text{f}} \, T^A)
+ \text{H.c.} ,
\label{spinelecopfin}
\end{equation}
where $C$ is a positive constant.

The interaction $\mathcal{H}_{\text{H}}$ among the localized spins of the SU($N$) KHM \eqref{eqn:SUN-KHM} is described by the
SU($N$) antiferromagnetic Heisenberg spin chain, which is known to be integrable by the Bethe ansatz \cite{Sutherland-75} 
and displays a quantum critical behavior in the SU($N$)$_1$ universality class \cite{Affleck-NP86,Affleck-88}. 
The low-energy description is obtained by expressing the SU($N$) spin operators in terms of SU($N$)$_{1,\text{s}}$ chiral currents 
$J^{A}_{\text{s}, \text{R/L}}$ (with ``s'' standing for the local spins) and the ``spin'' WZNW $g_{\text{s}}$ field 
with the scaling dimension $(N-1)/N$ \cite{Affleck-NP86,Affleck-88,Assaraf-A-C-L-99,James-K-L-R-T-18}:
\begin{equation}
\begin{split}
S^{A}_{n}/a_0 \simeq& J^{A}_{\text{s}, \text{L}} +  J^{A}_{\text{s}, \text{R}} 
+ \left\{  i  \lambda \, \be^{\frac{ i 2 \pi }{Na_0}x} \; {\rm Tr} ( g_{\text{s}}  T^A) + \text{H.c.}  \right\}
\\
&+ \sum_{p=2}^{N-2} \be^{ i \frac{2 \pi p}{N a_0}x } n_{\text{s}, p}^{A},
\end{split}
\label{spinop}
\end{equation}
 where $ \lambda$ is a non-universal constant 
that stems from the averaging of the underlying charge degrees of freedom which are frozen in the insulating phase of the SU($N$) KHM. As discussed in Appendix \ref{sec:umklappsun}, $ \lambda$ turns out to be a complex number whose argument depends on $N$: $\lambda = | \lambda | \be^{i \theta_0}$ with
\begin{equation}
\begin{split}
 \theta_0  =& 0, \frac{\pi}{N} \; \; \; \; \; \; N = 2p  \\
 \theta_0  =&  \pm \frac{\pi}{2N}  \; \;  \; \; N = 2p +1 .
\end{split}
\label{phasespinop}
\end{equation} 

The higher-harmonics ($2 \pi p/N$) parts of the decomposition \eqref{spinop} are related to the SU($N$)$_{1,\text{s}}$ primary fields 
$\Phi_{\text{s}, p}$ ($p=2, \ldots, N-2$) with the scaling dimension $p(N-p)/N$ which transform
in the fully antisymmetric representations of SU($N$) represented by Young diagrams with a single column and $p$ rows 
(the representations $\mathcal{R}_{p}$ in Appendix~\ref{sec:Young-diag}):
\begin{equation}
 n_{\text{s}, p}^{A} := i \alpha_p  \; {\rm Tr} ( \Phi_{\text{s}, p} T_p^A),
\label{spinopprimaryappen}
\end{equation}
where $T_p^A$ are SU($N$) generators in the same representations and $\alpha_p$ are the corresponding non-universal constants.  
By the hermiticity of $S^{A}_{n}$, the $2pk_{\text{F}}$ component of the spin-density (\ref{spinop}) satisfies the constraint: 
$n_{\text{s}, p}^{A \dagger} = n_{\text{s}, N-p}^{A}$.
The low-energy properties of $\mathcal{H}_{\text{H}}$, i.e., the SU($N$) Heisenberg spin chain, is then described by the Hamiltonian density:
\begin{equation}
\mathcal{H}_{\rm H} = \frac{2\pi v^{\text{(s)}}_\text{s}}{N + 1} \left[ : J^A_{\text{s}, \text{R}} J^A_{\text{s}, \text{R}}: 
+ : J^A_{\text{s},\text{L}} J^A_{\text{s}, \text{L}}: \right] - \gamma  J^A_{\text{s}, \text{R}} J^A_{\text{s}, \text{L}},
\label{sutherlandham}
\end{equation}
where $v^{\text{(s)}}_\text{s}$ is the spin velocity [the superscript ``(s)'' implies the local spins]  
and the positive coupling constant $\gamma$ accounts for the logarithmic corrections 
to the SU($N$)$_1$ quantum critical behavior \cite{Affleck-G-S-Z-89,Itoi-K-97,Majumdar-M-02}. 
Combining Eqs.~\eqref{contfreehambis} and \eqref{sutherlandham}, we see that when $J_{\text{K}}=0$, the continuum limit 
of the KHM \eqref{eqn:SUN-KHM} is made of three CFTs $\mathcal{H}_\text{hop}+\mathcal{H}_{\rm H}$ corresponding to 
$\text{U(1)}_{\text{c}} \times \text{SU($N$)}_{1,\text{f}} \times \text{SU($N$)}_{1,\text{s}}$.  
\subsection{Constraints from the existence of a mixed global anomaly}
\label{sec:mixednanomaly}

We now discuss the LSM argument based on the translation and (on-site) SU($N$) symmetries within the field-theory description.  
To this end, it is crucial to correctly identify how the two symmetries are implemented in the low-energy effective field theories  
paying particular attention to the existence of two different low-energy descriptions (corresponding to two different conformal embeddings). 

\begin{figure}[htb]
\begin{center}
\includegraphics[width=\columnwidth,clip]{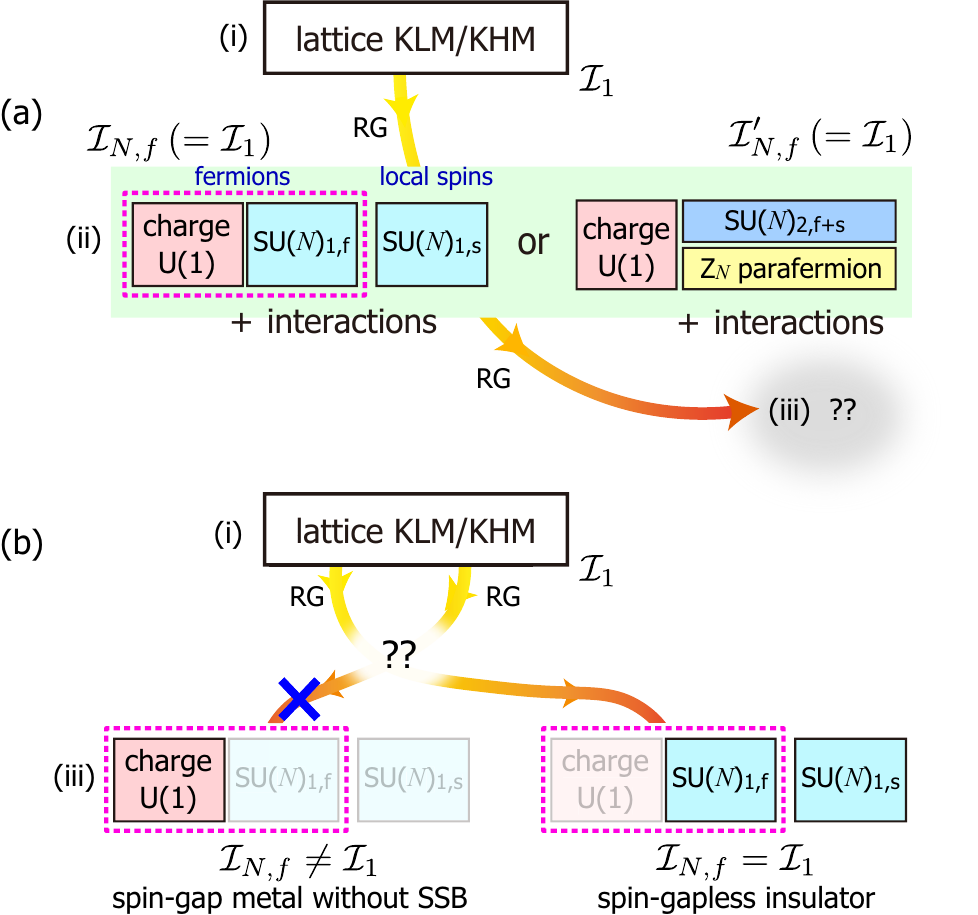}
\end{center}
\caption{(a) The original lattice model [at stage-(i)] and low-energy effective field theories [(ii) and (iii)] connected by RG flow (indicated by the arrows) 
must share the same index $\mathcal{I}$ in common.  The two descriptions shown in the stage-(ii) are explained 
in Secs.~\ref{sec:mixednanomalybasis1} and \ref{sec:mixednanomalybasis2}.    
(b) Situations allowed (right) and forbidden (left) by the anomaly-matching argument 
at commensurate fillings $f=m/N$ ($m=1,\ldots, N-1$) [see the discussion in Sec.~\ref{sec:mixednanomalybasis1}].  
\label{fig:RG-invariant}
}
\end{figure}

As is illustrated in Fig.~\ref{fig:RG-invariant}(a), there are three stages in the RG flow: 
(i) the original lattice model to which the LSM argument apply, 
(ii) the intermediate scale at which the two in-chain parts $\mathcal{H}_\text{hop}$ and $\mathcal{H}_\text{H}$ 
are interacting only weakly and described by a set of gapless CFTs 
with some interactions allowed by symmetries and the filling $f$ (now the flow is in the vicinity of the CFT fixed-point), 
and (iii) ``far-infrared (IR)'' region in which the Kondo coupling fully renormalizes the system (the system flowing toward the real IR-fixed point).  
To determine the ground state, we need to know the effective field theory at stage-(iii).  
The 't Hooft anomaly-matching condition applies to all these stages [in the true ultraviolet (UV)-limit (i), anomalies are replaced with the corresponding LSM indices].  
For instance, one may use the matching between the indices at (i) and (ii) to check the validity of low-energy field theories 
on which the standard RG approach is developed especially when the continuum limit is not obvious 
[as in the case of the KLM \eqref{eqn:SUN-fermion-spin}].  

As we will see below, the UV index is matched by the IR one [at (iii)] in different ways depending on $N$, the filling $f$, 
and low-energy (IR) effective theories at the stage-(iii).  
There may be several candidate scenarios of the ground-state phase for a given set of parameters.  
However, whatever scenario we take, the corresponding IR effective theory [(iii)] must share the same anomaly with the original lattice model [(i)].  
The constraints obtained this way is independent of the details of the model (such as the sign and strength of $J_{\text{K}}$).  
Therefore, we need case-by-case analyses to find the actual ground state as will be done in Sec.~\ref{sec:Weakcoupling}.  
Below, we introduce two different formulations of the low-energy effective theories associated to  
the two conformal embeddings and check how anomaly arises in each formulation. 

\subsubsection{SU($N$)$_1$
$\times$ SU($N$)$_1$ basis}
\label{sec:mixednanomalybasis1}
In the first, we describe the system in terms of the charge boson $\Phi_{\text{c}}$ [$\text{U(1)}_{\text{c}}$] and the SU($N$)$_{1,\text{f}}$ WZNW CFT 
which originate from the conduction electrons as well as the SU($N$)$_{1,\text{s}}$ WZNW CFT from the local moments 
(see Sec.~\ref{sec:Contlimit}). 
The one-site translation symmetry T$_{a_0}$ introduced in Sec.~\ref{sec:LSM-momenum-counting} 
is translated into a crucial {\em on-site internal} symmetry in the effective field theories   
that governs the low-energy properties of the model \eqref{eqn:SUN-KHM}.  
The form of the one-site translation T$_{a_0}$ for the low-energy fields can be read off directly from the correspondence 
\eqref{spinelecopfin} and \eqref{spinop} as:
\begin{equation}
\begin{split}
& \Phi_{\text{c}} \xrightarrow{\text{T$_{a_0}$}}   \Phi_{\text{c}}  +  \sqrt{ \frac{N}{\pi}} k_{\text{F}} a_0 
=  \Phi_{\text{c}}  +  \sqrt{ N \pi} f    \\
& g_{\text{f}}  \xrightarrow{\text{T$_{a_0}$}}  g_{\text{f}}  \; , \quad   
g_{\text{s}}  \xrightarrow{\text{T$_{a_0}$}}  \be^{\frac{ i 2 \pi }{N}}  g_{\text{s}}  \; .
\end{split}
\label{transfield}
\end{equation}
The original T$_{a_0}$ symmetry on a lattice translates to a filling-dependent shift of the charge bosonic field $\Phi_{\text{c}}$,  
whereas it acts on the spin WZNW  $g_{\text{s}}$ field 
as a discrete  $\mathbb{Z}_N$ symmetry which is the center of the SU($N$) group.
The UV limit, i.e., non-interacting limit of the SU($N$) KHM \eqref{eqn:SUN-KHM}, is then described by a CFT which is built 
from U(1)$_{\text{c}}$ $\times$ SU($N$)$_{1,\text{f}}$ $\times$ SU($N$)$_{1,\text{s}}$ CFTs enriched by the on-site internal symmetry $\text{T}_{a_0}$ \eqref{transfield}.   

Mixed global anomaly for the SU($N$)$_k$ WZNW CFT enriched with a discrete $\mathbb{Z}_p$ symmetry has been studied 
over the years in different contexts \cite{Felder-G-K-88,Numasawa-Y-18,Tanizaki-S-18,Yao-H-O-19}.
It is known that there is a mixed global anomaly between  $\mathbb{Z}_N$  and PSU($N$) $=$ SU($N$)/$\mathbb{Z}_N$ symmetry of the SU($N$)$_1$ WZNW model \cite{Tanizaki-S-18, Numasawa-Y-18, Yao-H-O-19}.
Coupling the WZNW model to a nontrivial background gauge field ${\cal A}_{\text{PSU}(N)}$ gives a non-trivial phase ambiguity in the partition function 
of the theory (${\cal Z}_{\text{WZNW}}$) under the action of the $\mathbb{Z}_N$ group
\cite{Yao-H-O-19}:
\begin{equation}
{\cal Z}_{\text{WZNW}} [ {\cal A}_{\text{PSU}(N)} ] \rightarrow  \exp{(i 2 \pi/N)} {\cal Z}_{\text{WZNW}} [ {\cal A}_{\text{PSU}(N)} ] .
\label{partitionfuncWZNW}
\end{equation}
This phase ambiguity in the description reveals the existence of a mixed global anomaly 
which, according to the 't Hooft-anomaly-matching argument \cite{tHooft-anomaly-80}, should be present non-perturbatively 
in the low-energy effective field theory 
with PSU($N$) $\times$ $\mathbb{Z}_N$ symmetry which governs the IR physics of the lattice model.\footnote{%
At this point, one may wonder if the modular anomaly, which is known \cite{Felder-G-K-88} to exist in the IR field theories, plays some roles in restricting 
the low-energy behavior of the lattice (UV) model.  However, since the modular invariance is an emergent symmetry which does not exist 
on a lattice, we cannot use it with the 't Hooft anomaly-matching condition.}

First let us consider the situation where all the three degrees of freedom [$\text{U(1)}_{\text{c}}$, SU($N$)$_{1,\text{f}}$, 
and SU($N$)$_{1,\text{s}}$] are gapless as is the case for the RG stage-(ii) or $J_{\text{K}}=0$.  
Then, there are two contributions to the entire anomaly.  First, a combination of the charge-conservation U(1)$_{\text{c}}$ and 
T$_{a_{0}}$ [which is a subgroup of chiral-U(1)] leads to a chiral anomaly: $\be^{i 2\pi f}$, while PSU($N$) and $\mathbb{Z}_{N}$ 
give $\be^{i \frac{2\pi}{N} }$ as is seen in \eqref{partitionfuncWZNW}.    
The total phase 
\begin{equation}
\exp[ i 2\pi \mathcal{I}^{(1)}_{N,f}] = \be^{i 2\pi(f+1/N) } 
\end{equation}
originating from the mixed anomaly perfectly coincides with the phase $\be^{i 2\pi \mathcal{I}_{1}}$ associated 
with the first LSM index $\mathcal{I}_{1}$ \eqref{eqn:1st-LSM-index}.  
The second LSM index $\mathcal{I}_{2}$ \eqref{eqn:2nd-LSM-index} is related to the following internal U(1) acting only on the charge sector:
\begin{equation}
\Phi_{\text{c}}  \to  \Phi_{\text{c}}  +  N\sqrt{ N \pi} f   \; , 
\label{transfield-1b}
\end{equation}
for which we obtain another anomaly index:
\begin{equation}
\exp[ i 2\pi \mathcal{I}^{(2)}_{N,f}] = \be^{i 2\pi N f }   \; .
\label{eqn:2nd-anomaly-index}
\end{equation}

According to recent identification of the LSM indices as the lattice counterparts of anomalies in the underlying 
field theories \cite{Cheng-Z-B-V-B-16,Cho-H-R-17,Furuya-O-17,Jian-B-X-18,Metlitski-T-18,Else-T-20,Cheng-S-23,Aksoy-M-F-T-23,Seifnashri-24}, 
this coincidence may be viewed as the manifestation of the 't Hooft anomaly matching \cite{tHooft-anomaly-80}  
between the lattice model \eqref{eqn:SUN-KHM} that may be considered as the extreme UV limit 
and the IR effective field theory $\mathcal{H}_\text{hop} + \mathcal{H}_{\rm H}$ [see Eqs.~\eqref{contfreehambis} and \eqref{sutherlandham}] 
with $\text{U(1)}_{\text{c}} \times \text{SU($N$)}_{1,\text{f}} \times \text{SU($N$)}_{1,\text{s}}$.  
Put it another way, fully-gapless metallic phase described by $c=1+(N-1)+(N-1)=2N-1$ CFT is allowed by 
't Hooft anomaly matching {\em regardless of filling} $f$ [see Fig.~\ref{fig:RG-invariant}(a)]. 
Although it is not straightforward to derive the effective field theory for the KLM \cite{Tsvelik-94} 
(due to the absence of the direct interaction $J_{\text{H}}$ among the local moments that lifts the huge degeneracy in the spin part), 
the coincidence between the lattice LSM and 't Hooft anomaly 
in the effective field theory suggests that the two lattice models (KLM and KHM) share the same low-energy effective theory 
$\mathcal{H}_\text{hop}+\mathcal{H}_{\rm H}$ at the stage-(ii). 

Suppose now the system is in an insulating phase in which the charge boson $\Phi_{\text{c}}$ is pinned and 
the charge sector becomes fully gapped [see Fig.~\ref{fig:RG-invariant}(b)].  
One can then integrate out this charge field to obtain the low-energy theory described solely 
by the $\text{SU($N$)}_{1,\text{f}} \times \text{SU($N$)}_{1,\text{s}}$ CFT.   
Now it is clear that we can no longer use \eqref{transfield} to represent T$_{a_0}$.  
After averaging over the charge field fluctuations in \eqref{spinelecopfin}, one sees that, provided that filling is 
commensurate $f=m/N$ [$m\,( \in \mathbb{Z})$ fermions per site], 
the one-site translation T$_{a_0}$ symmetry can also be implemented as:
\begin{equation}
\begin{split}
& g_{\text{f}} \xrightarrow{\text{T$_{a_0}$}}  \be^{\frac{ i 2 m\pi }{N}} g_{\text{f}}   \\
& g_{\text{s}}  \xrightarrow{\text{T$_{a_0}$}} \be^{\frac{ i 2 \pi }{N}}  g_{\text{s}} \; .
\end{split}
\label{transfieldaverage}
\end{equation}
The $\text{T}_{a_0}$ symmetry acts as a $\mathbb{Z}_N$ (respectively $\mathbb{Z}_{N/\text{gcd}(N,m)}$) symmetry 
for the $g_{\text{s}}$ (respectively $g_{\text{f}}$) WZNW field. 
Note that for $m \notin \mathbb{Z}$, the transformed $g_{\text{f}}$ is no longer an SU($N$) matrix and \eqref{transfieldaverage} is not 
allowed as a legitimate internal symmetry of the WZNW CFT. 
In the SU($N$)$_{1,\text{f}}$ $\times$ SU($N$)$_{1,\text{s}}$ formulation with the one-site translation action (\ref{transfieldaverage}), 
both $\text{SU($N$)}_{1}$ factors contribute non-trivial phases and the total phase ambiguity is given by: 
\begin{equation}
\exp[ i 2\pi \mathcal{I}^{(1)}_{N,f}] = 
\be^{i \frac{2m \pi}{N}} \be^{i \frac{2\pi}{N}} =\exp[ i 2 \pi (1/N+ f)] \; ,
\label{eqn:mixed-anomaly-SUNxSUN}
\end{equation}  
thereby correctly reproducing the first LSM index $\mathcal{I}_{1}$ \eqref{eqn:1st-LSM-index} at $f=m/N$ even after the charge sector is gapped out 
[the second one \eqref{eqn:2nd-LSM-index} which is $\mathcal{I}_{2} =m=0$ (mod $1$) does not give any constraint].  
Therefore, translation-invariant insulators are possible {\em only} at the commensurate fillings $f=m/N$ [$m \in \mathbb{Z}$; 
see Fig.~\ref{fig:RG-invariant}(b)]. 

For the other rational fillings $f=p/q$ [$p$ and $q\,(\neq N)$ being coprime], 
the implementation \eqref{transfieldaverage} of $\text{T}_{a_0}$ is no longer applicable and we need to go back to 
\eqref{transfield}.  Then, the set of UV (LSM) indices $(\mathcal{I}_{1}, \mathcal{I}_{2})= (f+1/N, Nf)$ 
and that of the IR insulating phase $(\mathcal{I}^{(1)}_{N,f},\mathcal{I}^{(2)}_{N,f})=(1/N,0)$ (mod $1$) never match (unless $f = 0$), 
and consequently opening a charge gap is precluded unless $\text{T}_{a_{0}}$ is broken.  
By matching $\mathcal{I}_{2}$ with its IR value, we can conclude that any insulating phase at generic rational fillings $f=p/q$ 
must spontaneously break $\text{T}_{a_{0}}$ and possess $q/\text{gcd}(N,q)$ degenerate ground states.   The behavior of 
the remaining spin sector is constrained by the new index $q \mathcal{I}_{1}/\text{gcd}(N,q)$ (mod $1$); when it is non-zero, the spin sector 
is either gapless or gapped accompanied by further breaking of $\text{T}_{a_{0}}$.  
The spin-charge dimerized insulator with algebraic spin-spin correlation proposed in Refs.~\onlinecite{Xavier-P-M-A-03,Xavier-M-08,Huang2020}  
for the $N=2$ KLM at $f=1/4$ [$q/\text{gcd}(N,q)=2$] perfectly fits the above scenario.  

Similarly, we can discuss the possibility of metallic phases with fully gapped spin excitations.  
When these happen, the LSM index $f+1/N$ and the mixed 't Hooft anomaly $f$ at stage-(iii) never match for generic filling $f$.  
This implies that spin-gapped metals are forbidden in general 
unless translation symmetry $\text{T}_{a_0}$ is broken [see Fig.~\ref{fig:RG-invariant}(b)]; if they are realized, the ground state must be at least $N$-fold degenerate and break $\text{T}_{a_0}$ spontaneously. 

Note that anomaly is absent if $\mathcal{I}^{(1)}_{N,f} (=\mathcal{I}_{1}) = f +1/N =0$ and $\mathcal{I}^{(2)}_{N,f} =Nf=0$ (mod $1$), 
which means that there is no obstruction to gapping out {\em all} the (spin and charge) degrees of freedom while preserving T$_{a_{0}}$; 
trivial IR theories, i.e., uniform fully gapped insulating ground states, whatever they may be, are possible {\em only} at the filling $f = (N-1)/N$.  
This is nothing but the special filling at which we find the SU($N$)-singlet Kondo insulator at strong coupling \cite{Totsuka-23}.  
In Sec.~\ref{sec:Weakcoupling}, we will identify two different types of such insulating phases depending on the sign of $J_{\text{K}}$.  
In contrast, for other fillings with ${\cal I}^{(1)}_{N,f}  \notin \mathbb{Z}$, the existence of a mixed global anomaly \eqref{eqn:mixed-anomaly-SUNxSUN} 
prevents the stabilization of a non-degenerate (translationally-invariant) fully-gapped ground states.  
In order to fulfill the constraint from the 't Hooft anomaly matching, the insulating ground states must either support gapless spin excitations 
or spontaneously break translation symmetry.  As we will see in the next section, both possibilities occur in the model \eqref{eqn:SUN-KHM} 
depending on $N$ and $f$. 
One thus reproduces the constraint from the LSM theorem in Sec.~\ref{sec:LSM-predictions} by exploiting the existence 
of a mixed global anomaly in the underlying field theory $\mathcal{H}_{\text{hop}} + \mathcal{H}_{\text{H}}$.  

\subsubsection{SU($N$)$_2$
$\times$ $\mathbb{Z}_N$ basis}
\label{sec:mixednanomalybasis2}
Next, we consider yet another  CFT embedding (see, for instance,  Refs. \onlinecite{Griffin-N-89,Lecheminant-T-15}) 
to single out the low-energy SU($N$) spin degrees of freedom of the original lattice model \eqref{eqn:SUN-KHM} \footnote{%
Intuitively, we move from the tensor-product basis of a pair of spins to a new basis in which the total spin is diagonal.}:
\begin{equation}
\text{SU($N$)}_{1} \times \text{SU($N$)}_{1} \sim  \text{SU($N$)}_2 \times \mathbb{Z}_N ,
\label{embedding}
\end{equation}
where ${\mathbb{Z}}_N$ denotes the parafermionic CFT with central charge $c=2(N-1)/(N+2)$ which
describes the universal properties of the phase transition of the two-dimensional $\mathbb{Z}_N$
clock model \cite{Zamolodchikov-F-JETP-85}.
The SU($N$)$_2$ CFT has central charge $c=2(N^2-1)/(N+2)$ and is generated by the currents  $I^A_{ \text{R},\text{L}}$ defined as follows:
\begin{equation}
I^A_{ \text{R}/\text{L} } = J^A_{ \text{f}, \text{R}/\text{L}} + J^A_{ \text{s}, \text{R}/\text{L}}  \; .
\label{currentsembedding}
\end{equation}
See Fig.~\ref{fig:RG-invariant}(a) for the relation between the two different low-energy descriptions in terms of 
$\text{SU($N$)}_1 \times \text{SU($N$)}_1$ and $\text{SU($N$)}_2 \times \mathbb{Z}_N$. 

When $N>2$, the two  SU($N$)$_1$  WZNW fields $g_{\text{f}}$ and $g_{\text{s}}$ can be expressed in the 
SU($N$)$_2$ $\times$ ${\mathbb Z}_N$ basis as \cite{Griffin-N-89,Lecheminant-T-15}:
\begin{equation}
\begin{split}
& (g_{\text{s}})_{\alpha \beta} \sim  G_{\alpha \beta} \, \sigma_1   \\
& (g_{\text{f}})_{\alpha \beta} \sim  G_{\alpha \beta}  \,  \sigma_1^{\dagger} , 
\end{split}
\label{ident}
\end{equation}
where $\alpha, \beta = 1, \ldots, N$, and $G$ is the SU($N$)$_2$  WZNW field with the scaling dimension $x_G = (N^2-1)/N(N+2)$ 
which transforms in the fundamental representation of SU($N$).  In Eq. (\ref{ident}), the first ${\mathbb Z}_N$ spin field $ \sigma_1$    
is one of the local order parameters $\sigma_k$ ($k=1,..,N-1$) which are primary fields 
of the $\mathbb{Z}_N$ CFT with the scaling dimension $x_{\sigma_{k}} = k(N-k)/N(N+2)$ and describe the low-temperature phase 
of the  two-dimensional $\mathbb{Z}_N$ clock model.  
When $N>2$, $\sigma_1$  and $\sigma^{\dagger}_1 = \sigma_{N-1}$ are independent fields with the same scaling dimension $x_{\sigma_{1}}$.

Again, the crucial step is to identify the translation symmetry T$_{a_{0}}$ as an internal symmetry in the SU($N$)$_2$
$\times$ $\mathbb{Z}_N$ basis.  We first try to implement \eqref{transfield} in the new basis.   
Using the identification \eqref{ident}, we immediately see that the following transformation
\begin{equation}
\begin{split}
&  \text{U(1)}: \;\; \Phi_{\text{c}} \rightarrow \; \Phi_{\text{c}}  +  \sqrt{ N \pi} f  \\
& \text{SU($N$)}_{2}:  \;\; G \rightarrow    \; \be ^{i \pi(1 - N)/N}  G  \\
& \mathbb{Z}_{N} : \;\;  \sigma_1  \rightarrow   \;  \be ^{i \pi(1 +N)/N} \sigma_1   
\end{split}
\label{transfield-2}
\end{equation}
works when $N=\text{odd}$, whereas, for even-$N$, there is no consistent way of translating \eqref{transfield} 
into the $\text{SU($N$)}_2 \times \mathbb{Z}_N$ language.  
It is easy to verify that the set of internal symmetries \eqref{transfield-2} ($N=\text{odd}$) 
leads to the same mixed anomaly $\mathcal{I}^{(1)}_{N,f}=f+1/N$.   

When the charge field $\Phi_{\text{c}}$ is fully gapped ($f$ is assumed to be commensurate, i.e., $f=m/N$), 
one can use Eq.~\eqref{transfieldaverage} instead to show that the $\text{T}_{a_0}$ symmetry can be consistently implemented only 
when $N$ is odd or when $N$ is even and $m$ is odd.  
For these cases, $\text{T}_{a_0}$ is implemented in the new [$\text{SU($N$)}_2 \times \mathbb{Z}_N$] basis as:
\begin{equation}
\begin{split}
& G \rightarrow   \; e ^{i \pi(1+m)/N} G  \\
& \sigma_1 \rightarrow    \; e ^{i \pi(1 - m)/N} \sigma_1  \; ,
\end{split}
\label{transoddm}
\end{equation}
when $m$ is odd ($N$ is arbitrary), and as
\begin{equation}
\begin{split}
& G \rightarrow    \; \be ^{i \pi(1+m-N)/N}  G  \\
& \sigma_1  \rightarrow   \;  \be ^{i \pi(1-m+N)/N} \sigma_1 \;  ,
\end{split}
\label{transevenm}
\end{equation}
when $N$ is odd and $m$ is even.  

There is no solution when both $N$ and $m$ are even such that $G$ is an SU($N$) matrix, i.e., ${\rm det} \, G =1$. It means that the T$_{a_0}$ symmetry cannot be consistently implemented as an internal symmetry. 
In such a case, the conformal embedding (\ref{embedding}) is not suitable to elucidate the low-energy properties of the SU($N$) KHM (\ref{eqn:SUN-KHM}). However, one can still use the SU($N$)$_1$ $\times$ SU($N$)$_1$ basis 
even when $N$ and $m$ are even as it will be the case for the half-field case.  

When the one-step translation symmetry T$_{a_0}$ can be consistently described as an internal symmetry in the
SU($N$)$_2$ $\times$ $\mathbb{Z}_N$ basis, one can derive a phase ambiguity as in Eq. (\ref{eqn:mixed-anomaly-SUNxSUN}) by exploiting the fact that the $\mathbb{Z}_N$ CFT is not anomalous \cite{Lin-S-21} and that the level-2 of the SU($N$)$_2$ CFT gives an extra factor $2$ 
in the phase of the partition function \eqref{partitionfuncWZNW} \cite{Numasawa-Y-18}.   
The implementations \eqref{transoddm} and \eqref{transevenm} give then respectively the total phase ambiguity: 
\begin{equation}
\begin{split}
& \exp[ i 2\pi \mathcal{I}^{(1)}_{N,f}] := 
\be^{i \frac{2\pi (m+1)}{N}} =\exp[ i 2 \pi (1/N+ f)] \; , \\
& \exp[ i 2\pi \mathcal{I}^{(1)}_{N,f}] := 
\be^{i \frac{2\pi (m+1 -N)}{N}} =\exp[ i 2 \pi (1/N+ f)] \; ,
\end{split}
\label{eqn:mixed-anomaly-SUNxZN}
\end{equation}  
thereby correctly reproducing the first LSM index $\mathcal{I}_{1}$ \eqref{eqn:1st-LSM-index} at $f=m/N$ as 
in the SU($N$)$_{1,\text{f}}$ $\times$ SU($N$)$_{1,\text{s}}$ formulation.


\section{Weak-coupling approach to the insulating phases of the SU($N$) Kondo-Heisenberg Model}
\label{sec:Weakcoupling}

In the previous sections, we have seen how the combination of the LSM indices of the lattice model and the mixed global anomalies in the IR effective 
theory constrains possible ground states for given $N$ and the filling $f$.   However, to identify the physical properties of 
the actual ground states, detailed case-by-case analyses are necessary.  
In this section, we focus on insulating phases at commensurate fillings 
and investigate them of the SU($N$) KHM (\ref{eqn:SUN-KHM}) by means of the low-energy approach of 
Sec. \ref{sec:Anomaly}. To this end, we consider a weak-coupling region where $|J_{\text{K}}| \ll t, J_{\text{H}}$ 
for commensurate fillings $f=m/N$ ($m = 1 \ldots N -1$). 
We focus only on the insulating phases, compatible with the LSM constraints, that can be stabilized 
in the zero-temperature phase diagram of the  SU($N$) KHM (\ref{eqn:SUN-KHM}). 

Let us first find the continuum expression of the Kondo coupling $\mathcal{H}_{\text{K}}$ that gives the interactions among 
the low-energy field effective theories \eqref{contfreehambis} and \eqref{sutherlandham}.  
To this end, we first plug the continuum limit of the SU($N$) spin operators of the conduction electron \eqref{spinelecopfin} 
and those of the localized moments \eqref{spinop} into the Kondo coupling $\mathcal{H}_{\text{K}}$, 
and then keep only the non-oscillatory terms satisfying 
$2 k_{\text{F}} + 2 p  \pi/(N a_0) \equiv 0 \; (\rm{mod} \; 2 \pi)$, with $p =1 , \dots N-1$ and $k_{\text{F}} = \pi m /(Na_0)$.  
 We thus find the following low-energy expression of the Kondo coupling:
\begin{equation}
\begin{split}
{\cal H}^{f=m/N}_{\text{K}} =& - J_{\text{K}} \, a_0 \alpha_{N-m} C \,  \be^{i \sqrt{ 4 \pi/N} \Phi_{\text{c}}}  \\
& \times  {\rm Tr} ( g_{\text{f}} \, T^A)
 \; {\rm Tr} (  \Phi_{\text{s},N-m} T_{N-m}^A)  + \text{H.c.} ,
 \end{split}
\label{contkondoleadinggenfilling}
\end{equation}
where $\Phi_{\text{s},p}$ denotes the $\text{SU($N$)}_{1,\text{s}}$ primary field appearing in Eq.~\eqref{spinopprimaryappen} 
and $ \alpha_{N-1} = \lambda$. 
We recall that the scaling dimension of the  $\Phi_{\text{s},p}$  field is $p (N-p)/N$ so that the interaction (\ref{contkondoleadinggenfilling}) 
has the scaling dimension 
\begin{equation}
\begin{split}
x_N (m) &= 1/N + (N-1)/N + m (N-m)/N \\
& = 1+ m (N-m)/N  
\end{split}
\label{eqn:scaling-dim-Kondo-int}
\end{equation}
and can be strongly relevant when $x_N (m) < 2$.  
The IR properties of the interaction (\ref{contkondoleadinggenfilling}) strongly depends on $m$ (i.e., filling $f$) 
leading to the stabilization of several different insulating phases as expected from the LSM argument.  
On top of this interaction, there is a marginal piece which stems from current-current interactions:
\begin{equation}
\mathcal{V}_{JJ} = J_{\text{K}}  \left( J^{A}_{\text{f},\text{L}} J^A_{\text{s}, \text{R}}  
+ J^{A}_{\text{f}, \text{R}} J^A_{\text{s},\text{L}}  \right) 
- \gamma J^A_{\text{s}, \text{R}} J^A_{\text{s}, \text{L}},
\label{contkondomarginal}
\end{equation}
where we have neglected current-current interactions made of currents of the same chirality (L/R) which just renormalize the ``light'' velocity 
$v_{\text{F}}$, and $\gamma$ in the marginally irrelevant coupling constant appearing in Eq.~\eqref{sutherlandham}. 

\subsection{$f =  \frac{N-1}{N}$}
\label{subsec:N-1/N}
We first consider the situation with $m=(N-1)$ fermions per site, i.e., filling $f= (N-1)/N$ since the LSM-anomaly argument 
predicts the possible formation of a featureless Kondo-insulating phase.   
In fact, a strong-coupling analysis is applicable when $J_{\text{K}} >0$ showing that the system is insulating 
as far as $J_{\text{K}}$ is sufficiently large \cite{Totsuka-23}.  
On the weak-coupling side, we start from the expression \eqref{contkondoleadinggenfilling} of the Kondo coupling, 
which simplifies for this filling as:
\begin{equation}
\begin{split}
& {\cal H}^{f=1-1/N}_{\text{K}}   \\
&  = - \frac{J_{\text{K}} a_0 C \lambda}{2}  \be^{i \sqrt{ 4 \pi/N} \Phi_{\text{c}}}  \left\{  \; {\rm Tr} ( g_{\text{f}}  g_{\text{s}}) 
-   \frac{1}{N} \; {\rm Tr} ( g_{\text{f}} ) \; {\rm Tr} (g_{\text{s}}) \right\}   \\
& \phantom{=} 
 + \text{H.c.}  \; .
\end{split}
\label{contkondoleading}
\end{equation}
This can also be expressed  in terms of the fields of the conformal embedding (\ref{embedding}):
\begin{equation}
\begin{split}
& {\cal H}^{f=1-1/N}_{\text{K}} = \mathcal{V}^{(1)}_{\text{K}} + \mathcal{V}^{(2)}_{\text{K}}   \\
&\mathcal{V}^{(1)}_{\text{K}}  = - J_{\text{K}} \lambda_1  \be^{i \sqrt{ 4 \pi/N} \Phi_{\text{c}}}  
\left\{  \mbox{Tr} \, G^2 +  \left( \mbox{Tr} \, G\right)^2  \right\} + \text{H.c.}   \\
& \mathcal{V}^{(2)}_{\text{K}}  = J_{\text{K}} \lambda_2  \be^{i \sqrt{ 4 \pi/N} \Phi_{\text{c}}}  \epsilon_1
\left\{ \left(\mbox{Tr} \, G\right)^2  - \mbox{Tr} \, G^2  \right\} + \text{H.c.} 
\end{split}
\label{kondoembedding}
\end{equation}
where the first thermal operator $ \epsilon_1$ of the ${\mathbb Z}_N$ CFT is singlet under the ${\mathbb Z}_N$ symmetry and has the scaling
dimension $x_{\epsilon_{1}} = 4/(N+2)$.    
In Eq.~\eqref{kondoembedding}, $\lambda_1$ and $\lambda_2$ are positive constants and the phase $\theta_0$ (\ref{phasespinop})
of the non-universal constant $\lambda$ has been absorbed in a redefinition of the charge field $\Phi_c$:
$\Phi_c \rightarrow \Phi_c +\sqrt{ N/4 \pi} \; \theta_0$.

The scaling dimension of the interaction in model \eqref{kondoembedding} is $2 -1/N <2$.  
It is thus a strongly relevant perturbation which couples the $\text{U(1)}_{\text{c}}$  charge degrees of freedom to the SU($N$)$_2$ 
and ${\mathbb Z}_N$ ones. A spectral gap $\Delta$ opens as: $\Delta \sim |J_{\text{K}}|^{N}$ regardless of the sign of $J_{\text{K}}$.  
A charge gap $\Delta_{\text{c}}$ is expected to open in the weak-coupling limit for either sign of $J_{\text{K}}$.
In the insulating phase, the charge field $\Phi_{\text{c}}$  is pinned.  As discussed in Appendix \ref{sec:umklappcharge},
one can determine the possible values of the pinning $ \langle \Phi_{\text{c}} \rangle$ by finding a pure umklapp operator which depends only on the charge  $\text{U(1)}_{\text{c}}$ degrees of freedom. 
In the even-$N$ case, we find $\langle \Phi_{\text{c}} \rangle = 0$ for either sign of $J_{\text{K}}$, while when $N$ is odd, 
one of the two inequivalent solutions $\langle \Phi_{\text{c}} \rangle = 0$ and $\sqrt{\frac{\pi}{4N}}$ must be chosen depending on the sign of the umklapp coupling.  
Below, we will keep only $\langle \Phi_{\text{c}} \rangle = 0$ for odd $N$, since the choice $\langle \Phi_{\text{c}} \rangle =\sqrt{\frac{\pi}{4N}}$ 
leads to physical results which are not consistent with those of the strong-coupling approach.  

Averaging over the charge degrees of freedom in the low-energy limit $E \ll \Delta_{\text{c}}$, 
the leading relevant contribution (\ref{kondoembedding}) becomes:
\begin{equation}
\begin{split}
& {\cal H}^{f=1-1/N}_{\text{K}} = \mathcal{V}^{(1)}_{\text{K}} + \mathcal{V}^{(2)}_{\text{K}}   \\
& \mathcal{V}^{(1)}_{\text{K}}  = - J_{\text{K}} \eta_1   \left\{ \mbox{Tr}\, G^2 +  \left(\mbox{Tr}\, G\right)^2 \right\} + \text{H.c.}  \\
& \mathcal{V}^{(2)}_{\text{K}}  = J_{\text{K}} \eta_2 \;   \epsilon_1
\left\{ \left(\mbox{Tr} \, G\right)^2  - \mbox{Tr} \,G^2  \right\} + \text{H.c.} ,
\end{split}
\label{kondoembeddingN=3}
\end{equation}
with $\eta_{1,2} >0$. When $E \ll \Delta_{\text{c}}$,  the one-site translation symmetry T$_{a_0}$ (\ref{transfieldaverage}) for $f=(N-1)/N$ 
acts on the spin degrees of freedom as:
\begin{equation}
\begin{split}
& G \rightarrow  G   \\
& \sigma_1 \rightarrow  \sigma_1  \, \be ^{i 2\pi/N} \; ,
\end{split}
\label{trans2}
\end{equation}
 so that T$_{a_0}$ is now realized as the global ${\mathbb{Z}}_N$ symmetry of the parafermionic CFT.  
 The model \eqref{kondoembeddingN=3} then reduces to the low-energy effective theory of the two-leg SU($N$) spin ladder with unequal spins, 
 one in the fundamental representation of SU($N$) and the other in its conjugate \cite{Capponi-F-L-T-20}.

\subsubsection{Kondo-singlet phase ($J_{\text{K}} > 0$)}
\label{sec:Kondo-singlet}

Let us first consider the $J_{\text{K}} > 0$ case in which we expect the singlet Kondo insulator 
when $J_{\text{K}}$ is large enough \cite{Totsuka-23}.  
The perturbation ${\cal V}^{(1)}_{\text{K}} $ is strongly relevant  
and opens a spin gap $\Delta_{\text{s}} \sim J_{\text{K}}^{N/2}$ for the SU($N$)$_2$ degrees of freedom.  
When $J_{\text{K}} >0$,  the WZNW $G$ matrix field is frozen to the ground-state configuration $G = \pm I$ (respectively $G =I$) 
if $N$ is even (respectively odd), with $I$ standing for the $N$-dimensional identity matrix.  
For the remaining ${\mathbb Z}_N$-sector, described the perturbation ${\cal V}^{(2)}_{\text{K}} $ in Eq.~\eqref{kondoembeddingN=3}, 
we get the following low-energy effective action when $E \ll \Delta_{\text{s}}$: 
\begin{equation}
{\cal S}_{\rm eff} = {\cal S}_{{\mathbb Z}_N} +  {\tilde \eta_2} \int d^2 x  \; \; \epsilon_1 ,
\label{effactionparaJperpF}
\end{equation}
with ${\tilde \eta_2} > 0$. In Eq.~\eqref{effactionparaJperpF}, $ {\cal S}_{{\mathbb Z}_N}$ is the Euclidean action 
of the ${\mathbb Z}_N$ parafermion CFT.   The ${\mathbb Z}_N$ effective action \eqref{effactionparaJperpF} is integrable 
and describes a massive field theory for either sign of ${\tilde \eta}_2$ \cite{Fateev-91}. 
Since $ {\tilde \eta_2} >0$ here, we have $\langle  \epsilon_1 \rangle <0$ and, in our convention, 
the underlying two-dimensional ${\mathbb Z}_N$ lattice model belongs 
to its high-temperature (paramagnetic) phase where the  ${\mathbb Z}_N$ symmetry is restored.  
It means that the one-site translation symmetry T$_{a_0}$ is preserved in the ground state [see Eq.~\eqref{trans2}];    
the resulting insulating phase is a fully gapped non-degenerate singlet phase which does not break any lattice symmetry.  
This may be physically identified as the SU($N$) Kondo singlet phase found in the strong-coupling analysis 
for $f = 1 -1/N$ and large positive $J_{\text{K}}$ in Ref.~\onlinecite{Totsuka-23},  
in which $(N-1)$ conduction electrons and one localized spin form an SU($N$) singlet on each site.  
A featureless fully gapped insulating phase predicted by the LSM argument for $f = 1 -1/N$ is thus realized for $J_{\text{K}} >0$ in the phase diagram 
of the SU($N$) KHM \eqref{eqn:SUN-KHM} from weak to strong positive $J_{\text{K}}$.
The Kondo singlet phase is illustrated in Fig.~\ref{fig:SU4-chiral-SPT-vs-Kondo-ins}(a).  

\subsubsection{Chiral symmetry protected topological phase ($J_{\text{K}}  < 0$)}

\begin{figure}[htb]
\begin{center}
\includegraphics[width=\columnwidth,clip]{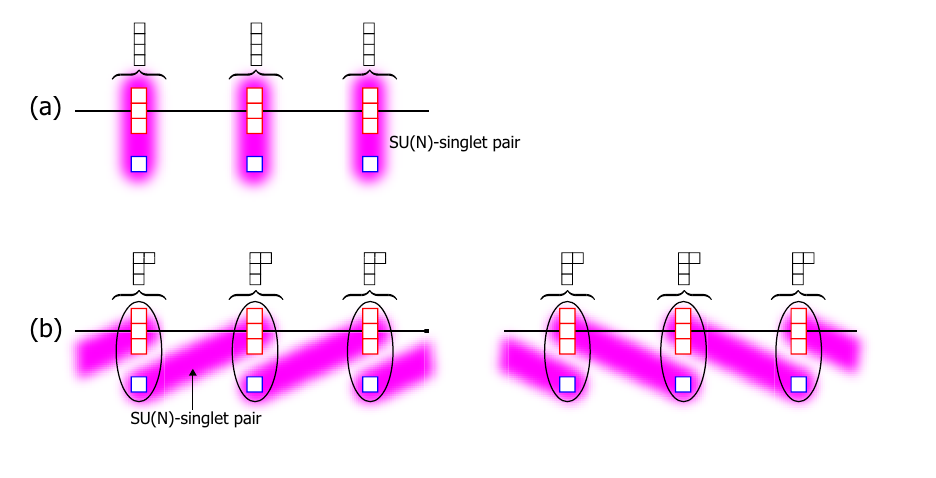}
\end{center}
\caption{Illustration of the two translation-invariant fully-gapped insulating states realized at $f=(N-1)/N$: 
(a) the SU(4) Kondo-singlet phase for $J_{\text{K}}>0$ and (b) the chiral SPT phase for $J_{\text{K}}<0$.   
The two states shown in (b) are parity partner of each other.  
The chiral SPT phase may be thought of as a {\em bond-centered} Kondo-singlet phase. 
\label{fig:SU4-chiral-SPT-vs-Kondo-ins}
}
\end{figure}

When $J_{\text{K}} < 0$, on the other hand, the minimization of the strongly perturbation ${\cal V}^{(1)}_{\text{K}} $ leads 
to the following solutions depending on the value of $N$ ($N > 2$) \cite{Capponi-F-L-T-20}:
\begin{equation}
\begin{split}
G= 
\begin{cases}
\pm i I  & (N=4p \geq 4)   \\
\be^{\pm i 2 p \pi/N} I  &  (N=4p+1 \geq 5)  \\
 \pm i \; {\rm diag} (1,1,1,1,1,-1) &  (N=6)   \\
\pm i \; \be^{\pm i  \pi /N}   I   &  (N=4p+2 \ge 10)   \\
\be^{\pm i (N+1) \pi/2N} I  &   (N=4p+3 \geq 3)  .
\end{cases}
\end{split}
\label{MinPhiS}
\end{equation}
Averaging over the $G$ field in Eq. (\ref{kondoembeddingN=3}), the effective action for the SU($N$)-singlet ${\mathbb Z}_N$ parafermion sector 
is still given by Eq. (\ref{effactionparaJperpF}) with ${\tilde \eta_2} >0$ and T$_{a_0}$  is unbroken as in the $J_{\text{K}} > 0$ case.  
In stark contrast to the SU($N$) Kondo-singlet phase found in the previous case, the insulating phase here breaks a discrete symmetry since the solutions (\ref{MinPhiS}) are two-fold degenerate ground states without breaking the one-site translation symmetry T$_{a_0}$. This discrete symmetry turns out to be the inversion symmetry or the site-parity symmetry $\mathcal{P}$:
\begin{equation}
\begin{split}
S^{A}_{i} &\xrightarrow{\mathcal{P}}  S^{A}_{-i}  \\
\hat{s}_{i}^{A} &\xrightarrow{\mathcal{P}}  \hat{s}_{-i}^{A}  \text{ ,}
\end{split}
\end{equation}
which is a symmetry of the lattice model (\ref{eqn:SUN-KHM}). Using the decompositions (\ref{spinop}) 
and (\ref{spinelecopfin}) and by averaging over the  $\text{U(1)}_{\text{c}}$ charge field, 
we find the identification of the inversion symmetry on the $g_{\text{s}}$ and $g_{\text{f}}$  WZNW fields:
\begin{equation}
\begin{split}
& g_{\text{s}} (x) \xrightarrow{\mathcal{P}}   - \be^{- i 2 \theta_0} g_{\text{s}}^{\dagger} (-x)  \\
& g_{\text{f}}(x) \xrightarrow{\mathcal{P}}   - \be^{i 2 \theta_0} g_{\text{f}}^{\dagger} (-x)  \text{ ,}
\end{split}
 \label{siteparityg}
\end{equation}
where the phase $\theta_0$ is defined by Eq. (\ref{phasespinop}). Using the conformal embedding (\ref{ident}), we observe that  $G (x)\xrightarrow{\mathcal{P}}   G^{\dagger} (-x) $
under the inversion symmetry.
The latter is thus spontaneously broken in the solutions (\ref{MinPhiS}) and two-fold degenerate ground states are formed 
without breaking the one-site translation symmetry T$_{a_0}$.  
In this respect, the resulting insulating phase for $J_{\text{K}} <0$ corresponds to the chiral SPT phase found in two-leg spin ladders with unequal spins or other 1D SU($N$) models in the adjoint representation of the SU($N$) group \cite{Rachel-S-S-T-G-10,Morimoto-U-M-F-14,Ueda-M-M-18,Roy-Q-18,Fromholz-C-L-P-T-19,Capponi-F-L-T-20}.  This fully gapped topological phase which is protected by the on-site projective SU($N$) [PSU($N$)] symmetry 
preserves T$_{a_0}$ but breaks the inversion symmetry spontaneously.  
In contrast to the Kondo-singlet phase in which SU($N$) singlets are formed mainly on the $J_{\text{K}}$ bonds 
[see Fig.~\ref{fig:SU4-chiral-SPT-vs-Kondo-ins}(a)],  
the electron spins form SU($N$) singlets with the local moments on {\em neighboring} sites [see Fig.~\ref{fig:SU4-chiral-SPT-vs-Kondo-ins}(b)].   
In this sense, we may regard this phase as a {\em bond-centered} Kondo-singlet phase. 
A pair of two non-degenerate chiral SPT phases [the two states in Fig.~\ref{fig:SU4-chiral-SPT-vs-Kondo-ins}(b)] 
that are related to each other by inversion are degenerate.  
In open-boundary conditions, these chiral SPT phases have different sets of the left and right edge states, related by the conjugation symmetry, 
which transform either in the fundamental representation or the anti-fundamental one.

\subsection{$f =  \frac{1}{N}$}
\label{subsec:1/N}

We now consider the SU($N$) KHM (\ref{eqn:SUN-KHM})  with $f=1/N$  and thus $k_{\text{F}} =  \frac{\pi}{Na_0}$.
The leading contribution (\ref{contkondoleadinggenfilling}) of the continuum limit of the Kondo coupling is now given by:
\begin{equation}
\begin{split}
& {\cal H}^{f=1/N}_{\text{K}} \\  
& =  \frac{J_{\text{K}} C \lambda a_0 }{2}  \be^{i \sqrt{ 4 \pi/N} \Phi_{\text{c}} }   \left\{  \; {\rm Tr} ( g_{\text{f}} \, g_{\text{s}}^{\dagger}) 
-   \frac{1}{N} \; {\rm Tr} ( g_{\text{f}} ) \; {\rm Tr} (g_{\text{s}}^{\dagger}) \right\}  \\
& \phantom{=}  + \text{H.c.} 
\end{split}
\label{contkondoleading1/N}
\end{equation}
Using the results of Refs.~\onlinecite{Lecheminant-T-15,Herviou-C-L-23}, one can rewrite this in 
the U(1)$_{\text{c}} \times$ SU($N$)$_2 \times  \mathbb{Z}_N$  basis: 
\begin{equation}
\begin{split}
& {\cal H}^{f=1/N}_{\text{K}}  =  \tilde{\mathcal{V}}^{(1)}_{\text{K}} + \tilde{\mathcal{V}}^{(2)}_{\text{K}}   \\
& \tilde{\mathcal{V}}^{(1)}_{\text{K}}  
=  J_{\text{K}} \delta_1  \be^{i \sqrt{ 4 \pi/N} \Phi_{\text{c}} } \Psi_{1\text{L}} \Psi_{1\text{R}} + \text{H.c.}    \\
& \tilde{\mathcal{V}}^{(2)}_{\text{K}}  =  - J_{\text{K}} \delta_2 \,  \be^{i \sqrt{ 4 \pi/N} \Phi_{\text{c}} } \sigma_2 \, \text{Tr} \, \Phi_{\rm adj}    + \text{H.c.} , 
\end{split}
\label{kondoembedding1/N}
\end{equation}
where $\delta_{1,2} >0$ and $\Psi_{1\text{L},\text{R}}$ are the first ${\mathbb{Z}}_N$ parafermion currents 
with the conformal weights $h, {\bar h} = (N-1)/N$ which generate the ${\mathbb{Z}}_N$ parafermion algebra.  
In Eq.~\eqref{kondoembedding1/N}, $\Phi_{\rm adj} $ is the SU($N$)$_2$ primary field in the adjoint representation 
with the scaling dimension $2N/(N+2)$. 
In Eq.~\eqref{kondoembedding1/N},  as in the $f =  \frac{N-1}{N}$ case, the phase $\theta_0$ (\ref{phasespinop})
of the non-universal constant $\lambda$ has been absorbed in a redefinition of the charge field $\Phi_c$:
$\Phi_c \rightarrow \Phi_c +\sqrt{ N/4 \pi} \; \theta_0$.  

Though the two perturbations \eqref{contkondoleading} and \eqref{contkondoleading1/N} share the same scaling dimension  
$x_N(1) = x_{N}(N-1) = 2 -1/N <2$, the resulting IR phases are very different. 
A charge gap is expected to open since the interaction  (\ref{contkondoleading1/N}) which couples the charge degrees of freedom 
to the SU($N$)$_2$ and ${\mathbb{Z}}_N$ ones is strongly relevant.  As discussed in Appendix~\ref{sec:umklappcharge}, 
a pure umklapp process can be derived by considering higher-order terms in perturbation theory.  
In the even-$N$ case, the charge field $\Phi_{\text{c}}$ is pinned at the configuration $\langle \Phi_{\text{c}} \rangle = 0$ 
regardless of the sign of $J_{\text{K}}$.  
In the odd-$N$ case, on the other hand, we have two different solutions: 
$\langle \Phi_{\text{c}} \rangle = 0$  and $\langle \Phi_{\text{c}} \rangle =\sqrt{\frac{\pi}{4N}}$.   
For the consistency with the strong-coupling results \cite{Totsuka-23} [see Eq.~\eqref{eqn:eff-Heisenberg-sym-rank-2}], 
the solution has to be chosen as:
\begin{equation}
\langle \Phi_{\text{c}} \rangle =
\begin{cases}
\sqrt{\frac{\pi}{4N}} & \text{when $J_{\text{K}} > 0$}  \\
0 & \text{when $J_{\text{K}} < 0$}  \; .
\end{cases}
\label{eqn:phi_c-f-1oN}
\end{equation}

\subsubsection{Even-$N$ case}

We first consider the even-$N$ case ($N >2$) where $\langle \Phi_{\text{c}} \rangle = 0$ for both $J_{\text{K}} >0$ and $J_{\text{K}} < 0$. 
Averaging over the charge degrees of freedom in the low-energy limit $E \ll \Delta_{\text{c}}$, 
the leading relevant contribution \eqref{kondoembedding1/N} in the Kondo coupling reads:
\begin{equation}
\begin{split}
& {\cal H}^{f=1/N}_{\text{K}}  =  \tilde{\mathcal{V}}^{(1)}_{\text{K}} + \tilde{\mathcal{V}}^{(2)}_{\text{K}}  \\
& \tilde{\mathcal{V}}^{(1)}_{\text{K}}  =  J_{\text{K}}  {\tilde \delta}_1 \,  \left( \Psi_{1\text{L}} \Psi_{1\text{R}} + \text{H.c.}\right)    \\
& \tilde{\mathcal{V}}^{(2)}_{\text{K}}  =  - J_{\text{K}} {\tilde \delta}_2  \,   \text{Tr} \, \Phi_{\rm adj}   \; \left(  \sigma_2  + \text{H.c.}\right) , 
\end{split}
\label{kondoembeddingave1/N}
\end{equation}
where ${\tilde \delta}_{1,2}$ are positive constants. The one-site translation symmetry T$_{a_0}$ is now given by
[set $m=1$ in Eq.~\eqref{transoddm}]: 
\begin{equation}
\begin{split}
& G \rightarrow  G \; \be ^{i 2\pi/N}   \\
& \sigma_1  \rightarrow  \sigma_1 ,
\end{split}
\label{trans2even}
\end{equation}
so that T$_{a_0}$ acts as the center of the SU($N$) group.
The low-energy theory (\ref{kondoembeddingave1/N}) is very similar to that of the two-leg SU($N$) spin ladder with an interchain exchange interaction $J_{\text{K}}$ \cite{Lecheminant-T-15,Weichselbaum-C-L-T-L-18}.
The two perturbations in Eq. (\ref{kondoembeddingave1/N}) are strongly relevant with the same scaling dimension $2(N-1)/N < 2$. 
The analysis of the low-energy properties of model (\ref{kondoembeddingave1/N})  has been presented in details in Ref.~\onlinecite{Lecheminant-T-15}.  

\begin{figure}[htb]
\begin{center}
\includegraphics[width=\columnwidth,clip]{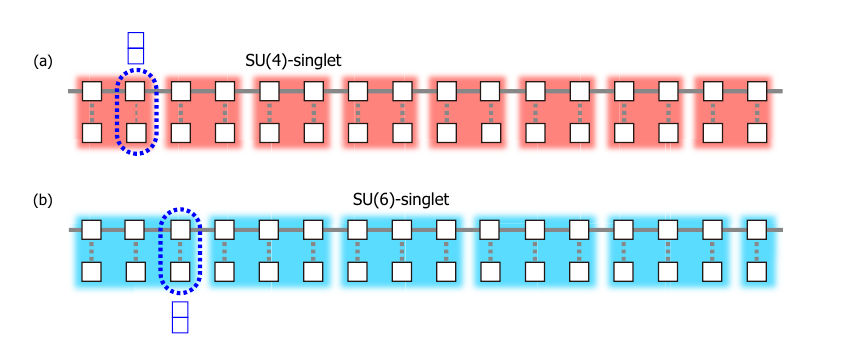}
\end{center}
\caption{The insulating ground states ({\em plaquette phase}) with finite spin gaps formed when $J_{\text{K}} >0$ and $N=\text{even}$: 
(a) $N=4$ and (b) $N=6$.  The $N$ spins enclosed by each colored square form an SU($N$) singlet.   
There are $N/2$ degenerate states related by translation.   
\label{fig:SUN-GS-positive-Jk}
}
\end{figure}

When $J_{\text{K}} >0$, a plaquette phase with a ground-state degeneracy $N/2$ which breaks spontaneously T$_{a_0}$ has been found.  
It corresponds to the formation of $4k_{\text{F}}$-valence-bond solid (VBS) and $4k_{\text{F}}$-CDW with order parameters:
\begin{equation}
\begin{split}
& {\cal O}_{4 k_{\text{F}} \text{-VBS}} \simeq  \be^{- \frac{i 4  \pi n}{N}} \; S^A_{n} S^A_{n+1}  \\
& {\cal O}_{4 k_{\text{F}} \text{-CDW}} \simeq   \be^{- i 4 k_{\text{F}} n } c^{\dagger}_{\alpha, n} c_{\alpha, n} \; .
\end{split}
\label{orderparameterplaquette}
\end{equation}
Physically, this SU($N$) Kondo-singlet plaquette phase is a product of SU($N$) singlets made from the hybridization 
of $N/2$ localized spins with $N/2$ fermion ones  (see Fig.~\ref{fig:SUN-GS-positive-Jk}; note that there is one fermion per site on average).  

\begin{figure}[htb]
\begin{center}
\includegraphics[width=\columnwidth,clip]{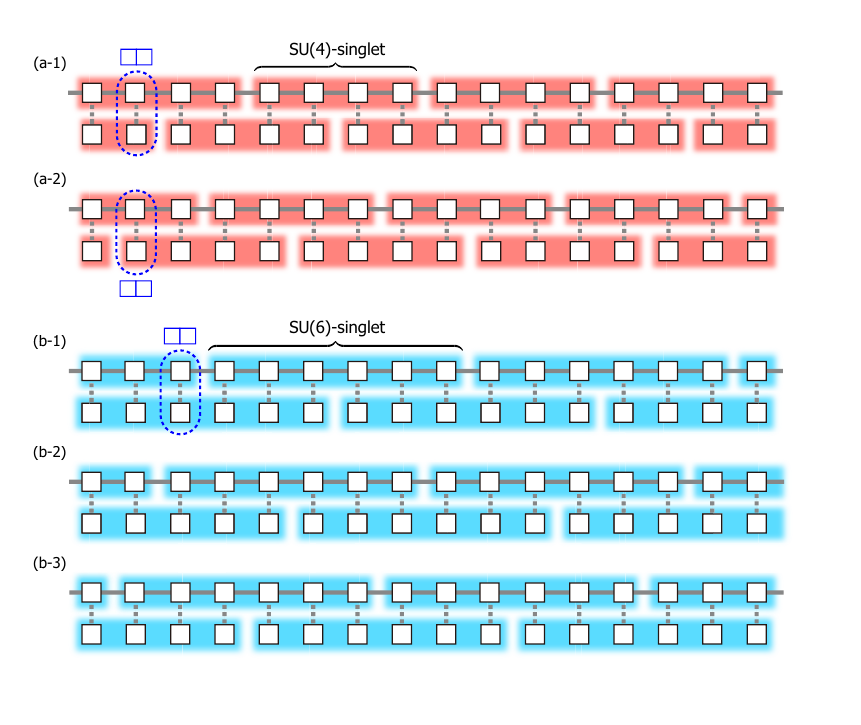}
\end{center}
\caption{The $N/2$-fold degenerate insulating ground states ({\em staggered-singlet phase}) with finite spin gaps formed when $J_{\text{K}} <0$ and $N=\text{even}$: (a-1,2) $N=4$ and (b-1,2,3) $N=6$.  The $N$ spins contained in each colored square form an SU($N$) singlet.     
\label{fig:SUN-GS-negative-Jk}
}
\end{figure}

When $J_{\text{K}}  < 0$, the perturbation $\tilde{\mathcal{V}}^{(1)}_{\text{K}}$ in Eq. (\ref{kondoembeddingave1/N}) 
acts only in the ${\mathbb{Z}}_N$ sector and corresponds to a massive integrable deformation 
of the ${\mathbb{Z}}_N$ parafermion \cite{Fateev-91,Fateev-Z-Zn-91}. 
The ${\mathbb{Z}}_N$ sector thus acquires a mass gap. In the low-energy limit, we can integrate out the $\mathbb{Z}_{N}$ degrees of freedom 
in $\tilde{\mathcal{V}}^{(2)}_{\text{K}}$ to derive the effective interaction in the SU($N$)$_2$ sector \cite{Lecheminant-T-15}:
\begin{eqnarray}
\mathcal{V}^{f=1/N}_{\text{eff}} \simeq    {\tilde \gamma} \;   \text{Tr} \, \Phi_{\rm adj},
\label{Heff1/NNodd}
\end{eqnarray} 
with ${\tilde \gamma} > 0$.  The effective field theory describes the SU($N$)$_2$ CFT perturbed by the adjoint operator which 
is a strongly relevant perturbation with the scaling dimension $2N/(N+2) < 2$ thereby opening a spectral gap in the spin sector 
when $N$ is even \cite{Lecheminant-15,Kikuchi-22}.  
Therefore, the ground state for $J_{\text{K}}  < 0$ is a $N/2$-fold degenerate full-gap insulator that is characterized by a staggered pattern 
of SU($N$)-singlets ({\em staggered-singlet phase}; see Fig.~\ref{fig:SUN-GS-negative-Jk}) similar to what was found 
in Ref.~\onlinecite{Lecheminant-T-15} for the two-leg SU($N$) spin ladder with ferromagnetic interchain interaction.  
In Appendix \ref{sec:flagsigma}, we argue the connection between the IR-limit of the effective field theory \eqref{Heff1/NNodd}  
with negative $J_{\text{K}}$ and the non-linear sigma model on the flag manifold  SU($N$)/U(1)$^{N-1}$ 
with $N-1$ topological angles $\theta_a = 4\pi a/N$ ($a = 1, \ldots, N-1$)  
which is known to describe the IR properties of the SU($N$) Heisenberg spin chain 
in the symmetric rank-$2$ tensor representation \cite{Affleck-B-W-22,Wamer-L-M-A-20,Wamer-A-20}. 
The latter spin chain is fully gapped with ground-state degeneracy $N/2$ \cite{Yao-H-O-19,Wamer-A-20}.  
Interestingly enough, as seen in Eq.~\eqref{eqn:eff-Heisenberg-sym-rank-2}, the SU($N$) KLM for $f=1/N$ is described 
by such a model in the strong-coupling regime $J_{\text{K}} \rightarrow - \infty$ \cite{Totsuka-23}.  
Thus we have arrived at consistent descriptions both for weak and strong couplings.   
\subsubsection{Odd-$N$ case}
Now let us consider the cases with odd-$N$. 
We begin with the $J_{\text{K}}  < 0$ case where the charge-field condenses such that $\langle \Phi_{\text{c}} \rangle =0$ 
[see Eq.~\eqref{eqn:phi_c-f-1oN}].  
The strongly relevant perturbation \eqref{kondoembeddingave1/N} describes the physical properties of the underlying insulating phase 
of the lattice model.   
As in the even-$N$ case, the ${\mathbb{Z}}_N$ sector is fully gapped, and by averaging over these massive degrees of freedom, 
we find the same effective interaction $\mathcal{V}^{f=1/N}_{\text{eff}}$ in Eq.~\eqref{Heff1/NNodd}.   
However, there is a striking difference from the previous case in the IR properties since $N$ is now odd.   
The SU($N$)$_2$ WZNW CFT perturbed by the effective interaction \eqref{Heff1/NNodd} has been investigated 
in Refs.~\onlinecite{Lecheminant-15,Kikuchi-22}; 
while the adjoint perturbation is a strongly relevant perturbation with the scaling dimension $2N/(N+2)$, a massless RG  flow 
from SU($N$)$_2$  to SU($N$)$_1$ CFT is predicted when $N$ is odd and ${\tilde \gamma} > 0$.  
Explicit proofs in the $N=3$ case have been given 
by mapping the model \eqref{Heff1/NNodd} with $N=3$ onto the ${\mathbb{Z}}_3$ Gepner's parafermions \cite{Lecheminant-15} 
or by exploiting a semiclassical analysis \cite{Herviou-C-L-23} (See also Appendix~\ref{sec:flagsigma}).   
Since the current-current interaction (\ref{contkondomarginal}) is marginally irrelevant and scales to zero when $J_{\text{K}} <0$, 
we find that  the SU($N$) KHM at $1/N$ filling belongs to an insulating phase with gapless SU($N$) spin degrees of freedom when $N$ is odd and $J_{\text{K}} <0$.   It describes a C0S$(N-1)$ insulating phase phase in full agreement with the LSM constraints for $f=1/N$  
and the strong-coupling result of the SU($N$) KHM since for odd $N$ the effective spin model \eqref{eqn:eff-Heisenberg-sym-rank-2} 
is believed to be gapless in the SU($N$)$_1$ universality class \cite{Rachel-T-F-S-G-09,Yao-H-O-19,Wamer-A-20}.

In the $J_{\text{K}} >0$ case, the $\Phi_{\text{c}}$-field is expected to be pinned at a different  
value $\langle \Phi_{\text{c}} \rangle = \sqrt{\pi/4N}$ and 
the low-energy interaction \eqref{kondoembedding1/N} reduces, after averaging over the charge degrees of freedom, to: 
\begin{equation}
\begin{split}
& {\cal H}^{f=1/N}_{\text{K}} = \tilde{\mathcal{V}}^{(1)}_{\text{K}} + \tilde{\mathcal{V}}^{(2)}_{\text{K}}   \\
& \tilde{\mathcal{V}}^{(1)}_{\text{K}}  = J_{\text{K}} {\tilde \delta}_1  \left( \be^{ \frac{i \pi}{N}}  \Psi_{1\text{L}} \Psi_{1\text{R}} + \text{H.c.} \right)    \\
& \tilde{\mathcal{V}}^{(2)}_{\text{K}} = - J_{\text{K}} {\tilde \delta}_2  \left( \be^{\frac{i \pi}{N}}  \text{Tr} \, \Phi_{\text{adj}}   \;   \sigma_2  + \text{H.c.}  \right) ,
\end{split}
\label{kondoembeddingave1/Ntotherpin}
\end{equation}
with ${\tilde \delta}_{1,2} >0$.  When $N$ is odd, one can absorb the phase factor $\be^{ \frac{i \pi}{N}}$ in Eq.~\eqref{kondoembeddingave1/Ntotherpin} 
by the following redefinition on the  ${\mathbb Z}_N$ parafermion currents:
\begin{equation}
\begin{split}
& \Psi_{k \text{L}} \rightarrow  {\tilde \Psi}_{k \text{L}} =\left( -1\right)^k \be^{i k \pi/N} \Psi_{k \text{L}}  \\
& \Psi_{k \text{R}} \rightarrow {\tilde \Psi}_{k \text{R}} =  \Psi_{k \text{R}} ,
\end{split}
\label{paracurrtransNodd}
\end{equation}
${\tilde \Psi}_{k\text{L}}$ being still a parafermionic current (${\tilde \Psi}_{k \text{L}}^{N} \sim I$) when $N$ is odd. 
The  transformation of the ${\mathbb{Z}}_N$ spin fields $\sigma_k$ should be consistent with the fusion rules of the  ${\mathbb{Z}}_N$ parafermionic theory \cite{Zamolodchikov-F-JETP-85}:
$\sigma_k \mu_k \sim \Psi_{k \text{L}}$ and $\sigma_k \mu_k^{\dagger} \sim \Psi_{k \text{R}}$ ($\mu_k$ being
the ${\mathbb{Z}}_N$  disorder fields). 
We thus deduce:
\begin{equation}
\begin{split}
& \sigma_k  \rightarrow  {\tilde \sigma}_k  = \be^{i k \pi/2N \pm i k \pi/2} \sigma_k  \\
& \mu_k  \rightarrow   {\tilde \mu}_k  = \be^{i k \pi/2N  \pm i k \pi/2}   \mu_k ,
\end{split}
\label{sigmatransfo}
\end{equation}
where the sign $+$ (respectively $-$) is chosen when $N = 4p +3$ (respectively $N = 4p +1$). 
After this transformation, the low-energy interaction \eqref{kondoembeddingave1/Ntotherpin} reads: 
\begin{equation}
\begin{split}
& {\cal H}^{f=1/N}_{\text{K}} = \tilde{\mathcal{V}}^{(1)}_{\text{K}} + \tilde{\mathcal{V}}^{(2)}_{\text{K}}  \\
& \tilde{\mathcal{V}}^{(1)}_{\text{K}}  = - J_{\text{K}} {\tilde \delta}_1  \left( {\tilde \Psi}_{1 \text{L}}  {\tilde \Psi}_{1 \text{R}}+ \text{H.c.} \right)    \\
& \tilde{\mathcal{V}}^{(2)}_{\text{K}} = J_{\text{K}} {\tilde \delta}_2  \, \text{Tr} \, \Phi_{\rm adj}   \,  \left(  {\tilde \sigma}_2   + \text{H.c.}  \right) ,
\end{split}
\label{kondoembeddingave1/Ntotherpinfin}
\end{equation}
which is identical to Eq. (\ref{kondoembeddingave1/N}) except for the sign flip: $J_{\text{K}} \rightarrow - J_{\text{K}}$.   
Now we can borrow the results obtained above for {\em negative} $J_{\text{K}}$ and $\langle \Phi_{\text{c}} \rangle = 0$; 
we again find a massless RG flow SU($N$)$_2$  $\rightarrow$ SU($N$)$_1$ 
which might imply an insulating C0S$(N-1)$ phase with a gapless spin sector described 
by the SU($N$)$_1$ CFT for {\em positive} $J_{\text{K}}$ as well.   

However, this is not the end of the story.  
In fact, for $J_{\text{K}} >0$, one has to be very careful about the marginal interaction \eqref{contkondomarginal}   
since along the massless RG flow $\text{SU($N$)}_2$ $\rightarrow$ $\text{SU($N$)}_1$, the SU($N$)$_2$  currents ${I}_{\text{L},\text{R}}^A$  
are transmuted to the SU($N$)$_1$ currents $\mathcal{J}_{\text{L},\text{R}}^A$ in the far-IR limit.  
The current-current interaction (\ref{contkondomarginal}) then gives a residual contribution in the low-energy effective Hamiltonian 
for the SU($N$)$_{1}$ spin sector:
\begin{equation}
  \mathcal{H}^{f=1/N}_{\rm IR}  = \frac{2\pi v}{N + 1} \left( : {\cal J}^A_{\text{R}} {\cal J}^A_{\text{R}}: 
  + : {\cal J}^A_{\text{L}} {\cal J}^A_{\text{L}}: 
\right)  + \lambda_{\rm eff}  {\cal J}_{\text{R}}^{A} {\cal J}_{\text{L}}^{A} ,
\label{HcontJ1J2trans}
\end{equation}
where $\lambda_{\rm eff}  > 0$ when $J_{\text{K}}$ is positive. 
As is well known \cite{James-K-L-R-T-18}, the effective Hamiltonian \eqref{HcontJ1J2trans} with positive $\lambda_{\rm eff}$ 
is a massive integrable field theory suggesting a fully gapped phase (C0S0) with: 
\begin{equation}
\begin{split}
& \langle {\cal O}_{ 2k_{\text{F}} \text{-VBS}} \rangle  := \langle  \be^{- i 2 \pi n/N} \, S^A_{n} S^A_{n+1} \rangle  \ne  0   \\
& \langle  {\cal O}_{2 k_{\text{F}} \text{-CDW}} \rangle   :=  \langle   \be^{- i 2 \pi n/N}  c^{\dagger}_{\alpha, n} c_{\alpha, n}\rangle  \ne  0 \; .
\end{split}
\label{orderparameterNevenJKpositif}
\end{equation}
These imply the coexistence of $2k_{\text{F}}$-VBS order and $2k_{\text{F}}$-CDW 
with a $N$-fold ground-state degeneracy that results from the 
spontaneous-breaking of T$_{a_0}$. 

\subsection{Other commensurate fillings}
\label{subsec:otherfilling}

We now consider general commensurate fillings $f =  \frac{m}{N}$ with $m \ne 1, N-1$.  The low-energy properties of the SU($N$) KHM (\ref{eqn:SUN-KHM})  are governed by the interaction (\ref{contkondoleadinggenfilling})  and the marginal piece (\ref{contkondomarginal}).
The interacting part \eqref{contkondoleadinggenfilling} has the scaling dimension $x_N (m) = 1+ m (N-m)/N$ 
[see Eq.~\eqref{eqn:scaling-dim-Kondo-int}] and can be relevant, marginal or irrelevant depending on filling, i.e., the value of $m$.  
For instance, at half-filling ($m=N/2$) the interaction is irrelevant when $N>4$ and 
the Kondo coupling  $\mathcal{H}_{\text{K}}$ (\ref{eqn:SUN-KHM})  is strongly oscillating and averages to zero in the low-energy limit when $N$ is odd.   
Nevertheless, a charge gap might be generated in higher-orders of perturbation theory.  
Therefore, we tentatively assume here the formation of a charge gap and discuss  the nature of the resulting insulating phase which emerges  within our low-energy approach. A comparison will be done with the LSM predictions summarized in Table \ref{tab:KHM-phases-by-LSM}.  
Detailed numerical analyses of the lattice model are called for to check the existence of the postulated charge gap 
for particular commensurate fillings and $J_{\text{K}}$. 

When the interaction (\ref{contkondoleadinggenfilling}) is strongly irrelevant,  the current-current contribution (\ref{contkondomarginal}) governs the IR properties of the SU($N$) KHM model:  
\begin{equation}
\mathcal{V}_{JJ} = g_1 \left( J^{A}_{\text{s},\text{L}} J^A_{\text{f}, \text{R}}  + J^{A}_{\text{s},\text{R}} J^A_{\text{f}, \text{L}}  \right) + g_2
J^A_{\text{s}, \text{R}} J^A_{\text{s}, \text{L}},
\label{contkondomarginalN}
\end{equation}
with initial conditions $g_1 (0) = J_{\text{K}}$ and $g_2 (0) = - \gamma < 0$.  

The one-loop RG equations for the perturbation \eqref{contkondomarginalN} are:
\begin{equation} 
{\dot g}_{1,2} = \frac{ N g^2_{1,2}}{4 \pi} \; . 
\label{1loopcurr}
\end{equation}
When $J_{\text{K}}  < 0$, the perturbation \eqref{contkondomarginalN} is marginally irrelevant and scales to zero in the far-IR limit. 
The resulting insulating phase supports gapless spin excitations and corresponds to a multicomponent Luttinger liquid phase C0S$2(N-1)$. 
 
 When $J_{\text{K}}  >0$, on the other hand, the interaction $g_{1}$ is marginally relevant 
and one finds $g_1 \rightarrow \infty$ and $g_2 \rightarrow 0$ in the far-IR limit.  
The low-energy theory that governs the strong-coupling behavior of the spin sector is then: 
\begin{equation}
\begin{split}
 {\cal H}_{\text{IR}} = & \frac{2\pi v^{\text{(f)}}_\text{s} }{N + 1} \left( : J^A_{\text{f}, \text{R}} J^A_{\text{f}, \text{R}}: 
+ : J^A_{\text{f}, \text{L}} J^A_{\text{f}, \text{L}}: \right)    \\
&+ \frac{2\pi v^{\text{(s)}}_\text{s}}{N + 1} \left( : J^A_{\text{s}, \text{R}} J^A_{\text{s}, \text{R}}: 
+ : J^A_{\text{s}, \text{L}} J^A_{\text{s}, \text{L}}: \right)   \\
& + g_{*} \left( J^{A}_{\text{s}, \text{L}} J^A_{\text{f}, \text{R}}  + J^{A}_{\text{s}, \text{R}} J^A_{\text{f}, \text{L}}  \right),
\end{split}
\label{HIRNevenhalfilling}
\end{equation}
with $g_{\ast} = g_1(t^{*}) >0$ ($t^{*}$ being the RG time when the strong-coupling regime is reached). 

One can solve this theory using a trick exploited in Ref.~\onlinecite{White-A-96} in the study of the two-leg zigzag spin ladder. 
Following the trick, we first perform a transformation on the set of the SU($N$)$_1$ currents 
$\{ J^{A}_{\text{f}, \text{L/R}}, J^{A}_{\text{s}, \text{L/R}} \}$ 
and introduce a new set $\{ J^{A}_{1, \text{L/R}}, J^{A}_{2, \text{L/R}} \}$:
\begin{equation}
\begin{split}
& J^{A}_{1, \text{L}} := J^A_{\text{f}, \text{L}} \; , \quad J^{A}_{1, \text{R}} := J^A_{\text{s}, \text{R}},   \\ 
& J^{A}_{2, \text{L}} := J^A_{\text{s}, \text{L}} \; , \quad  J^{A}_{2, \text{R}} := J^A_{\text{f}, \text{R}} \;  .
\end{split}
\label{currtransform}
\end{equation}
By neglecting the velocity anisotropy $|v^{\text{(s)}}_\text{f} - v^{\text{(s)}}_\text{s}|$, 
the IR Hamiltonian density (\ref{HIRNevenhalfilling}) separates into two commuting SU($N$)$_1$ Thirring models:
\begin{equation}
\begin{split}
& {\cal H}_{\text{IR}}  = {\cal H}_{1} +  {\cal H}_{2},   \\
& {\cal H}_{i} :=  \frac{2\pi v}{N + 1} \left( : J^A_{i, \text{R}} J^A_{i,\text{R}}: 
+ : J^A_{i, \text{L}} J^A_{i, \text{L}}: \right) + g_{*}  J^{A}_{i, \text{L}} J^A_{i, \text{R}}  \; , \\
& \left[{\cal H}_{1} , {\cal H}_{2} \right] = 0  \; . 
\end{split}
\label{Thirringmodels}
\end{equation}
The SU($N$) Thirring model ${\cal H}_{i}$ is exactly solvable and develops a non-perturbative spectral gap \cite{James-K-L-R-T-18} 
when $J_{\text{K}}  >0$ (i.e., $g_{\ast}>0$).   
Therefore, we conclude that the resulting insulating phase is fully gapped (C0S0). 

The next step is to identify the nature of this phase.  To this end, we introduce the SU($N$)$_1$ WZNW fields $G_{1,2}$ associated 
to the new set of currents \eqref{currtransform}.  
In the ground states of ${\cal H}_{1,2}$,  we have the long-range ordering of $\langle {\rm Tr} \, G_{1,2} \rangle$:
\begin{equation}
\begin{split}
& \langle {\rm Tr} \, G_1 \rangle =  \langle {\rm Tr} \left( g_{\text{f} \text{L}} \, g_{\text{s} \text{R}} \right) \rangle \ne 0 \; ,   \\
& \langle {\rm Tr} \, G_2 \rangle = \langle {\rm Tr} \left( g_{\text{s} \text{L}} \, g_{\text{f} \text{R}} \right) \rangle \ne 0 \; , 
\end{split}
\label{WZWav}
\end{equation}
where we have introduced the left and right components $(g_{\text{f} \text{L/R}},g_{\text{s} \text{L/R}})$ of the original WZNW fields 
$g_{\text{f}}$ and $g_{\text{s}}$. 
The non-zero expectation values of the composite order parameters $\langle {\rm Tr} \, G_{1,2} \rangle$ \eqref{WZWav} 
indicate that there is a strong hybridization between the SU($N$) spins of the itinerant fermion ($g_{\text{f}}$) 
and the local moment ($g_{\text{s}}$) in the ground state of the model ${\cal H}_{\text{IR}}$ \eqref{Thirringmodels}.  
With this in mind, we introduce a spin-polaron which is a bound-state formed by the conduction electron and the localized spin moment  
as in Refs.~\onlinecite{Danu-L-A-R-21,Zhang-V-22}:
\begin{equation}
{\tilde c}^{\dagger}_{\alpha,n} :=  c^{\dagger}_{\beta,n} T^A_{\beta \alpha}  S_{n}^A .
\label{dressedfermion}
\end{equation}
Out of the spin-polaron ${\tilde c}^{\dagger}_{\alpha,n}$ and the itinerant fermion, we could then define a composite-CDW order parameter 
with oscillations at $2k^{*}_{\text{F}}$ ($k^{*}_{\text{F}} = \tfrac{m\pi}{Na_0} +  \tfrac{\pi}{N a_0} $):
\begin{equation}
{\cal O}_{\text{c-CDW}} \simeq   \be^{- i  2 k^{*}_{\text{F}} n }   {\tilde c}^{\dagger}_{\alpha,n} c_{\alpha,n} 
= \be^{- i  2 k^{*}_{\text{F}} n } \hat{s}_{n}^{A} S^A_{n} ,
\label{compositeCDW}
\end{equation}
which couples the dominant fluctuation component of the conduction-electron spin to that of the localized moment  
[note that $\tfrac{2m\pi}{Na_0}$ and $\tfrac{2\pi}{N a_0}$ are from the conduction electrons and the local moments, respectively].  
The characteristic momentum of the resulting CDW gets renormalized and shifted 
from the value $2k_{\text{F}}= \tfrac{2m\pi}{Na_0}$ expected from the fermion filling to $2 k^{*}_{\text{F}}$ 
by the momentum of the localized-spin fluctuations.  
This order parameter (\ref{compositeCDW}) with a large Fermi surface associated with its composite nature has already been introduced 
in the context of the 1D SU(2) KHM for incommensurate fillings in Refs.~\onlinecite{Zachar-01,Zachar-T-01,Berg-F-K-10}.  
The continuous description of the composite-CDW order parameter (\ref{compositeCDW}) can be obtained  
by means of the identities \eqref{spinelecopfin} and  \eqref{spinop}:
\begin{equation}
\begin{split}
{\cal O}_{\text{c-CDW}} &\simeq - \lambda C \be^{ i \sqrt{ 4 \pi/N} \Phi_{\text{c}}}  \; 
{\rm Tr} ( g_{\text{f}} \, T^A)  \; {\rm Tr} ( g_{\text{s}} \, T^A) \\
&\sim - \frac{\lambda C}{2} \left\{   {\rm Tr} ( g_{\text{f}}\,  g_{\text{s}} ) - \frac{1}{N} {\rm Tr} ( g_{\text{f}})  \; {\rm Tr} ( g_{\text{s}})  \right\},  
\; E \ll \Delta_c \; ,
\end{split}
\label{compositeCDWcont}
\end{equation}
where the charge degrees of freedom have been averaged over around $\langle \Phi_{\text{c}} \rangle = 0$ . 
In the ground-state of the low-energy Hamiltonian \eqref{Thirringmodels}, we find:
\begin{equation}
\langle {\cal O}_{\text{c-CDW}} \rangle \sim   \langle  {\rm Tr} ( G_1 )   \rangle  \langle  {\rm Tr} ( G_2 )   \rangle  \ne 0  .
\label{compositeCDWvev}
\end{equation}

This phase breaks T$_{a_0}$ spontaneously leading to degenerate ground states \footnote{%
Note that after T$_{a_0}$, the right-hand side acquires a phase $\be^{\frac{ i 2 (m+1)\pi }{N}}$ by Eq.~\eqref{transfieldaverage}.}.  
For instance, in the half-filled case $f=1/2$ 
($N$ is assumed even), the momentum of the composite CDW (\ref{compositeCDW}) is $2k^{*}_{\text{F}} = \tfrac{\pi}{a_0} +  \tfrac{2\pi}{N a_0} $.  The degeneracy depends on the parity of $N/2$; when $N=4p +2$ (respectively $N = 4p > 4$) the ground-state degeneracy is $N/2$ (respectively $N$). 
We thus find, at half-filling $f=1/2$, the emergence of a fully gapped $2k^{*}_{\text{F}}$-composite CDW phase for $J_{\text{K}}  >0$ 
with ground-state degeneracy which is consistent with the LSM  prediction \eqref{eqn:LSM-GSD-half-filling}.  

Finally, a remark is in order about the treatment of the interactions in this section.  
In the above argument, we have assumed that the first part \eqref{contkondoleadinggenfilling} of the Kondo coupling is irrelevant so that 
the marginal part \eqref{contkondomarginal} plays a crucial role.  
However, the interaction \eqref{contkondoleadinggenfilling} can be strongly relevant in some particular cases.  
For instance, for $N=8$ with $m=2$, we have the scaling dimension $x_8 (2) = 7/4 < 2$, while at half-filling with $N=4$ the interaction is marginal 
[$x_4(2) = 2$] and competes with the current-current interaction (\ref{contkondomarginal}).  
In such situations, a special analysis of the interaction (\ref{contkondoleadinggenfilling}) is required which is 
beyond the scope of this paper and will be addressed elsewhere.

In Fig.~\ref{fig:SUN-KLM-phase-diag} and Table~\ref{tab:KHM-phases}, we summarize the properties of the insulating phases 
at commensurate fillings discussed above. 

\begin{widetext}
\begin{center}
\begin{table}[H]
\caption{\label{tab:KHM-phases} Insulating phases of the SU($N$) Kondo-Heisenberg Hamiltonian \eqref{eqn:SUN-KHM} 
for commensurate fillings $f=m/N$ ($m=1,\ldots, N-1$).  Featureless insulators [SU($N$) Kondo insulator and chiral SPT] occur 
only at $f=1-1/N$ as is predicted by the LSM argument.}
\begin{ruledtabular}
\begin{tabular}{lcc}
filling ($f$) &  $J_{\text{K}}>0$  &  $J_{\text{K}}<0$  \\
\hline
$1/N$ &  
\begin{tabular}{ll}
$N=\text{even}$: & $N/2$-fold degenerate, full-gap (Fig.~\ref{fig:SUN-GS-positive-Jk}) \\
$N=\text{odd}$:  & $N$-fold degenerate, full-gap 
\end{tabular} 
& 
\begin{tabular}{ll}
$N=\text{even}$: &   $N/2$-fold degenerate, full-gap (Fig.~\ref{fig:SUN-GS-negative-Jk}) \\
$N=\text{odd}$:  & spin gapless 
\end{tabular} 
\\
\hline
$m/N$ ($m \neq 1,N-1$) & full-gap with composite-CDW 
& spin gapless [C0S$2(N-1)$]
\\
\hline
$1-1/N$ & T$_{a_0}$-inv. full-gap SU($N$) Kondo-singlet insulator (Fig.~\ref{fig:SU4-chiral-SPT-vs-Kondo-ins})
 &  
parity-broken (T$_{a_0}$-inv.) full-gap chiral SPT (Fig.~\ref{fig:SU4-chiral-SPT-vs-Kondo-ins})\\
\hline
\begin{tabular}{l}
$1/2$ (half-filling)\\
(only when $N$ even $N>4$)
\end{tabular}  
& full-gap composite-CDW 
\begin{tabular}{ll}
$N/2=\text{odd}$: & $N/2$-fold degenerate \\
$N/2=\text{even}$:  & $N$-fold degenerate 
\end{tabular} 
&  spin gapless [C0S$2(N-1)$]
\end{tabular}
\end{ruledtabular}
\end{table}
\end{center}
\end{widetext}

\section{Concluding remarks}
\label{sec:Conclusion}
To summarize, in this paper, we identified various possible insulating phases of the SU($N$) Kondo-lattice model [KLM; \eqref{eqn:SUN-KLM}] 
and Kondo-Heisenberg model [KHM; \eqref{eqn:SUN-KHM}] by means of several complementary analytical approaches. Non-perturbative constraints 
based on the LSM argument, that depend only on the kinematical information (e.g., filling $f$, the type of local moments, etc.),   
were derived by exploiting the translational and global SU($N$) symmetries of the lattice models. 
Specifically, two different indices \eqref{eqn:LSM-index-gen} were introduced for the original lattice models 
in which the local SU($N$) moments transform in a representation specified by a Young diagram with 
$n_{\text{yng}}$ boxes.  Depending on $N$, the filling $f$, and $n_{\text{yng}}$, the general constraints strongly restrict the phase structure, 
especially the possible insulating phases of these models as summarized in Table.~\ref{tab:KHM-phases-by-LSM} for $n_{\text{yng}} =1$ 
[i.e., for the local moments in the defining representation of SU($N$)].  

For instance, the symmetric Kondo insulator with a spin gap [like the one found in the SU(2) KLM at half-filling] 
can occur only at filling $f = 1 - 1/N$ [see Eq.~\eqref{eqn:filling-Kondo-ins-general} for generic local moments]. 
For other commensurate fillings $f=m/N$ ($m=1,\ldots,N-2$), several different insulating phases with gapless spin degrees of freedom 
or multiple ground states with spontaneously broken translational symmetry can appear depending on $f$ and $N$ (see 
Table.~\ref{tab:KHM-phases-by-LSM}).

In the case of the SU($N$) KHM (\ref{eqn:SUN-KHM}) where a field-theory analysis can be derived, the LSM argument 
was shown to be equivalent to the 't Hooft anomaly matching condition of the resulting low-energy effective field theory.  
The existence of a mixed global anomaly between $\mathbb{Z}_N$ (the representation of the one-step translational symmetry $\text{T}_{a_{0}}$ in the continuum) and SU($N$) symmetries gives strong constraints on the possible insulating phases which emerge in the far IR limit. 
For example, when an anomaly-related index $\mathcal{I}^{(1)}_{N,f} = f +1/N$, which is to be identified with the first index 
$\mathcal{I}_{1}$ \eqref{eqn:1st-LSM-index} in the LSM argument, satisfies $\mathcal{I}^{(1)}_{N,f} \in \mathbb{Z}$ 
(when this is the case, $\mathcal{I}^{(2)}_{N,f} \in \mathbb{Z}$ automatically), 
no anomaly exists and uniform fully gapped insulating phases are allowed.  
In contrast, for other fillings with ${\cal I}^{(1)}_{N,f}  \notin \mathbb{Z}$, a mixed global anomaly is present thereby excluding symmetric 
full-gap insulators; from the 't Hooft anomaly matching, the resulting insulating ground states must then either support gapless spin excitations 
or be degenerate due to the spontaneous breaking of the translation symmetry, in full agreement with the LSM approach.

A weak-coupling approach to the SU($N$) KHM (\ref{eqn:SUN-KHM}) for commensurate fillings $f=m/N$ ($m=1,\ldots, N-1$) 
enables us to identify the nature of the insulating phases allowed by the LSM and 't Hooft anomaly-matching constraints.
By assuming the existence of a charge gap, we found several insulating phases depending on $f$, $N$, and the sign of the Kondo coupling $J_{\text{K}}$ (see Table \ref{tab:KHM-phases}).  As is suggested by the non-perturbative arguments, translation-invariant full-gap insulators occur only at $f=1-1/N$; 
the usual Kondo insulator with local (site-centered) Kondo singlets for antiferromagnetic $J_{\text{K}}$ and 
the chiral SPT insulator for ferromagnetic $J_{\text{K}}$ with bond-centered Kondo singlets that break inversion symmetry 
[see Figs.~\ref{fig:SU4-chiral-SPT-vs-Kondo-ins}(a) and (b)].   
For other commensurate fillings, we generically found spin-gapless insulators when $J_{\text{K}} < 0$ (for odd-$N$ and $f=1/N$) 
or fully gapped ones with ground-state degeneracy when $J_{\text{K}} > 0$.  
In the latter case ($J_{\text{K}} > 0$), a variety of degenerate insulating states have been found depending on the filling $f$ such as 
the plaquette phase (Fig.~\ref{fig:SUN-GS-positive-Jk}), 
the staggered-singlet phase (Fig.~\ref{fig:SUN-GS-negative-Jk}), 
and the long-range-ordered composite-CDW phase with the hybridization between the itinerant and local spin moments [see Eq.~\eqref{compositeCDW}]. 

The combination of the analytical approaches of this paper together with the strong-coupling study of Ref.~\onlinecite{Totsuka-23} led us 
to conjecture a (schematic) global phase diagram of the SU($N$) KLM as function of the filling $f$ and 
the Kondo coupling $J_{\text{K}}$ which is presented in Fig.~\ref{fig:SUN-KLM-phase-diag}.  
Though the insulating phases were derived explicitly for the SU($N$) KHM, we believe that the identified phases should be present 
in the SU($N$) KLM as well, since most of our arguments are based on non-perturbative constraints which rely only on kinematical information 
common to both models.  Clearly, large-scale numerical simulations are called for to shed further light on the zero-temperature phase 
diagrams of these models.

\section*{Acknowledgements}
The authors would like to thank Sylvain Capponi, Chang-Tse Hsieh, Shinsei Ryu, and Yuya Tanizaki for helpful discussions 
and correspondences.  
The authors are supported by the IRP project ``Exotic Quantum Matter in Multicomponent Systems (EXQMS)'' from 
CNRS. 
The author (KT) is supported in part by Japan Society for the Promotion of Science (JSPS) KAKENHI Grant No. 21K03401.  

\appendix
\section{A crash course on Young diagrams and SU($N$) representations}
\label{sec:Young-diag}
This appendix quickly summarizes the minimal knowledge on the Young diagrams and its relation to the irreducible representations of SU($N$).   
Let us first introduce the fundamental representations that are building blocks of all possible irreducible representations.  
There are $(N-1)$ fundamental representations $\mathcal{R}_{p}$ 
each of which is realized by a fixed number $p \, (=1,\ldots, N-1)$ 
of $N$-colored fermions $c^{\dagger}_{\alpha}$ ($\alpha=1,\ldots, N$) 
[the two cases $m=0,N$ correspond to SU($N$)-singlet and are trivial].  
The $n$-fermion representation $\mathcal{R}_{p}$ is spanned by the states of the form 
(the bracket $[\cdots]$ stands for anti-symmetrization):
\begin{equation}
\begin{split} 
& |{}_{[ \alpha_{1} , \ldots , \alpha_{p} ] } \rangle := 
c^{\dagger}_{\alpha_{1}} c^{\dagger}_{\alpha_{2}} \cdots c^{\dagger}_{\alpha_{p}} |0\rangle_{\text{F}} 
\end{split}
\end{equation}
and has dimensions $\frac{N!}{(N-p)!p!}$.   
We assign the following single-column Young diagrams:
\begin{equation}
\mathcal{R}_{p}: \quad 
\text{\scriptsize $p$} \left\{ 
{\tiny \yng(1,1,1,1)  }
\right.  
\quad (m=1,\ldots, N-1)  
\end{equation}
to these representations.  By construction, the $n$ boxes in the same column are anti-symmetrized.  

\subsection{Defining representation and its conjugate}
\label{sec:SUN-defining-rep}
The simplest of them is the $N$-dimensional (defining) representation ($\mathcal{R}_{1}$; ${\tiny \yng(1) }$)  
which is spanned by the following $N$ single-fermion ($p=1$) states:
\[
|{}_{\alpha}\rangle := c^{\dagger}_{\alpha} |0\rangle_{\text{F}}  \quad (\alpha=1,\ldots, N)  
\]
and has been used for the local spins of the models \eqref{eqn:SUN-KLM} and \eqref{eqn:SUN-KHM}.  

The conjugate representation $\overline{\mathcal{R}}_{p}$ of $\mathcal{R}_{p}$ is obtained by applying the particle-hole 
transformation:
\begin{equation*}
\begin{split}
& |{}^{[ \alpha_{1},\ldots , \alpha_{p} ] } \rangle :=  
c_{\alpha_{p}} \cdots c_{\alpha_{1}} |\text{f}\rangle_{\text{F}}  \\
& = \frac{1}{(N-p)!} \sum_{\{ \beta_{i} \}} \epsilon^{\alpha_{1}\cdots \alpha_{n} \beta_{p+1}\cdots \beta_{N}}   
| \underbrace{ {}_{[ \beta_{p+1} ,\cdots , \beta_{N}] }}_{N-p}  \rangle  \\
& \left( |\text{f}\rangle_{\text{F}} = c^{\dagger}_{1} \cdots c^{\dagger}_{N} |0\rangle_{\text{F}}   \right) \; .
\end{split}
 \end{equation*}
 As the right-hand side transforms like $\mathcal{R}_{N {-}p}$, the conjugation transforms the Young diagram as: 
 \begin{equation}
\text{\scriptsize $p$} \left\{ 
{\tiny \yng(1,1,1,1)  }
\right.  \; (\mathcal{R}_{p} )  \; \xrightarrow{\text{conjugate}} \; 
\text{\scriptsize $N {-} p$} \left\{ 
{\tiny \yng(1,1)  }
\right.   \; (\overline{\mathcal{R}}_{p}  = \mathcal{R}_{N {-}p} )
\;   .
\label{eqn:conjugation-fundamental-rep}
\end{equation}
Clearly, the following $N$ one-hole states 
\[
|{}^{\alpha} \rangle =  c_{\alpha} |\text{f}\rangle_{\text{F}} 
= (-1)^{\alpha-1} \prod_{\beta \neq \alpha} c_{\beta}^{\dagger} |0\rangle_{\text{F}}
\quad (\alpha=1,\ldots, N)
\]
span the conjugate $\overline{\mathcal{R}}_{1}$ 
of the one-fermion representation $\mathcal{R}_{1}$ (${\tiny \yng(1)}$).   

\subsection{General representations}
\label{sec:SUN-gen-irrep}
\begin{figure}[hbt]
\begin{center}
\includegraphics[scale=0.4]{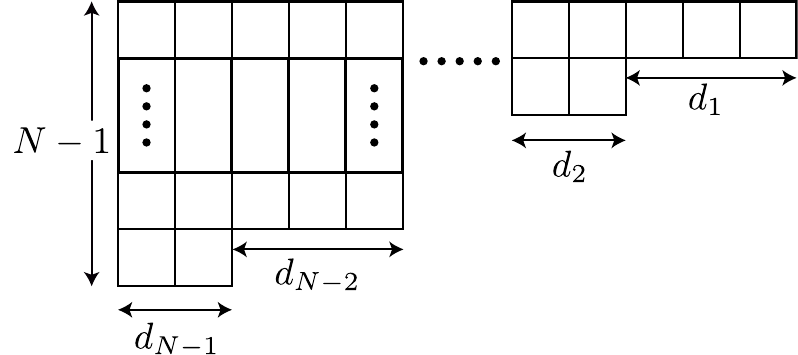}
\end{center}
\caption{The Young diagram corresponding to the SU($N$) irreducible representation specified by the Dynkin labels 
$(d_1,d_2,\ldots, d_{N-1})$.
\label{fig:Dynkin-Young}}
\end{figure}

The generic irreducible representations are constructed by tensoring the $N-1$ fundamental representations 
$\mathcal{R}_{n}$:
\begin{equation}
\mathcal{R}_{1}^{\otimes d_{1}} \otimes \cdots \otimes \mathcal{R}_{N-1}^{\otimes d_{N-1}} \; .
\end{equation}
In doing so with fermions, we need to introduce an additional degree of freedom (``flavor'') on top of the color 
$\alpha(=1,\ldots,N$).   
The set of non-negative integers (Dynkin labels) $( d_{1}, \ldots, d_{N-1} )$ uniquely specifies the irreducible representation.  
The Young diagram corresponding to a generic representation $( d_{1}, \ldots, d_{N-1} )$ is made of 
$d_{1}$ length-1 columns, $d_{2}$ length-2 ones, and so on (see Fig.~\ref{fig:Dynkin-Young}).  

For example, the diagram
\[
{\tiny \yng(3,1)} 
\]
stands for the representation $(2,1,0,\ldots,0)$, while the adjoint representation $(1,0,\ldots, 0,1)$ 
under which the SU($N$) generators transform is specified as:
\begin{equation}
\text{\scriptsize $N {-} 1$} \left\{ 
{\tiny
\yng(2,1,1,1,1)
}
\right.  \; .
\label{eqn:Young-adjoint}
\end{equation} 
The conjugate of a given representation is obtained by applying the rule \eqref{eqn:conjugation-fundamental-rep} 
to each column of the corresponding Young diagram and then rearranging the columns into the correct form.  
For instance, the adjoint representation \eqref{eqn:Young-adjoint} is self-conjugate.  
\section{LSM twist for Kondo coupling}
\label{sec:LSM-for-Kondo}
The integer $m^{\text{(S)}}$ that determines the relative phase between the charge and spin twists [see Eq.~\eqref{eqn:SUN-Heisenberg-term-variation}] 
can be fixed by considering the energy cost from the Kondo coupling:
\begin{equation}
\begin{split}
& \mathcal{H}_{\text{K}} = 
J_{\text{K}}  \sum_{j} \left(  \sum_{A=1}^{N^{2}-1}\hat{s}_{j}^{A} S_{j}^{A} \right) \\
& \phantom{ \mathcal{H}_{\text{K}}  } 
= J_{\text{K}}  \sum_{j} \left\{ 
\sum_{\mu,\nu=1}^{N} \widehat{\mathcal{S}}^{\mu\nu}_{j} \mathcal{S}^{\nu\mu}_{j}  
- \frac{1}{N} \sum_{j} \hat{n}_{j} \hat{n}^{(e)}_{j}   \right\}     
\end{split}
\end{equation}
to which a product of the two twists 
$\widehat{\mathcal{U}}_{\alpha}^{(\text{F})} \widehat{\mathcal{U}}_{\alpha}^{(\text{S})} \! (2\pi m^{\text{(S)}})$ acts.  

Using Eqs.~\eqref{eqn:twist-fermion-spin} and \eqref{eqn:twist-local-spin}, we obtain:
\begin{equation}
\begin{split}
& \sum_{j=1}^{L} \sum_{\mu,\nu}
\left\{ 
\widehat{\mathcal{U}}_{\alpha}^{(\text{F})}{}^{\dagger}
 \widehat{\mathcal{S}}^{\mu\nu}_{j}  
\widehat{\mathcal{U}}_{\alpha}^{(\text{F})}{} 
\right\}
\left\{
\widehat{\mathcal{U}}_{\alpha}^{(\text{S})} \! (2 \pi m^{(\text{S})})^{\dagger}
\mathcal{S}^{\nu\mu}_{j}
 \widehat{\mathcal{U}}_{\alpha}^{(\text{S})} \! (2 \pi m^{(\text{S})}) 
\right\}  \\
&= \sum_{j=1}^{L} \sum_{\mu\neq \alpha} 
\left\{ 
\be^{-i \frac{2\pi}{L}(1+m^{\text{(S)}}) j} \widehat{\mathcal{S}}^{\alpha\mu}_{j} \mathcal{S}^{\mu\alpha}_{j}
+ \be^{+ i \frac{2\pi}{L}(1+m^{\text{(S)}}) j} \widehat{\mathcal{S}}^{\mu\alpha}_{j} \mathcal{S}^{\alpha\mu}_{j}
\right\} \\
& \phantom{=} 
+ \sum_{j=1}^{L} \sum_{\mu,\nu \neq \alpha} \widehat{\mathcal{S}}^{\mu\nu}_{j}  \mathcal{S}^{\nu\mu}_{j}  \; .
\end{split}
\label{eqn:variation-Kondo-coupling}
\end{equation}
It is important to note that, in contrast to the variation of the other parts \eqref{eqn:SUN-hopping-term-variation} ($\mathcal{H}_{\text{hop}}$) and  
\eqref{eqn:SUN-Heisenberg-term-variation} ($\mathcal{H}_{\text{H}}$), explicit site($j$)-dependence does not cancel in the exponent, which means 
that the increase of the Kondo energy created by the twist 
$\widehat{\mathcal{U}}_{\alpha}^{(\text{F})} \widehat{\mathcal{U}}_{\alpha}^{(\text{S})} \! (2\pi m^{\text{(S)}})$ is 
of the order $\text{O}(L)$ [$L^{-n} \sum_{j} (1+m^{\text{(S)}})^{n} j^{n} \sim L$].   
Therefore, we see that the only way to avoid this large $\text{O}(L)$ energy cost and create low-lying excitations is to take 
$m^{\text{(S)}} = -1$ and consider the following particular combination \eqref{eqn:elementary-twist-SUN-KHM}:
\begin{equation}
\begin{split}
\widehat{\mathcal{U}}_{\alpha} & := 
\widehat{\mathcal{U}}_{\alpha}^{(\text{F})}  \widehat{\mathcal{U}}_{\alpha}^{(\text{S})} (- 2 \pi )  
= \exp\left\{  i \frac{2\pi}{L} \sum_{j=1}^{L} j (\hat{n}_{\alpha,j} - Q_{\alpha,j} ) \right\}  \\
& \quad (\alpha=1,\ldots, N) \; .
\end{split}
\end{equation}

\section{Bosonization of fermion part}
\label{sec:bosonization-fermions}
\subsection{Orthogonal transformation to spin-charge basis}
\label{sec:orthogonal-tr}

At low energies, the $N$ species of lattice fermions $c_{\alpha,j}$ ($\alpha=1,\ldots,N$) are expressed 
by the left ($L_{\alpha}$) and right-moving ($R_{\alpha}$) Dirac fermions as in Eq.~\eqref{contlimitDirac}.   
Then, these $2N$ Dirac fermions are bosonized using a set of scalar fields $\varphi_{\alpha, \text{L/R}}$ as:
\begin{equation}
\begin{split}
 L_{\alpha} & = \frac{\kappa_{\alpha}}{\sqrt{2 \pi a_0}} \; e^{ - i \sqrt{4 \pi}\varphi_{\alpha, \text{L}} }, \\
 R_{\alpha} & = \frac{\kappa_{\alpha}}{\sqrt{2 \pi a_0}} \; e^{ i \sqrt{4 \pi}\varphi_{\alpha, \text{R}} }  \; ,
 \end{split}
\label{bosoabeleqApp}
\end{equation}
where $[\varphi_{\alpha, \text{R}}, \varphi_{\beta, \text{L}} ] = i \delta_{\alpha\beta}/4$ and 
$\kappa_{\alpha}\,(=\kappa^{\dagger}_{\alpha})$ are the Klein factors that satisfy $\{ \kappa_{a},  \kappa_{b}\} = 2 \delta_{ab}$.   
As in the usual electron systems, we now move on from the color($\alpha$)-based basis 
$\vec{\varphi}_{\text{L/R}} = (\varphi_{1 \text{L/R}},\ldots, \varphi_{N \text{L/R}})^{\text{T}}$ to the spin-charge separated ones 
\begin{equation}
\begin{split}
& \vec{\Phi} := ( \Phi_{\text{c}} ,  \Phi_{\text{s},1}, \ldots,   \Phi_{\text{s},N-1} )^{\text{T}} \; , \\
& \vec{\Theta} := (  \Theta_{\text{c}} ,  \Theta_{\text{s},1} , \ldots,   \Theta_{\text{s},N-1} )^{\text{T}} 
\end{split}
\end{equation}
[the first elements ($\Phi_{\text{c}}$ and $\Theta_{\text{c}}$) describe the charge sector and the remaining ones 
are associated to the SU($N$)-spin] by the following transformation:
\begin{equation}
\begin{split}
\left( 
\begin{array}{c}
\vec{\Phi} \\
\hline 
\vec{\Theta} 
\end{array}
\right) 
= \left( 
\begin{array}{c|c}
\mathcal{R} & \mathcal{R} \\
\hline
\mathcal{R} & - \mathcal{R} 
\end{array}
\right)
\left( 
\begin{array}{c}
\vec{\varphi}_{\text{L}} \\
\hline 
\vec{\varphi}_{\text{R}} 
\end{array}
\right)  \; , 
\end{split}
\label{eqn:orthogonal-tr}
\end{equation}
where the $N$-dimensional orthogonal matrix $\mathcal{R}$ is defined using the $N$ weights $\{ \vec{\mu}_{\alpha} \}$ 
in the defining representation as:
\begin{equation}
\begin{split}
& \mathcal{R} := 
\begin{pmatrix}
1/\sqrt{N} & 1/\sqrt{N} & \cdots & 1/\sqrt{N} \\
\sqrt{2} \vec{\mu}_{1} & \sqrt{2} \vec{\mu}_{2} & \cdots & \sqrt{2} \vec{\mu}_{N}  
\end{pmatrix}  
\\
& [ \vec{\mu}_{\alpha}{\cdot} \vec{\mu}_{\beta} = (\delta_{\alpha\beta} - 1/N) /2 \, , \; 
\mathcal{R}^{\text{T}} \mathcal{R} = \mathbf{1} ] \; .
\end{split}
\end{equation}

If we plug the expressions \eqref{bosoabeleqApp} into the Hamiltonian \eqref{HamcontDirac} and carry out the change of basis 
\eqref{eqn:orthogonal-tr}, we arrive at:
\begin{equation}
\begin{split}
\mathcal{H}_{\text{hop}} = &
\frac{\pi v^{\text{(f)}}_\text{c} }{N}\left[
: j_{\text{c,R}}^{2} :  + : j_{\text{c,L}}^{2} :
\right]  \\
& + \frac{v^{\text{(f)}}_\text{s}}{2}  \sum_{a=1}^{N-1} \left[
:(\partial_{x}\Phi_{\text{s},a})^{2}: +:(\partial_{x}\Theta_{\text{s},a})^{2} :
\right]   \; ,
\end{split}
\end{equation}
where the charge current is defined as:
\[
j_{\text{c,L/R}} := 
\frac{1}{\sqrt{\pi}} \sum_{\alpha=1}^{N} \partial_{x} \varphi_{\alpha,\text{L/R}}  \; .
\]
This is the free-boson representation of the Hamiltonian \eqref{contfreehambis}.  

\subsection{Gauge redundancy}
\label{sec:redundancy}
The $2N$ bosons $\varphi_{\alpha \text{L/R}}$ introduced in Eq.~\eqref{bosoabeleqApp} are defined only modulo $\sqrt{\pi}$ 
and any shifts of the form: 
\begin{equation}
\varphi_{\alpha, \text{L/R}} \sim \varphi_{\alpha, \text{L/R}} + \sqrt{\pi} n_{\alpha, \text{L/R}}  \quad (\alpha=1,\ldots, N, \; 
n_{\alpha, \text{L/R}} \in \mathbb{Z}) 
\end{equation}
do not affect physics ({\em gauge redundancy}).  This property is crucial in correctly counting the number of inequivalent ground states 
in multi-component systems (see,  e.g., Refs.~\cite{Lin-B-F-98,Lecheminant-T-06-SU4,Imamura-T-H-19}).  
In fact, from Eq.~\eqref{eqn:orthogonal-tr}, one can immediately see that whenever the difference between a pair of 
$\vec{\Phi}$$(\vec{\Theta}$)-fields are written as:
\begin{equation}
\begin{split}
& 
\begin{pmatrix}
\delta \vec{\Phi} \\  \delta \vec{\Theta}  
\end{pmatrix} 
= \sqrt{\pi} 
\left( 
\begin{array}{c|c}
\mathcal{R} & \mathcal{R} \\
\hline
\mathcal{R} & - \mathcal{R} 
\end{array}
\right)
\begin{pmatrix} \vec{n}_{\text{L}}  \\ \vec{n}_{\text{R}}  \end{pmatrix}
= \sqrt{\pi} 
\begin{pmatrix}
\mathcal{R} ( \vec{n}_{\text{L}} + \vec{n}_{\text{R}} ) \\
\mathcal{R} ( \vec{n}_{\text{L}} - \vec{n}_{\text{R}} ) 
\end{pmatrix}  
\\
& \vec{n}_{\text{L/R}} := ( n_{1, \text{L/R}} ,\ldots, n_{N, \text{L/R}} ) 
\; ,
\end{split}
\label{eqn:gauge-redundancy-cond}
\end{equation}
they must be regarded as physically equivalent.   Suppose we are given a pair of semi-classical ground states 
in which $\vec{\Phi}$-fields are pinned to 
$\vec{\Phi}_{\text{cl}}$ and $\vec{\Phi}^{\prime}_{\text{cl}}$.   If there exist integral vectors $\vec{n}_{\text{L/R}}$ satisfying
\begin{equation}
\delta \vec{\Phi} = \vec{\Phi}_{\text{cl}} -  \vec{\Phi}^{\prime}_{\text{cl}} 
=    \sqrt{\pi} \mathcal{R} ( \vec{n}_{\text{L}} + \vec{n}_{\text{R}} ) .
\label{eqn:equivalent-Phi}
\end{equation}
(since $\vec{\Theta}$ is indefinite in this case, we have only to consider the first set of equations), 
the two ground states are physically equivalent.   

\section{LSM in the continuum}
\label{sec:continuum-LSM}
To find the continuum counterpart of \eqref{eqn:U-fermion-twist}, we bosonize 
the local fermion density $\hat{n}_{\alpha,j}=c_{\alpha,j}^{\dagger} c_{\alpha,j}$ as:
\[
\hat{n}_{\alpha,j} \simeq \frac{1}{\sqrt{\pi}} \partial_{x} \phi^{(\text{f})}_{\alpha} (x) \; , 
\]
where we have introduced the Bose fields $\phi^{(\text{f})}_{\alpha}$ and $\theta^{(\text{f})}_{\alpha}$ by: 
$\phi^{(\text{f})}_{\alpha} := \varphi_{\alpha, \text{L}} + \varphi_{\alpha, \text{R}}$ and 
$\theta^{(\text{f})}_{\alpha} := \varphi_{\alpha, \text{L}} - \varphi_{\alpha, \text{R}}$.  
Then, it is easy to find the continuum counterpart of the fermion twist \eqref{eqn:U-fermion-twist}:\footnote{%
A more careful treatment suggests that we need to include the surface term to obtain the correct expression 
given in Ref.~\onlinecite{Aligia-B-05} (see also Ref.~\cite{Yao-F-19}), 
whereas the naive expression \eqref{eqn:Uf-continuum-1} given here suffices to our purposes.  
We thank Y.~Fukusumi for pointing this subtlety out. }
\begin{equation}
\widehat{\mathcal{U}}_{\alpha}^{(\text{f})} 
= \exp\left[ i \frac{2}{L} \sqrt{\pi} \int_{0}^{L}\! dx \, x \partial_{x} \phi^{(\text{f})}_{\alpha}(x) \right]   
 \; .
 \label{eqn:Uf-continuum-1}
\end{equation}
In fact, using $[ \partial_{x}\phi^{(\text{f})}_{\alpha}(x) , \theta^{(\text{f})}_{\beta}(y) ] = -i \delta_{\alpha\beta} \delta(x - y)$, 
we can readily check that the above $\widehat{\mathcal{U}}_{\alpha}^{(\text{f})}$ correctly 
adds $x$-dependent phases to the left and right movers [see Eq.~\eqref{eqn:c-on-U-fermion-twist}]:
\begin{equation*}
\begin{split}
& R^{\dagger}_{\beta} \sim \be^{-i \sqrt{\pi}(\phi_{\beta}-\theta_{\beta})}
 \xrightarrow{\hat{\mathcal{U}}^{(\text{f})}_{\alpha}} 
\be^{-i \frac{2\pi}{L} x \delta_{\alpha\beta} } R^{\dagger}_{\beta}  \; , \\
& L^{\dagger}_{\beta} \sim \be^{ i \sqrt{\pi}(\phi_{\beta} + \theta_{\beta})}  
\xrightarrow{\hat{\mathcal{U}}^{(\text{f})}_{\alpha}} 
\be^{-i \frac{2\pi}{L} x \delta_{\alpha\beta} } L^{\dagger}_{\beta} \; ,
\end{split}
\end{equation*}
thereby reproducing Eq.~\eqref{eqn:c-on-U-fermion-twist} in the continuum limit.  
Using relations similar to \eqref{eqn:fermion-density-color-resolved}, we can rewrite \eqref{eqn:Uf-continuum-1} as:
\begin{equation}
\begin{split}
\widehat{\mathcal{U}}_{\alpha}^{(\text{f})} 
=&  \exp\left[ i \frac{2}{L} \sqrt{ \frac{\pi}{N} } \int_{0}^{L} \! dx \, x \partial_{x} \Phi_{\text{c}} (x) \right]  \\
& \times \exp\left[ i \frac{2}{L} \sqrt{\pi} \sum_{a=1}^{N-1} [ \vec{\mu}_{\alpha}]_{a} 
\int_{0}^{L} \! dx \, x \partial_{x} \Phi^{(\text{f})} _{\text{s}, a} (x) \right]     \; .
\end{split}
\label{eqn:LSM-UF-in-continuum}
\end{equation}

For the spin twist, we plug the continuum expression of $Q_{\alpha,j}$ 
\begin{equation*}
Q_{\alpha,j} = \sum_{\beta=1}^{N} [Q_{\alpha}]_{\beta\beta} \, n^{(\text{s})}_{\beta,j} 
\rightarrow 
- \frac{1}{\sqrt{\pi}} \sum_{a=1}^{N-1} [\vec{\mu}_{\alpha}]_{a} \, \partial_{x} \Phi_{\text{s},a}^{(\text{s})}  
\end{equation*}
into \eqref{eqn:SUN-LSM-spin-twist-tentative} to obtain:
\begin{equation}
\widehat{\mathcal{U}}_{\alpha}^{(\text{s})} (- 2\pi)
= \exp\left[ i \frac{2}{L} \sqrt{\pi} \sum_{a=1}^{N-1} [ \vec{\mu}_{\alpha}]_{a} 
\int_{0}^{L} \! dx \, x \partial_{x} \Phi^{(\text{s})}_{\text{s}, a} (x) \right]   
\; .
\label{eqn:LSM-US-in-continuum}
\end{equation}
The elementary twists are obtained by combining \eqref{eqn:LSM-UF-in-continuum} and \eqref{eqn:LSM-US-in-continuum}.  

Finally, a generic twist operation $\widehat{\mathcal{U}}_{(m_{1},\ldots, m_{N})}$ in the continuum splits 
into the charge and spin parts:
\begin{subequations}
\begin{equation}
\widehat{\mathcal{U}}_{(m_{1},\ldots, m_{N})} 
= \widehat{\mathcal{U}}^{(\text{c})}_{M} \cdot \widehat{\mathcal{U}}^{(\text{s})}_{(\overline{m}_{1},\ldots, \overline{m}_{N})} 
\end{equation}
with 
\begin{equation}
\begin{split}
& \widehat{\mathcal{U}}^{(\text{c})}_{M} 
: = \exp\left\{ i \frac{2\pi }{L}  \frac{M}{N} 
\int_{0}^{L} \! dx \, x \sqrt{ \frac{N}{\pi} } \partial_{x} \Phi_{\text{c}} (x) \right\}   \\
& \widehat{\mathcal{U}}^{(\text{s})}_{(\overline{m}_{1},\ldots, \overline{m}_{N})} := 
\exp\Biggl\{ i \frac{2\pi }{L}  \sum_{a=1}^{N-1} \biggl[ 
\left( \sum_{\alpha=1}^{N} \overline{m}_{\alpha} \vec{\mu}_{\alpha} \right)_{a}  \\
& \qquad \qquad  \times   \int_{0}^{L} \! dx \, x \frac{1}{\sqrt{\pi} } 
\left( \partial_{x} \Phi^{(\text{f})} _{\text{s},a} (x) + \partial_{x} \Phi^{(\text{s})} _{\text{s},a} (x) \right) 
\biggr] \Biggr\}  \; ,
\end{split}
\end{equation}
\end{subequations}
which is to be compared with the lattice expression \eqref{eqn:LSM-twist-generic-2}.\footnote{%
Note that $\partial_{x} \Phi^{(\text{F})} _{\text{c}} (x)$ counts the density of itinerant fermions 
\[ 
\hat{n} \sim \sqrt{ \frac{N}{\pi} } \partial_{x} \Phi^{(\text{F})} _{\text{c}} (x) 
\]
and that $\left( \partial_{x} \Phi^{(\text{F})} _{a} (x) + \partial_{x} \Phi^{(\text{S})} _{a} (x) \right)/\sqrt{\pi}$ 
gives the local SU($N$) weight $\vec{\lambda}$ (including the that from both the itinerant and local fermions).}   

Now suppose that spin[SU($N$)]-charge separation occurs at low energies.  
Then, $\widehat{\mathcal{U}}^{(\text{c})}_{M}$ that involves only the charge boson $\Phi^{(\text{F})} _{\text{c}}$ 
of the itinerant fermions affects only the charge sector, 
while $\widehat{\mathcal{U}}^{(\text{s})}_{(\overline{m}_{1},\ldots, \overline{m}_{N})}$ twists the entire spin sector 
that includes both the itinerant ($\Phi^{(\text{f})} _{\text{s},a}$) and local ($\Phi^{(\text{s})} _{\text{s},a}$) spins.  

To get more insight into the low-energy spectral structure, let us calculate the energy shift due to LSM twists 
using the Luttinger-liquid Hamiltonian.  
Plugging all these into the low-energy expressions \eqref{contfreehambis} and \eqref{sutherlandham}, 
we obtain the following Luttinger-liquid expression 
of the $\text{O}(L^{-1})$ energy shift:
\begin{equation}
\begin{split}
& \Delta E_{(m_{1},\ldots, m_{N})}  \\
& = 
\frac{2\pi}{L} v^{\text{(f)}}_\text{c}  \frac{M^{2}}{N} K_{\text{c}} 
+ 
\frac{2\pi}{L} \left( v^{\text{(f)}}_\text{s} + v^{\text{(s)}}_\text{s}  \right) 
2 \left( \sum_{\alpha=1}^{N} \overline{m}_{\alpha} \vec{\mu}_{\alpha} \right)^{2}  
\; ,
\end{split}
\label{eqn:energy-increase-KH-LL}
\end{equation}
where $K_{\text{c}}$ is the Luttinger-liquid parameter introduced in Eqs.~\eqref{sineGordonchargeNodd} and \eqref{sineGordonchargeNeven} 
that encodes the effects of marginal interactions.   
The first term corresponds to the excitations in the charge sector, while the second to the spin [i.e., SU($N$)] excitations.  

The simplest choice 
\[  (m_{1} , \ldots, m_{N} ) = (1, 0, \ldots, 0)  \]
corresponds, despite its simple looking, to the following {\em spin-charge entangled} twist:
\begin{equation*}
\begin{split}   
& \text{charge:} \quad M=1, \\
& \text{spin:} \quad   (\overline{m}_{1} , \ldots, \overline{m}_{N} ) = (1-1/N, -1/N, \ldots, -1/N)  , \\
& \left( \sum_{\alpha=1}^{N} \overline{m}_{\alpha} \vec{\mu}_{\alpha} 
= \vec{\mu}_{1} \right)    
\end{split}
 \end{equation*}
 and increases the energy of the system as:
 \[
 \Delta E_{(1,0,\ldots, 0)} = \frac{2\pi}{L} v^{\text{(f)}}_\text{c}  \frac{K_{\text{c}}}{N} 
+ 
\frac{2\pi}{L} \left( v^{\text{(f)}}_\text{s}  + v^{\text{(s)}}_\text{s}  \right) \frac{N-1}{N}  \; .
\]
This indicates that the $(1, 0, \ldots, 0)$-twist creates a spin excitation corresponding to the primary states of the two SU($N$)${}_{1}$ CFTs 
(second term) as well as the charge excitation proportional to $K_{\text{c}}$.  

 If the charge sector gets gapped by forming some sort of charge-ordered phases (e.g., Mott, CDW, etc.) with $K_{\text{c}} \to 0$, 
 $\Phi _{\text{c}}$ is almost pinned, whereas the conjugate $\Theta _{\text{c}}$ disappears 
 at low energies [see, e.g., Eqs.~\eqref{sineGordonchargeNodd} and \eqref{sineGordonchargeNeven}].   
Then, as is seen in Eq.~\eqref{eqn:energy-increase-KH-LL}, 
the twist $\widehat{\mathcal{U}}_{(1,0,\ldots, 0)}$ 
excites only the spin sector leaving the gapped charge sector intact (as it affects only $\Theta _{\text{c}}$).  
 
 On the other hand, the {\em uniform} twist with vanishing zero-mean part 
$(m_{1} , \ldots, m_{N} ) = (1, \ldots, 1)$ corresponds to 
 \begin{equation}
 \begin{split}
& \widehat{\mathcal{U}}_{(1 ,\ldots, 1)} 
= \exp\left\{  i \frac{2\pi}{L} \sum_{j=1}^{L} j  \left( \sum_{\alpha=1}^{N} \hat{n}_{\alpha,j} \right)  \right\}  \\
& \;\; \longrightarrow \;\;
\exp\left\{ i \frac{2\pi }{L} \int\! dx \, x \sqrt{ \frac{N}{\pi} } \partial_{x} \Phi_{\text{c}} (x) \right\} 
\end{split}
\end{equation}
that excites {\em only} the charge part leaving the spin sector intact
\[
\Delta E_{(1,\ldots, 1)} = \frac{2\pi}{L} v^{\text{(f)}}_\text{c}  N K_{\text{c}}   \; .
\]
In the charge-ordered phases where 
$\Phi_{\text{c}}$ is locked (and $K_{\text{c}} \to 0$), $\widehat{\mathcal{U}}_{(1 ,\ldots, 1)}$ does not create excitations at all 
as is suggested intuitively (note that $\widehat{\mathcal{U}}_{(1 ,\ldots, 1)}$ does not change a charge-ordered state 
$\otimes_{i} | n_{i} \rangle$).

\section{Umklapp interaction in the 1D SU($N$) Hubbard model}
\label{sec:umklappsun}
In this Appendix, we discuss the values of the phase $\theta_0$ \eqref{phasespinop} of the non-universal constant $\lambda$ 
which occurs in the low-energy expression of the SU($N$) spin operator \eqref{spinop} of the localized spin.  
This coupling constant stems from the averaging of the charge degrees of freedom in the Mott-insulating phase 
of the 1D U($N$) Hubbard chain at $1/N$-filling:
\begin{equation}
\begin{split}
{\cal H}_{\text{Hubbard}} =&  - t \sum_i \sum_{\alpha=1}^{N} \left( c^{\dagger}_{\alpha,i+1} c_{\alpha,i}  + \text{H.c.}\right)
 \\
&+ \frac{U}{2} \sum_{i, \alpha,\beta} n_{\alpha,i}   n_{\beta,i} \left( 1 - \delta_{\alpha\beta} \right).
\end{split}
\label{hubbardSUN}
\end{equation}
In the limit of large repulsive $U$, this model (\ref{hubbardSUN}) reduces, at low energies, to the SU($N$) Heisenberg spin chain 
${\cal H}_{\text{H}}$ \eqref{eqn:SUN-KHM}. 

The SU($N$) spin operator assumes a form similar to Eq.~\eqref{spinelecopfin} except that now $\be^{i \sqrt{ 4 \pi/N} \Phi_\text{c}}$ 
is replaced with its expectation value as the charge degrees of freedom are fully gapped in the large-$U$ limit:
\begin{equation}
S^{A}_{n}/a_0 \simeq J^{A}_{\text{s,L}}  +  J^{A}_{\text{s,R}}  + i  C e^{\frac{ i 2 \pi x}{Na_0}}
\langle \be^{i \sqrt{ 4 \pi/N} \Phi_\text{c}}  \rangle_{\text{c}}   \; {\rm Tr} (  g_{\text{s}}  T^A)   + \text{H.c.},
\label{spinSUNapp}
\end{equation}
where $C = \frac{\sqrt{N}}{2 \pi a_0^{1/N}}$ and the charge degrees of freedom have been averaged in the Mott-insulating phase.  
As has been seen in Sec.~\ref{sec:Weakcoupling}, the actual expectation value of the charge bosonic field $\langle \Phi_\text{c}  \rangle$ is crucial.  
To determine how the charge boson $\Phi_\text{c}$ is pinned, we revisit here the argument of Ref.~\onlinecite{Assaraf-A-C-L-99} 
on the generation of the umklapp term which opens a charge gap in the large-$U$ regime of the  U($N$) Hubbard model (\ref{hubbardSUN}). 

We first use the continuum limit (\ref{contlimitDirac}) with the Fermi momentum $k_\text{F} = \pi/(Na_0)$ of the lattice fermion $c_{\alpha,i}$ 
of model (\ref{hubbardSUN}).
In stark contrast to the $N=2$ case, the umklapp term for $N>2$ does not appear in the naive continuum limit of the  U($N$) Hubbard model \eqref{hubbardSUN} but requires higher-order perturbation that generates a $2Nk_\text{F}$ non-oscillating 
piece \cite{Assaraf-A-C-L-99}. One can find its expression by exploiting the symmetries of model (\ref{hubbardSUN}).  
Namely, the umklapp operator should be U($N$)-singlet, and invariant under the one-step translation T$_{a_0}$ and the site-parity P$_{\text{s}}$ 
($c_{\alpha,i} \xrightarrow{{\text P}_{\text{s}}}  c_{\alpha,-i}$) symmetries 
that act on the left-right moving Dirac fermions \eqref{contlimitDirac} as follows:
\begin{equation}
\begin{split}
& L_{\alpha}   \xrightarrow{{\text T}_{a_0}}  e^{\frac{ - i  \pi }{N}} L_{\alpha} \; , \quad  
R_{\alpha}   \xrightarrow{{\text T}_{a_0}}  e^{\frac{i \pi }{N}} R_{\alpha}  \\
& L_{\alpha} (x)   \xrightarrow{{\text P}_{\text{s}}}   R_{\alpha} (-x) \; , \quad  
R_{\alpha} (x)   \xrightarrow{{\text P}_{\text{s}}}   L_{\alpha} (-x)  \; .
\end{split}
\label{latticesymDiracApp}
\end{equation}
The umklapp operator of the lowest scaling dimension which is a U($N$) singlet and invariant under \eqref{latticesymDiracApp} is: 
\begin{equation}
{\cal O}_{\text{umklapp}} = \prod_{\alpha=1}^{N} L^{\dagger}_{\alpha} R_{\alpha} + \text{H.c.} \; .
\label{umklappop}
\end{equation}
The next step is to obtain a bosonized expression of Eq.~\eqref{umklappop}.
To this end, one uses the Abelian bosonization rules \eqref{bosoabeleqApp} of the Dirac fermions given in Appendix~\ref{sec:bosonization-fermions}.  
The umklapp term (\ref{umklappop}) can be expressed in terms of the charge field $\Phi_{\text{c}}$ and its expression depends on the parity of $N$:
\begin{equation}
\begin{split}
{\cal O}_{\text{umklapp}}^{\text{even-}N} & = \frac{(-1)^{N/2}}{2^{N-1} (\pi a_0)^N}  \cos(\sqrt{ 4 \pi N} \Phi_{\text{c}})  \\
{\cal O}^{\text{odd-}N}_{\text{umklapp}} &= - \frac{(-1)^{(N-1)/2}}{2^{N-1} (\pi a_0)^N} \sin(\sqrt{ 4 \pi N} \Phi_{\text{c}} ) ,
 \end{split}
\label{umklappchargeApp}
\end{equation}
where $\Phi_{\text{c}} = \sum_{\alpha} \varphi_{\alpha}/\sqrt{N}$. 
The charge degrees of freedom are thus described by a $\beta^2 = 4 \pi N$  sine-Gordon model whose explicit form depends on the parity of $N$:
\begin{equation}
\begin{split}
{\cal H}^{\text{even-} N}_{\text{c}} =&  \frac{ v^{\text{(f)}}_\text{c} }{2} \left\{ \frac{1}{K_{\text{c}}} \left(\partial_x \Phi_{\text{c}} \right)^2 
+ K_{\text{c}} \left(\partial_x \Theta_{\text{c}} \right)^2 \right\} \\
& - \lambda_{\text{c}} \cos  \left(\sqrt{ 4 \pi N} \Phi_{\text{c}} \right)   \\
{\cal H}^{\text{odd-} N}_{\text{c}} = & \frac{ v^{\text{(f)}}_\text{c} }{2} \left\{ \frac{1}{K_{\text{c}}} \left(\partial_x \Phi_{\text{c}} \right)^2 
+ K_{\text{c}} \left(\partial_x \Theta_{\text{c}} \right)^2 \right\}  \\
& - \lambda_{\text{c}} \sin  \left(\sqrt{ 4 \pi N} \Phi_{\text{c}} \right) ,
 \end{split}
 \label{sineGordoncharge1overNfillingApp}
\end{equation}
where $v_{\text{c}}$ and $K_{\text{c}}$ are respectively the charge and Luttinger parameter, and $\lambda_{\text{c}}$ is an unknown coupling constant.

In the Mott-insulating phase when $K_{\text{c}} = 2/N$, the charge field $\Phi_{\text{c}}$ is pinned 
thereby forming the following ground states depending on the sign of $\lambda_{\text{c}}$:
\begin{subequations}
\begin{equation}
\begin{split}
\langle \Phi_{\text{c}} \rangle &=  n \sqrt{\frac{\pi}{N}}  \quad (\lambda_{\text{c}}  > 0) ,  \\
\langle \Phi_{\text{c}} \rangle &= \sqrt{\frac{\pi}{4N}}+ n \sqrt{\frac{\pi}{N}} \quad (\lambda_{\text{c}} < 0 )  
\label{pinningchargefieldeven1/NApp}
\end{split}
\end{equation}
in the even $N$ case ($n$ being arbitrary integers), and  
\begin{equation}
\begin{split}
\langle \Phi_{\text{c}} \rangle &=  \sqrt{\frac{\pi}{16N}} + \ell \sqrt{\frac{\pi}{N}} \quad (\lambda_{\text{c}}  > 0)  ,  \\
\langle \Phi_{\text{c}} \rangle &= - \sqrt{\frac{\pi}{16N}} + \ell  \sqrt{\frac{\pi}{N}}  \quad  (\lambda_{\text{c}} < 0) 
\label{pinningchargefieldeodd1/NApp}
\end{split}
\end{equation}
\end{subequations}
in the odd $N$ case ($\ell$ being arbitrary integers).  

All these values do not necessarily represent physically inequivalent states since the bosons $\varphi_{\alpha, \text{L/R}}$ 
that express the physical fermions by Eq.~\eqref{bosoabeleqApp} are defined only modulo $\sqrt{\pi}$.   
In fact, as has been discussed in Appendix~\ref{sec:redundancy}, 
there is gauge redundancy in the bosonic charge field $\Phi_{\text{c}}$ which is  [see Eq.~\eqref{eqn:equivalent-Phi}]:
\begin{equation}
\Phi_{\text{c}}  \sim  \Phi_{\text{c}} + \sqrt{\frac{\pi}{N}} \; , 
\label{chargegaugeredundancyApp}
\end{equation}
and there are thus only two inequivalent ground states to consider: $\langle \Phi_{\text{c}} \rangle= 0, \sqrt{\frac{\pi}{4N}}$ $\left( 
\langle \Phi_{\text{c}} \rangle= \pm \sqrt{\frac{\pi}{16N}} \right)$ when $N$ is even (odd).
Averaging over the charge degrees of freedom in the large-$U$ limit,  the SU($N$) spin operator (\ref{spinSUNapp}) becomes:
\begin{equation}
S^{A}_{n}/a_0 \simeq J^{A}_{\text{s},\text{L}}  +  J^{A}_{\text{s}, \text{R}}  + i  C e^{\frac{ i 2 \pi x}{Na_0}}e^{i \theta_0} 
   \; {\rm Tr} (  g_{\text{s}}  T^A)   + \text{H.c.},
\label{spinSUNappfin}
\end{equation}
where the phase $\theta_0$ is given by Eq.~\eqref{phasespinop} depending on the pinning of the charge field (\ref{pinningchargefieldeven1/NApp}), (\ref{pinningchargefieldeodd1/NApp}).

\section{Charge-sector ground state at $f=1/N$ and $f=1-1/N$}
\label{sec:umklappcharge}
In this Appendix, we derive the umklapp operator of the KHM model (\ref{eqn:SUN-KHM}) for the two special fillings 
$f=\frac{N-1}{N}$ and $f=\frac{1}{N}$. 
As in Appendix~\ref{sec:umklappsun}, the umklapp process depends only on the  $\text{U(1)}_{\text{c}}$ charge degrees of freedom 
and can be obtained by considering higher-order processes in perturbation theory. 

We first consider the Kondo-interaction (\ref{kondoembedding}) for $f=\frac{N-1}{N}$ in the SU($N$)$_2$ 
 $\times$ ${\mathbb Z}_N$ basis. The derivation of the umklapp term depends on the parity of $N$.
 In the odd-$N$ case,  a contribution which depends only on  the $\text{U(1)}_{\text{c}}$ charge field $\Phi_{\text{c}}$ 
occurs at $N$-th order of $\mathcal{V}^{(1)}_{\text{K}}$ in Eq.~\eqref{kondoembedding}.  
Indeed, the latter term can be expressed in terms of the SU($N$)$_2$ primary field $ \Phi_{{\tiny \yng(2)}}$ 
which transforms in the symmetric rank-2 tensor representation ${\tiny \yng(2)}$ of SU($N$):
 \begin{equation}
 \mathcal{V}^{(1)}_{\text{K}}  \sim - J_{\text{K}} \, \be^{i \sqrt{ 4 \pi/N} \Phi_{\text{c}}}  \mbox{Tr} \,   \Phi_{{\tiny \yng(2)}}    + \text{H.c} .
\label{umklappNodd}
\end{equation}
By considering the fact that the identity operator appears in the OPE of $N$ 
$ \Phi_{{\tiny \yng(2)}}$ operators (note that the trivial SU($N$)-singlet appears in the decomposition of  
${\tiny \yng(2)}^{\otimes N} $), we find a umklapp operator which, together with the Luttinger-liquid part, gives 
the $\beta^2 = 4 \pi N$ sine-Gordon model:
\begin{equation}
\begin{split}
& {\cal H}^{\text{odd-}N}_{\text{c}} \\
& = \frac{v^{\text{(f)}}_\text{c} }{2} \left\{ \frac{1}{K_{\text{c}}} \left(\partial_x \Phi_{\text{c}} \right)^2 
+ K_{\text{c}} \left(\partial_x \Theta_{\text{c}} \right)^2 \right\}   
 - \mu_{\text{c}} \cos  \left(\sqrt{ 4  \pi N} \Phi_{\text{c}} \right),  
\end{split}
\label{sineGordonchargeNodd}
\end{equation}
where $\Theta_{\text{c}}$ is the dual charge field, $\mu_{\text{c}}$ is a coupling constant, $v^{\text{(f)}}_\text{c} $ and $K_{\text{c}}$ are respectively  the charge velocity 
and the Luttinger parameter, whose values as a function of $J_{\text{K}}$ are beyond the field theory analysis 
and requires complementary numerical approaches.

When $N$ is even, a umklapp contribution can be obtained at order $J_{\text{K}}^{N/2}$ of perturbation theory which stems from the second term $ \mathcal{V}^{(2)}_{\text{K}}  $ of the Kondo interaction (\ref{kondoembedding}). The latter can be expressed in terms of the SU($N$)$_2$ primary field $ \Phi_{{\tiny \yng(1,1)}}$ which transforms in the  representation ${\tiny \yng(1,1)}$ of the SU($N$) group:
\begin{equation}
 \mathcal{V}^{(2)}_{\text{K}}  \sim J_{\text{K}}  \be^{i \sqrt{ 4 \pi/N} \Phi_{\text{c}}}  \epsilon_1 \mbox{Tr} \,   \Phi_{{\tiny \yng(1,1)}}    + \text{H.c} .
\label{umklappNeven}
\end{equation}
Since we have $\epsilon_1^{N/2} \sim I$ in the OPE sense and the trivial irrep of SU($N$)  appears in the decomposition 
${\tiny \yng(1,1)} \otimes {\tiny \yng(1,1)}\otimes \ldots \otimes {\tiny \yng(1,1)}$ ($N/2$ times), one may conclude
that the sine-Gordon model with $\beta^2 = \pi N$ for the  $\Phi_{\text{c}}$ charge field emerges
at order $J_{\text{K}}^{N/2}$ of perturbation theory in the even-$N$ case:
\begin{equation}
\begin{split}
& {\cal H}^{\text{even-}N}_{\text{c}}  \\
& =  \frac{ v^{\text{(f)}}_\text{c} }{2} \left\{ \frac{1}{K_{\text{c}}} \left(\partial_x \Phi_{\text{c}} \right)^2 
+ K_{\text{c}} \left(\partial_x \Theta_{\text{c}} \right)^2\right\}  
- \mu_{\text{c}} \cos  \left(\sqrt{ \pi N} \Phi_{\text{c}} \right) \; .
\end{split}
\label{sineGordonchargeNeven}
\end{equation}

When $f=1/N$ a similar approach can be done from the Kondo interaction (\ref{kondoembedding1/N}).
The umklapp operator $\cos ( \sqrt{ 4 \pi N} \Phi_c)$ is generated in higher order of perturbation theory at order $J_{\text{K}}^N$ in the odd-$N$ case since $ \tilde{\mathcal{V}}^{(1)}_{\text{K}}$ in Eq. (\ref{kondoembedding1/N}) contains 
the  ${\mathbb{Z}}_N$ parafermion currents $\Psi_{1L} \Psi_{1R}$. Using  the fusion rule 
$ (\Psi_{1L} \Psi_{1R})^N \sim I$, which stems from the parafermion algebra \cite{Zamolodchikov-F-JETP-85}, one obtains the umklapp term (\ref{sineGordonchargeNodd}) in the odd-$N$ case. When $N$ is even, we now consider the operator 
$ \tilde{\mathcal{V}}^{(2)}_{\text{K}}$ in Eq. (\ref{kondoembedding1/N}) which contains the $\sigma_2$ term and 
the  SU($N$)$_2$ adjoint perturbation $\Phi_{\rm adj}$. Using the fact $\sigma_2^{N/2} \sim I$ and the fusion rule (\ref{fusionrule}), we get the $\beta^2 = \pi N$ sine-Gordon model (\ref{sineGordonchargeNeven}) as an umklapp term for the  $\Phi_{\text{c}}$ charge field in the even-$N$ case at order $J_{\text{K}}^{N/2}$ of perturbation theory.

As the scaling dimensions of the perturbation in Eqs.~\eqref{sineGordonchargeNodd} and \eqref{sineGordonchargeNeven} 
are $N K_{\text{c}}$ and $N K_{\text{c}}/4$, respectively, 
a charge gap opens when $K_{\text{c}} < 2/N$ (for odd-$N$) and $K_{\text{c}} < 8/N$ (for even-$N$).  
In the charge-gapped phase, the charge-bosonic field 
$\Phi_{\text{c}}$ of the sine-Gordon models \eqref{sineGordonchargeNodd} and \eqref{sineGordonchargeNeven} is pinned 
to one of the minima of the cosine potentials depending on the sign of $\mu_{\text{c}}$: 
\begin{subequations}
\begin{equation}
\begin{split}
& \langle \Phi_{\text{c}} \rangle =  \ell \sqrt{\frac{\pi}{N}}, \quad ( \mu_{\text{c}} > 0)   \\
& \langle \Phi_{\text{c}} \rangle = \sqrt{\frac{\pi}{4N}}+ \ell \sqrt{\frac{\pi}{N}}, \quad (\mu_{\text{c}} < 0),
\end{split}
\label{pinningchargefield-odd}
\end{equation}
for odd-$N$, and 
\begin{equation}
\begin{split}
& \langle \Phi_{\text{c}} \rangle = 2 \ell \sqrt{\frac{\pi}{N}}, \quad (\mu_{\text{c}} > 0)   \\
& \langle \Phi_{\text{c}} \rangle = \sqrt{\frac{\pi}{N}}+ 2 \ell \sqrt{\frac{\pi}{N}}, \quad ( \mu_{\text{c}} < 0)   
\end{split}
\label{pinningchargefield-even}
\end{equation}
\end{subequations}
for even-$N$ ($\ell$ being arbitrary integers).   Note that all these values do not necessarily represent physically inequivalent states.  
In fact, there is gauge redundancy \eqref{chargegaugeredundancyApp}: 
$\Phi_{\text{c}}  \sim  \Phi_{\text{c}} + \sqrt{\pi/N}$.  
Taking this into account in Eqs.~\eqref{pinningchargefield-odd} and \eqref{pinningchargefield-even}, 
we deduce that there is a single minimum $\langle \Phi_{\text{c}} \rangle = 0$ to consider in the even-$N$ case 
for either sign of $\mu_{\text{c}}$.  
In the odd-$N$ case, on the other hand, we have two different inequivalent solutions depending on the sign of $J_{\text{K}}$: 
$\langle \Phi_{\text{c}} \rangle = 0$ ($\mu_{\text{c}} > 0$) and $\sqrt{\frac{\pi}{4N}}$ ($\mu_{\text{c}} < 0$).  
Unfortunately, the precise $J_{\text{K}}$-dependence of the umklapp coupling $\mu_{\text{c}}$, which is crucial 
in selecting one of the two possible solutions for a given $J_{\text{K}}$, cannot be determined within our approach.  
This is why we have chosen, in the main text (Sec.~\ref{subsec:N-1/N}), one of the two in such a way that the physical conclusions 
drawn from the solution are consistent with those from the strong-coupling approach. 

\section{Mapping onto the non-linear sigma model on a flag manifold}
\label{sec:flagsigma}

In this Appendix, we connect the weak-coupling analysis for the $1/N$-filling with $J_{\text{K}}<0$ to the semiclassical description of the SU($N$) Heisenberg spin chain  in symmetric rank-$2$ tensor representation (\ref{eqn:eff-Heisenberg-sym-rank-2}) which describes the strong-coupling regime $J_{\text{K}} \rightarrow - \infty$  of the SU($N$) KLM for $f=1/N$ 
\cite{Totsuka-23}.

As described in Sec. \ref{subsec:1/N}, the low-energy effective theory which governs the properties of the 
SU($N$) KHM for one-fermion per site with $J_{\text{K}}<0$ is the SU($N$)$_2$ CFT perturbed by the adjoint operator (\ref{Heff1/NNodd}). 
In this respect, let us consider the most general SU($N$)$_2$  perturbed action compatible with the PSU($N$) symmetry and the one-site translation invariance, e.g., the ${\mathbb{Z}}_N$ symmetry (\ref{trans2even}), first introduced in Ref. \onlinecite{Ohmori-S-S-19}: 
\begin{eqnarray}
{\cal S} =  {\cal S}_{\rm WZNW} + \sum_{n=1}^{\left[N/2\right]} \int d^2 x  \; g_n {\rm Tr} \left[ G^n\right] {\rm Tr}  \left[\left(G^{\dagger}\right)^n\right],
\label{WZWint}
\end{eqnarray}
where ${\cal S}_{\rm WZNW}$ is the Euclidean action for the SU($N$)$_2$ CFT and $g_1 = {\bar \gamma} > 0$.
In Eq. (\ref{WZWint}), the $n = 3, \ldots \left[N/2\right]$ terms are actually irrelevant contributions and the $g_2$ term is a subleading relevant operator which is generated in the RG flow according to the fusion rules of SU($N$)$_2$  CFT:
\begin{equation}
\Phi_{\rm adj} \times \Phi_{\rm adj} \sim I + \Phi_{\rm adj} + \Phi^{'}+..
\label{fusionrule}
\end{equation}
where the dots describe terms that are marginal or irrelevant operators.  The SU($N$)$_2$ primary field $\Phi^{'}$  transforms in the self-conjugate representation of  SU($N$) with the Young tableau of $N$ boxes:
\begin{equation}
{\tiny
\text{\scriptsize $N-2$} \left\{ 
\yng(2,2,1,1)
 \right.  }   \; ,
 \end{equation}
and is relevant with the scaling dimension $x^{'} = 2 (N-1)/N < 2$ and translation invariant.  
Our approach cannot fix the sign of the coupling $g_2$ of this operator. We assume $g_2 >0$ to reproduce the strong-coupling result.

We can now consider analyse the field theory (\ref{WZWint}) by means of a strong-coupling limit.
When $g_n \rightarrow + \infty$, the potentiel term of Eq. (\ref{WZWint}) selects a SU($N$) matrix $G$ such that ${\rm Tr} \left[ G^n\right] = 0$ with $n=1,\ldots, \left[N/2\right]$. As shown in Ref. \onlinecite{Ohmori-S-S-19}, the latter condition can be extended to $n=1,\ldots, N - 1$ and
the SU($N$) $G$  field can be written as:
\begin{eqnarray}
G &=& U \Omega U^{\dagger} \nonumber \\
\Omega &=&  \omega^{- (N-1)/2} 
\begin{pmatrix}
\omega^{N-1}  & 0 & \cdots & 0 \\
0 & \omega^{N-2}   & \cdots & 0 \\
 \vdots & \cdots & \omega & 0\\
0 & \cdots & 0 & 1
\end{pmatrix} , \label{wzwfieldmanifold}\\
\nonumber  
\end{eqnarray}
$U$ being a general U($N$) matrix and $\omega =e ^{i 2\pi/N}$. The solution (\ref{wzwfieldmanifold}) describes a U($N$)/U(1)$^N$ $\sim$ SU($N$)/U(1)$^{N-1}$  flag manifold \cite{Affleck-B-W-22}.   Using the identification (\ref{wzwfieldmanifold}) in the action (\ref{WZWint}), it can be shown that the low-energy effective field theory is 
 a non-linear sigma model on the flag manifold  SU($N$)/U(1)$^{N-1}$ with $N-1$ topological $\theta$ terms $\theta_a = 4\pi a/N$ ($a = 1, \ldots, N-1$) \cite{Tanizaki-S-18,Ohmori-S-S-19}.  The flag sigma model with topological angles  $\theta_a = 2\pi p a/N$ is also known to control the IR properties of  SU($N$) Heisenberg spin chain  in symmetric rank-$p$ tensor representation \cite{Wamer-L-M-A-20}.  We thus deduce that the weak-coupling analysis for the SU($N$) KHM at $1/N$-filling with $J_{\text{K}}<0$ is connected to the physics of the SU($N$) Heisenberg spin chain  in symmetric rank-$2$  tensor representation. When $N$ is even, the flag sigma model with topological angles $\theta_a = 4\pi a/N$ ($a = 1, \ldots, N-1$) is fully gapped with a ground-state degeneracy $N/2$  whereas a gapless behavior is expected in the odd-$N$ case \cite{Wamer-A-PRB20,Wamer-A-20}.

%
\end{document}